\documentclass{aa}
\usepackage{graphicx}
\usepackage{amssymb}
\usepackage{amsmath}
\usepackage{natbib}
\usepackage{mathrsfs}
\usepackage{ulem}
\usepackage{txfonts}
\usepackage{lmodern}
\usepackage{booktabs}
\usepackage{subfig}
\usepackage{array}
\usepackage{color}
\usepackage{lscape}
\usepackage{afterpage}
\usepackage{pdflscape}

\usepackage[switch]{lineno} % switch=Put the line into the inner margin
\newcommand{\corot}{{\textsc{CoRoT}}}

\newcommand{\kepler}{\textit{Kepler}}
\newcommand{\plato}{\textsc{PLATO}}

\newcommand{\ind}[1]{_{\rm #1}}

\def\m2s2{\,m$^{2}$\,s$^{-2}$} %m2.s -2
\def\aov{\alpha\ind{ov}}

\newcommand{\vect}[1]{\boldsymbol{\rm #1}}

\newcommand{\cesam}{\textsc{Cesam2k}}
\newcommand{\mesa}{\textsc{MESA}}

\newcommand{\losc}{\textsc{losc}}

\newcommand\T{\rule{0pt}{2.6ex}}
\newcommand\B{\rule[-1.2ex]{0pt}{0pt}}

\begin{document}
\title{Measuring the extent of convective cores in low-mass stars using \kepler\ data: toward a calibration of core overshooting}
%\subtitle{}
\titlerunning{Measuring the extent of convective cores in low-mass stars using \kepler\ data}
\author{
S. Deheuvels\inst{1,2}
\and I. Brand\~ao \inst{3,4}
\and V. Silva Aguirre \inst{5} 
\and J. Ballot \inst{1,2}
\and E. Michel \inst{6}
\and M.~S. Cunha \inst{3,4}
\and Y. Lebreton \inst{6,7}
\and T. Appourchaux \inst{9}
}
\institute{Universit\'e de Toulouse; UPS-OMP; IRAP; Toulouse, France 
\and CNRS; IRAP; 14, avenue Edouard Belin, F-31400 Toulouse, France
\and Instituto de Astrof\'{i}sica e Ci\^{e}ncias do Espa\c{c}o, Universidade do Porto, CAUP, Rua das Estrelas, 4150-762 Porto, Portugal
\and Centro de Astrof\'{\i}sica e Faculdade de Ci\^encias, Universidade do Porto, Rua das Estrelas, 4150-762 Porto, Portugal
\and Stellar Astrophysics Centre, Department of Physics and Astronomy, Aarhus University, Ny Munkegade 120, DK-8000 Aarhus C, Denmark
\and LESIA, CNRS UMR 8109, Universit\'e Pierre et Marie Curie, Universit\'e Denis Diderot, Observatoire de Paris, 92195 Meudon, France
\and Observatoire de Paris, GEPI, CNRS UMR 8111, F-92195 Meudon, France
\and Institut de Physique de Rennes, Universit\'e de Rennes 1, CNRS UMR 6251, F-35042 Rennes, France
\and Univ. Paris-Sud, Institut d'Astrophysique Spatiale, UMR 8617, CNRS, B\^atiment 121, 91405 Orsay Cedex, France
}

\offprints{S. Deheuvels\\ \email{sebastien.deheuvels@irap.omp.eu}
}

%\date{Submitted ...}

\abstract{Our poor understanding of the boundaries of convective cores generates large uncertainties on the extent of these cores and thus on stellar ages. The detection and precise characterization of solar-like oscillations in hundreds of main-sequence stars by \corot\ and \kepler\ has given the opportunity to revisit this problem.}
%
%{It is known that asteroseismology can provide constraints on the size of convective cores through the sensitivity of oscillation modes to the sharp gradient of chemical composition that builds up at the core edge. 
{Our aim is to use asteroseismology to consistently measure the extent of convective cores in a sample of main-sequence stars whose masses lie around the mass-limit for having a convective core.}
%The sharp gradients of chemical composition that develop at the boundaries of convective core are seen as glitches by acoustic modes, and it is known that the mode frequencies depend on the size of the convective core. Our aim is to }
%
{We first test and validate a seismic diagnostic that was proposed to probe in a model-dependent way the extent of convective cores using the so-called $r_{010}$ ratios, which are built with $l=0$ and $l=1$ modes. We apply this procedure to 24 low-mass stars chosen among \kepler\ targets to optimize the efficiency of this diagnostic. For this purpose, we compute grids of stellar models with both the \cesam\ and \mesa\ evolution codes, where the extensions of convective cores are modeled either by an instantaneous mixing or as a diffusion process.}
%where ad-hoc extensions of the convective core are included over distances that are controlled by a free parameter. 
%
{We find that 10 stars or our sample are in fact subgiants. Among the other targets, we are able to unambiguously detect convective cores in eight stars and we obtain seismic measurements of the extent of the mixed core in these targets with a good agreement between the \cesam\ and \mesa\ codes. By performing optimizations using the Levenberg-Marquardt algorithm, we then obtain estimates of the amount of extra-mixing beyond the core that is required in \cesam\ to reproduce seismic observations for these eight stars and we show that this can be used to propose a calibration of this quantity. This calibration depends on the prescription chosen for the extra-mixing, but we find that it should be valid also for the code \mesa, provided the same prescription is used.}
{This study constitutes a first step towards the calibration of the extension of convective cores in low-mass stars, which will help reduce the uncertainties on the ages of these stars.}


\maketitle

\section{Introduction}

The extent of chemically mixed regions associated to stellar convective cores is notoriously uncertain. Several physical processes that remain challenging to describe theoretically are known to extend convective cores beyond the theoretical Schwarzschild limit. The most often cited among them is core overshooting. According to Schwarzschild's criterion, the boundary of a convective core corresponds to the layer above which upward-moving convective blobs are braked. However, this criterion neglects the inertia of the ascending blobs, which are expected to penetrate over a certain distance (overshoot) inside the radiative zone. The theoretical complexity of this phenomenon is well illustrated by the large number of developments that were proposed to describe it (e.g. \citealt{saslaw65}, \citealt{shaviv71}, \citealt{roxburgh78}, \citealt{zahn91} to quote only a few) and by the diversity of the predicted distances $d_{\rm ov}$ over which convective eddies are expected to overshoot in the stable region (predicted values for $d_{\rm ov}$ range from 0 to 2 $H_P$, where $H_P$ is the local pressure scale height). Current numerical simulations of overshooting are encouraging, but they are still far from reproducing the very high turbulence of stellar convection and cannot yet be used to obtain reliable prescriptions for core overshooting (see \citealt{dintrans09} for a review). Another complication arises from the fact that convective cores can also be extended due to rotationally-induced mixing (see \citealt{maeder09} and references therein). As a result, it is still an open issue to determine (1) over which distance convective cores are extended, (2) what the temperature stratification is like in these core extensions, and (3) how chemical elements are mixed in these regions. 

Since convective cores constitute reservoirs for nuclear reactions, the uncertainty on their sizes generates significant uncertainties on stellar ages, especially near the end of the main sequence (MS). \cite{lebreton14b} for instance estimated that an extension of convective cores over a typical distance of 0.2 $H_P$ can generate errors on stellar ages as large as 30\% at the turnoff. It also affects the isochrones that have turnoff masses above $\sim1.1M_\odot$, and thus the age of rather young clusters. 

To account for the combined effects of core overshooting and rotational mixing, 1D stellar models often consider an ad-hoc extra mixing at the edge of the convective core, which is either modeled as an \textit{instantaneous} mixing (simple extension of the mixed core) or as a diffusion process (\citealt{ventura98}), i.e. as a \textit{non-instantaneous} mixing (see \citealt{noels10} for a review). In both cases, the extent of the extra-mixing (usually known as the \textit{overshooting distance} $d_{\rm ov}$ even though overshooting may not be the only mechanism at work) depends on one free parameter. These models are clearly overly simplistic but current observations have not yet permitted to constrain more complex models. The overshooting distance has been observationally constrained by fitting isochrones to the color-magnitude diagrams of open clusters (e.g. \citealt{maeder81}, \citealt{vandenberg06}) and by performing calibrations using eclipsing binaries (e.g. \citealt{claret07}, \citealt{stancliffe15}). These studies typically pointed toward an instantaneous mixing over a distance $d_{\rm ov}\sim0.2 H_P$ (where $H_P$ is the local pressure scale height) with rather large star-to-star variations. The case of low-mass stars (typically $M \lesssim 1.5\,M_\odot$) is known to be problematic within this formalism. Indeed, for stars with small convective cores the overshooting region becomes unrealistically large because $H_P(r)\rightarrow\infty$ when $r$ goes to zero. This prompted several authors to consider an overshoot parameter $\aov$ that increases with stellar mass in the approximate mass range $1.1M_\odot\lesssim M\lesssim1.5M_\odot$ (e.g. \citealt{pietrinferni04}, \citealt{bressan12}). In these cases, an ad-hoc linear increase of $\aov$ as a function of $M$ was chosen, with some success in reproducing the turnoff of clusters with turnoff-masses around 1.3 $M_\odot$ (\citealt{pietrinferni04}). The problem remains however poorly constrained in this range of mass, and the use of eclipsing binary systems for this purpose is unfortunately of little help (\citealt{valle16}).

Recently, constraints on the extent of the extra mixing beyond convective cores has been obtained from asteroseismology. Sharp variations in the mean molecular weight profile at the boundary of the mixed core create a glitch to which oscillation modes are sensitive, which can be used to measure the extent of the mixed region associated to convective cores. This approach has been successfully applied to solar-like pulsators in the main sequence (\citealt{deheuvels10a},  \citealt{goupil11}, \citealt{silva13}, \citealt{guenther14}, \citealt{appourchaux15}), in the subgiant phase \citep{lanzarote,deheuvels11} and to several main-sequence B stars (e.g. \citealt{degroote10}, \citealt{neiner12}, \citealt{moravveji15}). All these studies reported the need for extended convective cores and confirmed the great potential of asteroseismology to measure this extension. However, we are still lacking consistent seismic studies of larger samples of stars, which are needed to better understand how the overshooting distance varies with stellar parameters.%and to obtain a calibration of this quantity.

In this paper, we took advantage of the detection of solar-like oscillations in hundreds of solar-like pulsators with an unprecedented level of precision by the space mission \kepler\ (\citealt{borucki10}) to consistently measure the extent of the convective core in a larger sample of stars. We have focused on stars whose masses lie around the mass-limit for having a convective core ($M\gtrsim1.1M_\odot$ at solar metallicity). For these stars, a large part of the core luminosity comes from the burning of $^3$He outside of equilibrium. Core overshooting can considerably increase the abundance of $^3$He in the core, and therefore also the core luminosity, size, and lifetime (\citealt{roxburgh85}, \citealt{deheuvels10a}). For instance an instantaneous overshooting over a distance of $0.1\,H_P$ in a 1.3-$M_\odot$ star generates an increase of as much as 50\% in the convective core radius during the main sequence\footnote{This is shown in Fig. \ref{fig_cc_cesam_mesa} of this paper, but note that this depends on the exact prescription that is adopted for core overshooting, as is discussed in Sect. \ref{sect_calibrate}.}. As a consequence, these stars are particularly good tracers of the existence and amount of core overshooting.

It has been shown in previous studies that the small separations built with $l=0$ and $l=1$ modes are particularly sensitive to the structure of the core (\citealt{provost05}, \citealt{deheuvels10a}, \citealt{silva11}), and that their ratios $r_{010}$ to the large separations are nearly insensitive to the so-called near-surface effects (\citealt{roxburgh03}). In Sect. \ref{sect_test_d01}, we show that this diagnostic can be used to obtain a model-dependent estimate of the extent of the mixed core by building a grid of models with the evolution code \cesam\ (\citealt{morel08}). We then select a subsample of 24 solar-like pulsators among \kepler\ targets that are the most likely to provide constraints on the amount of core overshooting based on the results of our grid of models, and we extract their mode frequencies from their oscillation spectra in Sect. \ref{sect_analysis}. In Sect. \ref{sect_grid}, we confront the observed ratios $r_{010}$ to those of two grids of models computed with \cesam\ and \mesa\ (\citealt{paxton11}). We consistently detect convective cores in eight of the selected targets and we obtain measurements of the extent of the mixed core in these stars. In Sect. \ref{sect_calibrate} we discuss the different existing prescriptions for core overshooting in low-mass stars, we show how our results can be used to calibrate the prescription used in the code \cesam, and we address the question whether such a calibration can be adapted in \mesa.

\section{Estimating the core size with seismology \label{sect_test_d01}}

\subsection{Asteroseismic diagnostics}

A sharp gradient of the mean molecular weight $\mu$ builds up at the boundary of the homogeneous convective core, which induces rapid variations in the sound speed profile, and even makes it discontinuous in the case of a growing core without microscopic diffusion. It is well known that such a glitch in $c(r)$ adds an oscillatory modulation to the expression of the mode frequencies as a function of the radial order. The period of this modulation is directly related to the depth of the glitch (\citealt{gough90}). This is not specific to the boundary of convective cores, and such acoustic glitches can also be produced by the base of convective envelopes or the helium ionization regions.

When the period of this oscillation is smaller than the frequency range of the observed frequencies, the acoustic depth of the glitch can be estimated in a model-independent way. I has been recently shown that the depth of the second helium ionization zone and the base of the convective envelope can be estimated with such a diagnostic (\citealt{lebreton12}, \citealt{mazumdar14}). Unfortunately, the glitch caused by convective cores induces a longer-period oscillation and only a fraction of the period can be observed. This makes it much more difficult to obtain model-independent information about the boundary of convective cores. \cite{cunha11} and \cite{brandao14} showed that the amplitude of the sound speed discontinuity at the core edge may be recovered in some favorable cases. It is not clear whether a model-independent estimate of the extent of the mixed core can be obtained. 

However, it has been shown by several studies that a model-dependent measurement of the core size can be obtained through seismology. Combinations of mode frequencies built with $l=0$ and $l=1$ modes are well suited for this type of study because they are particularly sensitive to the core structure (\citealt{provost05}, \citealt{deheuvels10a}). \cite{roxburgh03} advised to use the five-point separations $d_{01}$ and $d_{10}$ defined as
\begin{align}
d_{01}(n) & =  \frac{1}{8} \,(\nu_{0,n-1}-4\nu_{1,n-1}+6\nu_{0,n}-4\nu_{1,n}+\nu_{0,n+1})  \label{eq_d01} \\
d_{10}(n) & =  -\frac{1}{8} \, (\nu_{1,n-1}-4\nu_{0,n}+6\nu_{1,n}-4\nu_{0,n+1}+\nu_{1,n+1})  \label{eq_d10}.
\end{align}
They showed that the ratios between these small separations and the large separations constructed as 
\begin{eqnarray}
r_{01}(n) & = & \frac{d_{01}(n)}{\Delta\nu_1 (n)} \\
r_{10}(n) & = & \frac{d_{10}(n)}{\Delta\nu_0 (n+1)}
\end{eqnarray}
where $\Delta\nu_l (n) = \nu_{l,n}- \nu_{l,n-1}$ are largely insensitive to the structure of the outer layers, which makes them almost immune to the so-called near-surface effects. These ratios, referred to as $r_{010}$ when combined together, have been used e.g. to estimate the depth of the convective envelope and the second helium ionization zone in the Sun (\citealt{roxburgh09a}) or to establish the existence of a convective core in a \kepler\ target (\citealt{silva13}). We note that \cite{cunha07} proposed to use a combination of frequencies using modes of degrees up to 3 ($dr_{0213}$), which can interestingly be related to the intensity of the sound speed jump at the edge of growing cores. However, $l=3$ modes have low amplitudes in stars other than the Sun, and although several detections of such modes have been obtained (e.g. \citealt{deheuvels10b}, \citealt{metcalfe10}), it remains exceptional to reliably estimate their frequencies over several consecutive radial orders. In this study, we have tested and used the diagnostic based on the $r_{010}$ ratios.

\begin{figure}
\begin{center}
\includegraphics[width=9cm]{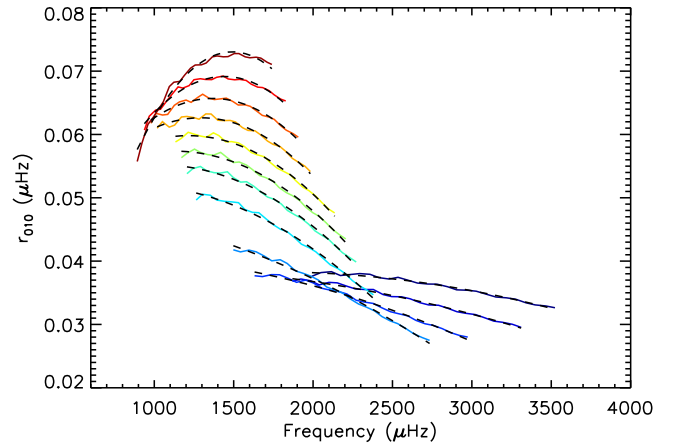}
\end{center}
\caption{Variations in the ratio $r_{010}$ around $\nu_{\rm max}$ as a function of frequency for models of 1.2 $M_\odot$ from the ZAMS (dark blue) to the beginning of the post main sequence (dark red). The dashed lines correspond to fits of $2^{\rm nd}$ order polynomials.
\label{fig_ratio_parabola}}
\end{figure}

Fig. \ref{fig_ratio_parabola} shows the behavior of the $r_{010}$ ratios for a model of 1.2 $M_\odot$ evolved from the zero-age main sequence (ZAMS) to the beginning of the subgiant phase. The ratios are represented only in the frequency range where modes are expected to be observed, i.e. over about 12 radial orders around the frequency of maximum power of the oscillations $\nu_{\rm max}$. As mentioned above, only a fraction of the period of the oscillation induced by the edge of the core can be observed, and the $r_{010}$ ratios can in fact be well approximated by second-order polynomials throughout the MS, as can be seen in Fig. \ref{fig_ratio_parabola}. 

Several studies have shown that the slope and mean value of $r_{010}$ ratios are a good indicator of the size of the mixed core (\cite{popielski05}, \citealt{deheuvels10a}, \citealt{silva11}). However, these previous studies either focused on a particular star or worked with models that share the same physical properties other than the mixing at the edge of the core. We know that several other parameters, such as the abundance of heavy elements, have a significant impact on the size of the convective core. We here aimed at testing the efficiency of this diagnostic tool.

%Is this diagnostic still efficient when a wide range of these parameters is taken into account? \cite{brandao14} addressed this question by computing grids of models with three different metallicities and found interestingly that if the slope of the $r_{010}$ ratio exceeds a certain limit, the star must have a convective core, regardless of the metal content. However, one is also interested in the possibility of measuring the size of the mixed region associated to the convective core, if it exists. Besides, the mean value of the $r_{010}$ ratio carries information about the star that are complementary to its slope (it depends in particular on the amount of hydrogen left in the core) and can advantageously be added to the diagnostic. Finally, this latter study focuses on stars that are in the linear domain and is thus biased toward young MS stars. 

%For this purpose, we computed a grid of models. The goal of this grid is twofold:
% \begin{enumerate}
%\item determine criteria to select among \kepler\ targets the stars that are the most promising; an analysis of the oscillation spectra of the selected targets will then provide estimates of their $r_{010}$ ratios 
%\item interpret these data by confronting them to the grid of models and obtain constraints on the amount of core overshooting whenever it is possible
%\end{enumerate}

\subsection{Testing the diagnostic of $r_{010}$ ratios \label{sect_test_diagnostic}}

\subsubsection{Description of the grid \label{sect_descript_grid}}

To determine in which circumstances the extent of the core can be estimated with the $r_{010}$ ratios, we computed a grid of models using the stellar evolution code \cesam\ (\citealt{morel08}). 

We used the OPAL 2005 equation of state and opacity tables as described in \cite{lebreton08}. The nuclear reaction rates were computed using the NACRE compilation (\citealt{angulo99}) except for the $^{14}N(p,\gamma)^{15}O$ reaction where we adopted the revised LUNA rate (\citealt{formicola04}). The atmosphere was described by Eddington's gray law. We assumed the classical solar mixture of heavy elements of \cite{asplund09} (hereafter AGSS09). Convection was treated using the Canuto-Goldman-Mazzitelli (CGM) formalism (\citealt{canuto96}). This description involves a free parameter, the mixing length, which is taken as a fraction $\alpha\ind{CGM}$ of the pressure scale height $H_P$.  We here assumed a value of $\alpha\ind{CGM}$ calibrated on the Sun ($\alpha_\odot=0.64$, \citealt{samadi06}). 

\begin{figure}
\begin{center}
\includegraphics[width=9cm]{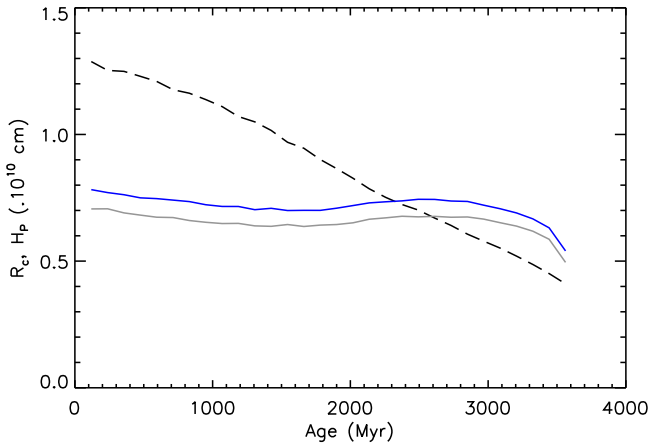}
\end{center}
\caption{Variations in the pressure scale height $H_P$ (dashed line) and the radius of the extended convective core $R_{\rm c}$ (solid blue line) with age for a 1.3 $M_\odot$ \cesam\ model with solar metallicity, a solar-calibrated value for the mixing length, $Y_0=0.26$, and $\aov=0.1$. The gray solid line indicates the Schwarzschild limit.
\label{fig_rc_hp}}
\end{figure}

To account for the physical processes that could increase the size of convective cores, we considered an instantaneous mixing beyond convective cores over a distance $d_{\rm ov}$ taken as a fraction $\aov$ of the pressure scale height $H_P$. The free parameter $\alpha_{\rm ov}$ is as often referred to as the \textit{overshoot parameter}. In order to avoid the overshooting region from unrealistically extending over a distance as large as the core itself, \cesam\ models define the overshooting distance as 
\begin{equation}
d_{\rm ov} = \aov \times\min(H_P, r_{\rm{s}})
\label{eq_dov}
\end{equation}
where $r_{\rm{s}}$ is the Schwarzschild limit of the core. We note that this is the case during most of the main sequence for stars with masses $\lesssim1.5M_\odot$, as shown by Fig. \ref{fig_rc_hp}. We have imposed the adiabatic temperature gradient in the overshoot region.

Microscopic diffusion is known to increase the abundance in heavy elements in the core as the star evolves, and thus to increase the size of convective cores. In this section, microscopic diffusion is not included in the models so that the core extension imposed by the overshoot parameter $\alpha_{\rm ov}$ can be partly attributed to its effects. The contribution from microscopic diffusion is addressed in Sect. \ref{sect_grid}.

The grid was computed with masses ranging from 0.9 to 1.5 $M_\odot$ (step 0.05 $M_\odot$), metallicities from $-0.4$ to 0.4 dex (step 0.1 dex), and two values of the initial helium abundance (0.26 or 0.30). Models were computed for values of $\aov$ ranging from 0 to 0.3 (step 0.05). For each evolutionary sequence, the mode frequencies were computed with the oscillation code \losc\ (\citealt{losc}) for about 60 models between the ZAMS and the beginning of the subgiant phase. We stopped the evolution as soon as mixed modes appear around $\nu_{\rm max}$, because these modes cause brutal variations in the $r_{010}$ ratios and prevent them from being directly used as a diagnostic for the core size.

\begin{figure*}
\begin{center}
\includegraphics[width=8cm]{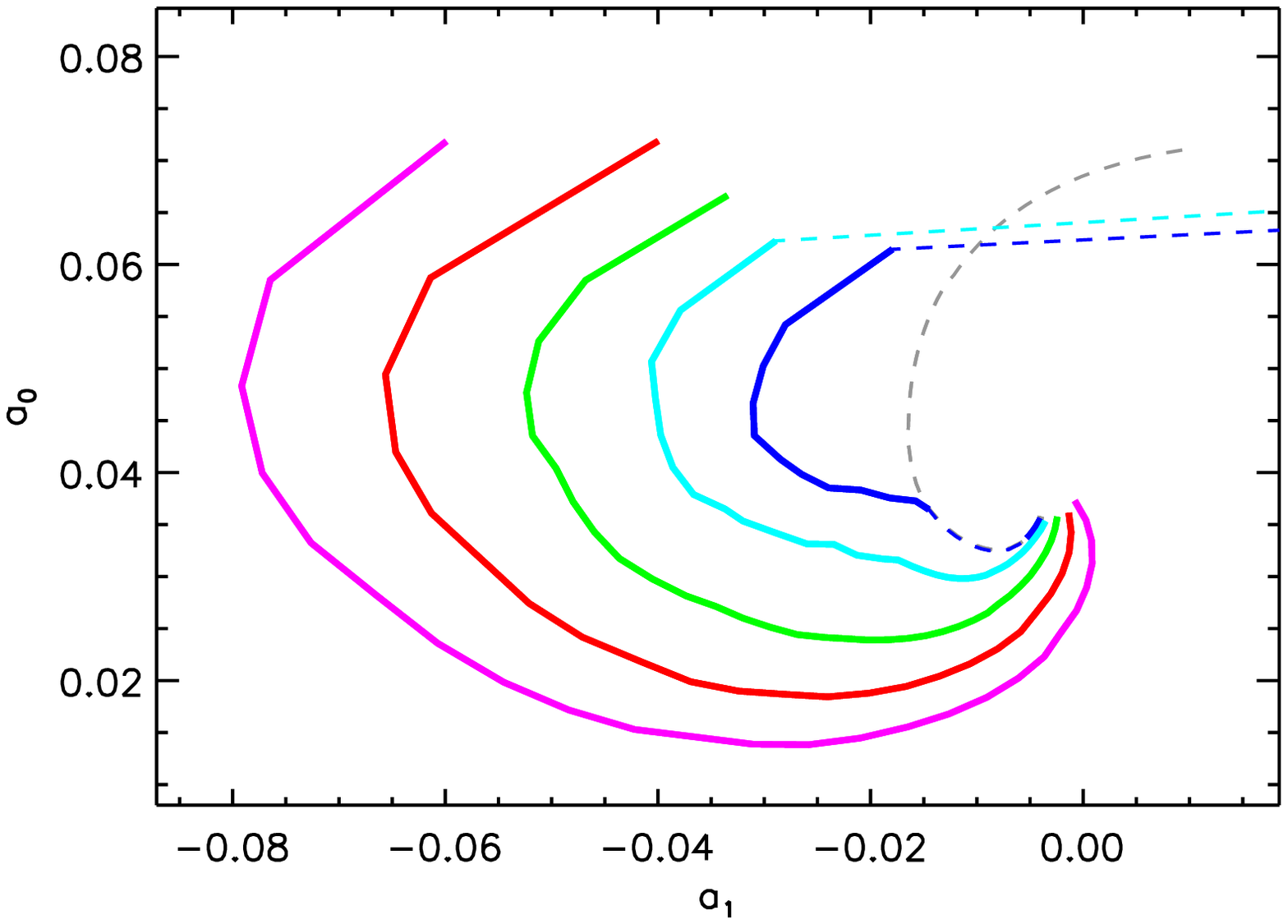}
\includegraphics[width=8cm]{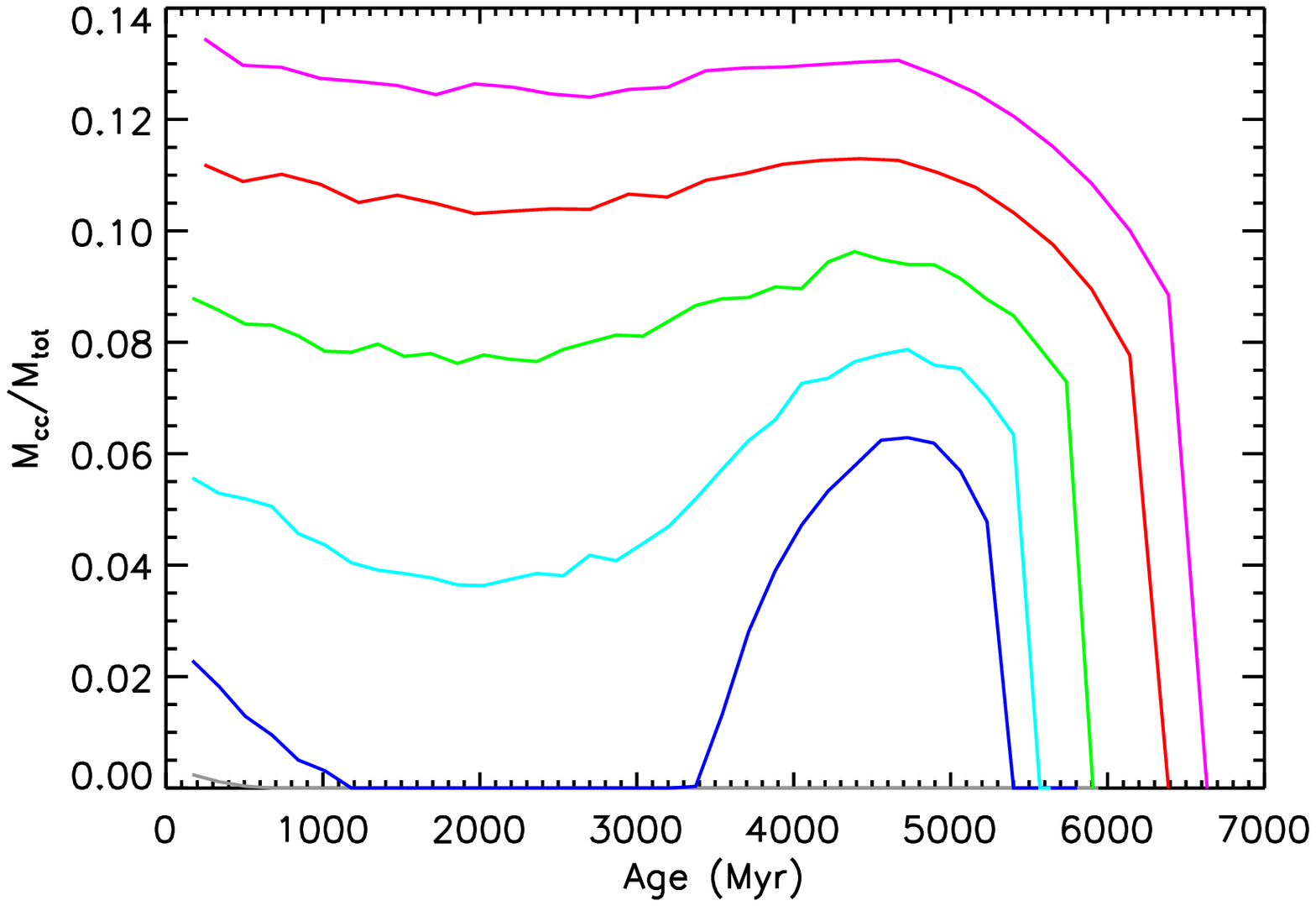}
\end{center}
\caption{\textbf{Left}: Evolutionary tracks of stellar models of 1.2 $M_\odot$ in the $(a_1,a_0)$ plane for different amounts of core overshooting: $\aov=0$ (gray), 0.1 (blue), 0.15 (cyan), 0.2 (green), 0.25 (red), and 0.3 (purple). Full (resp. dashed) lines indicate that the model has a convective (resp. radiative) core. \textbf{Right}: Variations in the size of the convective core as a function of age for the same models.
\label{fig_evol_ratio}}
\end{figure*}

For each of the models along the evolutionary tracks, we fitted 2$^{\rm nd}$ order polynomials of the type
\begin{equation}
P(\nu) = a_0 + a_1 (\nu-\beta) + a_2(\nu-\gamma_1)(\nu-\gamma_2)
\label{eq_poly}
\end{equation}
to the $r_{010}$ ratios. The parameters $\beta$, $\gamma_1$, and $\gamma_2$ were chosen to ensure that $P(\nu)$ is a sum of orthogonal polynomials for each model. The fits were performed in the approximate frequency range where modes are expected to be observed, i.e. about 12 orders around the frequency of maximum power of oscillations $\nu_{\rm max}$. This latter frequency was estimated for stellar models by assuming that it scales as the acoustic cutoff frequency. This assumption, which is the basis of the so-called seismic scaling relations, was observationally verified to work at the level of a few percent at least (\citealt{stello08}, \citealt{huber11}, \citealt{silva12}), and is gaining theoretical support (\citealt{belkacem11}). We note that during most of the MS, the $r_{010}$ ratios vary roughly linearly with frequency in the range of observed frequencies, so that the coefficient $a_2$ of the fit is negligible. 

\subsubsection{Evolutionary tracks in the $(a_1, a_0)$ plane}

Before commenting on the results of the grid, we show as an example the evolutionary tracks in the $(a_1, a_0)$ plane (slope versus mean value) of 1.2-$M_\odot$ models for different amounts of core overshooting (Fig. \ref{fig_evol_ratio}a). 
For comparison, the variations in the size of the convective core for the same models are shown as a function of age in Fig. \ref{fig_evol_ratio}b. 
As mentioned by \cite{silva11}, the trajectory of models in the $(a_1, a_0)$ plane depends in a complex way on the evolutionary stage, the size of the convective core, and the amplitude of the glitch in the sound speed. However, we can still broadly understand it. At the beginning of the MS, the stars with different $\alpha_{\rm ov}$ start roughly at the same point in the $(a_1, a_0)$ plane (bottom right corner in Fig. \ref{fig_evol_ratio}a). Indeed, the $\mu$-gradient at the edge of the core has not had time to build up yet, so the $r_{010}$ ratios are still nearly independent from the size of the convective core. As the star evolves, the glitch in the sound speed profile builds up, which causes the amplitude of the oscillations of the $r_{010}$ ratios to increase. Therefore both the mean value $a_0$ and the absolute value of the slope $|a_1|$ of the ratios increase. But also, as the star evolves, its $\nu_{\rm max}$ frequency decreases. As a result, the range of observable frequencies shifts to a different part of the oscillation produced by the glitch. As can be seen in Fig. \ref{fig_ratio_parabola}, when stars reach the end of the MS, the $r_{010}$ ratios lie around a maximum of this oscillation, which results in a decrease of the absolute value of the slope $|a_1|$. For post-main sequence stars, the mean slope $a_1$ even becomes positive. This explains why the evolutionary tracks of models in the $(a_1, a_0)$ plane are vaguely circular, as can be seen in Fig. \ref{fig_evol_ratio}. 

For models with larger amounts of overshooting, the convective core is larger. Therefore, the period of the oscillation caused by the glitch is shorter and the absolute mean slope $|a_1|$ of the $r_{010}$ ratios is larger. As a result, stars with larger $\aov$ are shifted to the left in the $(a_1, a_0)$ plane, and they draw larger circles. This confirms previous statements that the position in the $(a_1,a_0)$ plane is discriminant for the size of the core if all parameters other than $\aov$ are fixed.

%The value of the slope $a_1$ depends on both the size of the mixed core and the amplitude of the glitch in $c(r)$. Models show that the position in the $(a_1,a_0)$ plane is discriminant.

\subsubsection{Results of the grid}

\begin{figure*}
\begin{center}
\includegraphics[width=8cm]{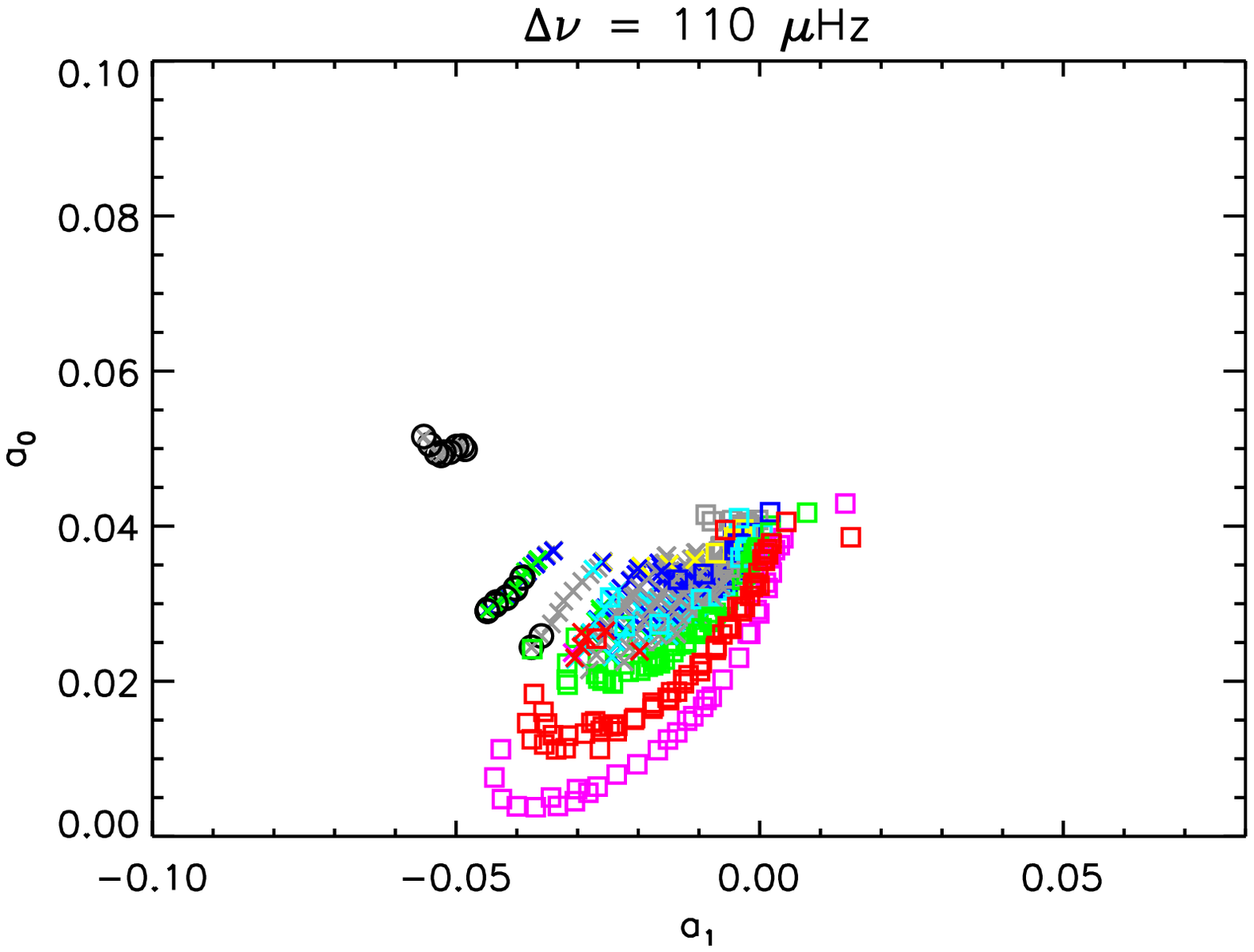}
\includegraphics[width=8cm]{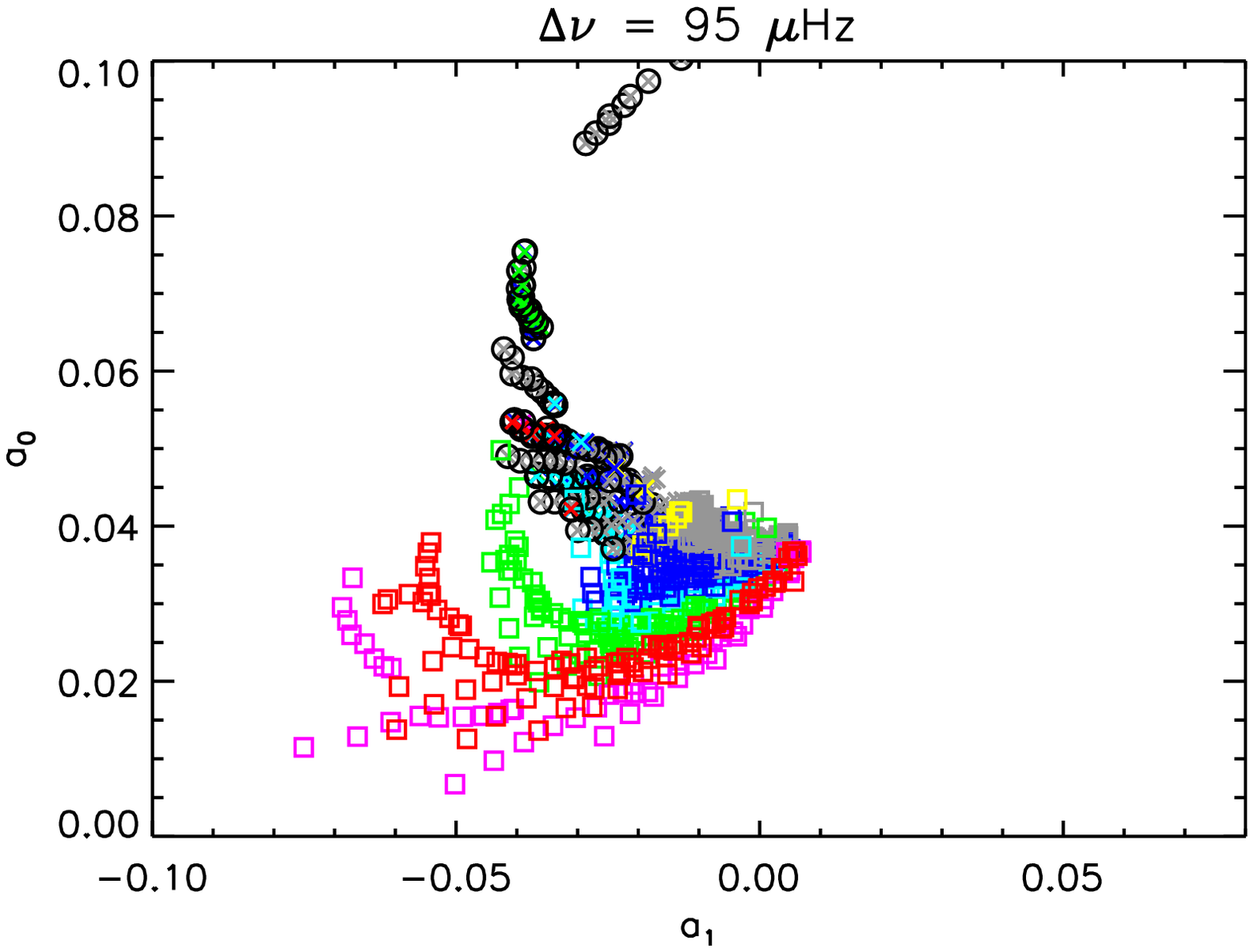}
\includegraphics[width=8cm]{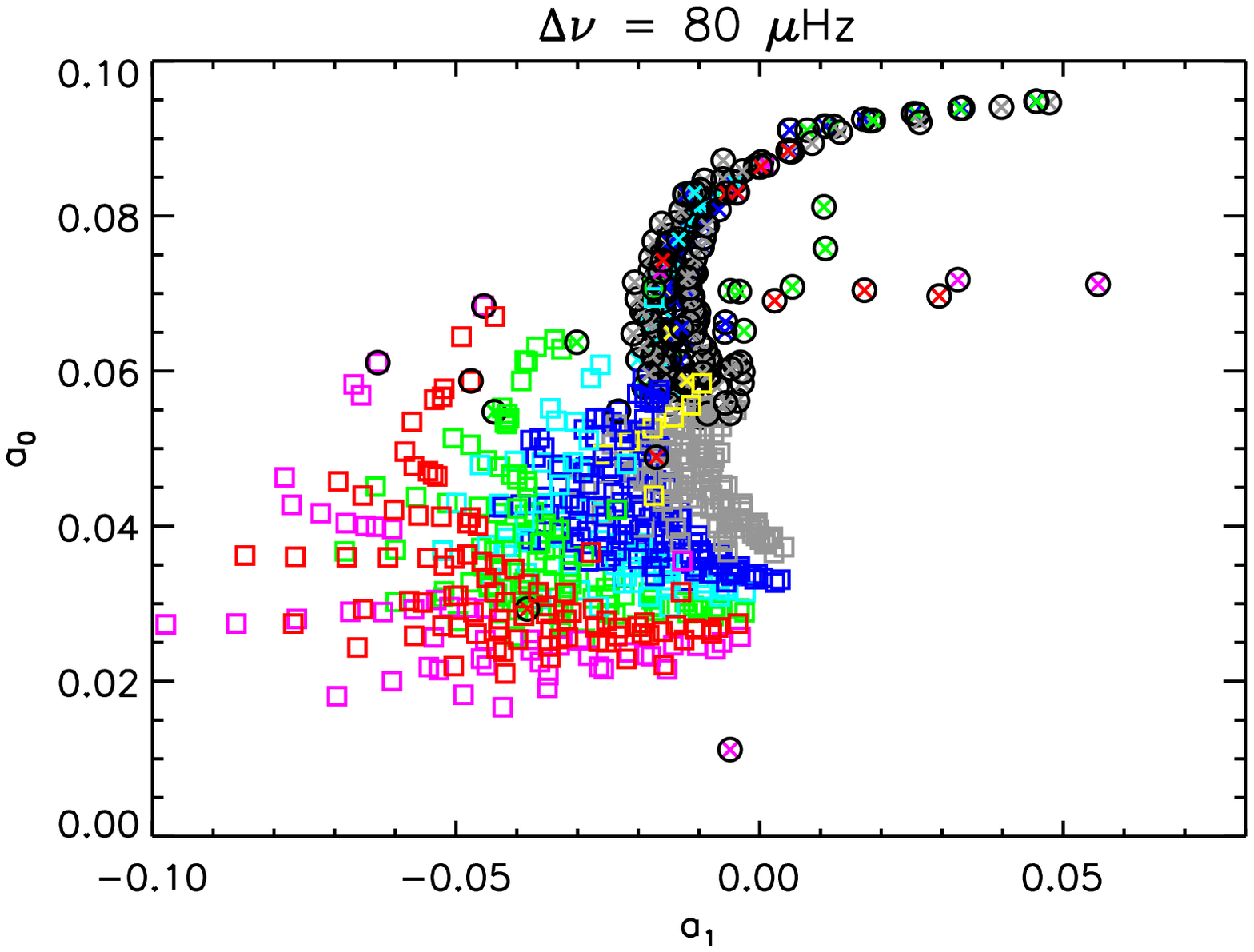}
\includegraphics[width=8cm]{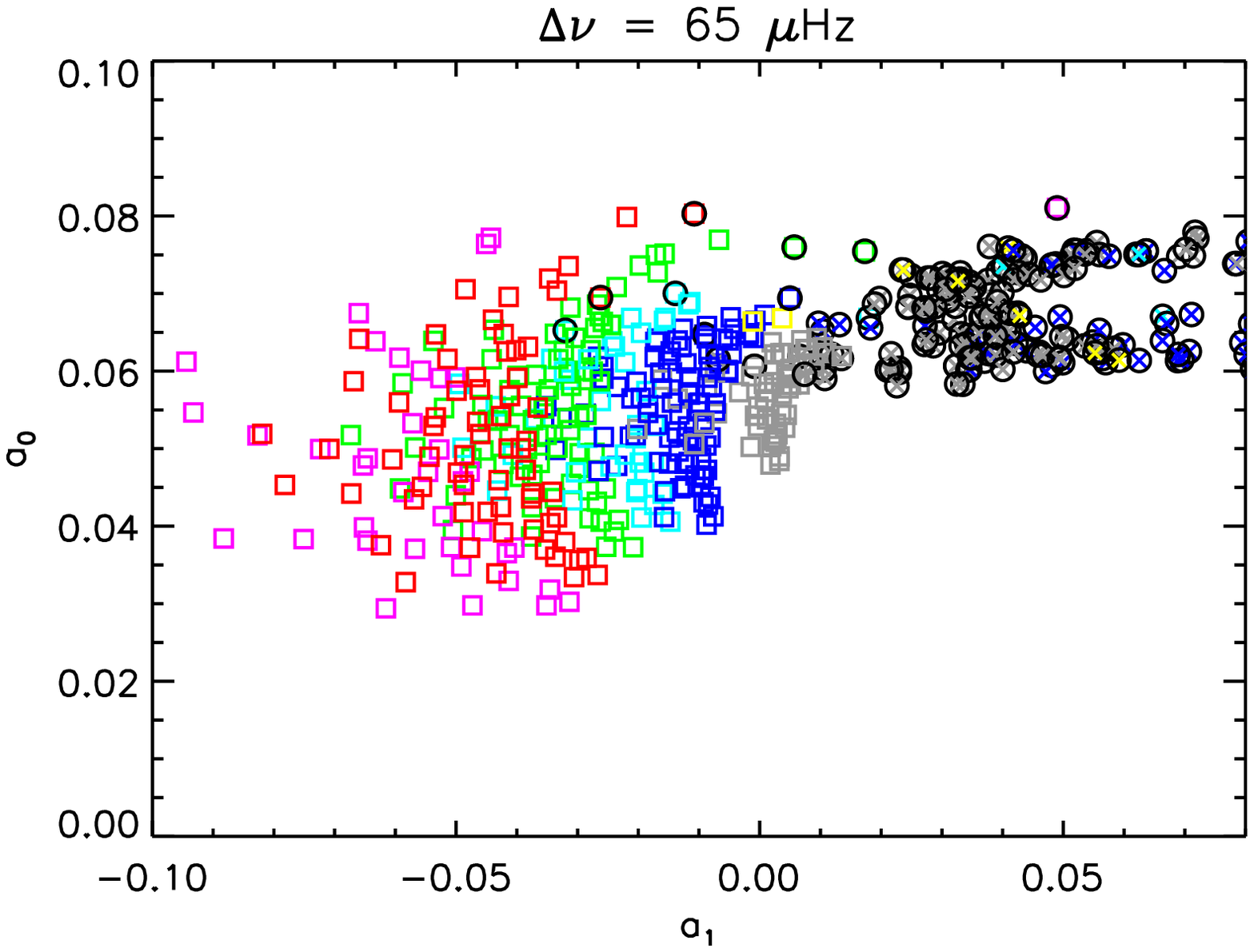}
\end{center}
\caption{Location of models in the $(a_1,a_0)$ plane at fixed $\Delta\nu$. Colors indicate the amount of core overshooting: $\aov=0$ (gray), 0.1 (blue), 0.15 (cyan), 0.2 (green), 0.25 (red), 0.3 (magenta). Open squares indicate models with a convective core, and crosses, models with radiative cores. The black open circles indicate models that are in the post-main-sequence ($X_{\rm c} < 10^{-2}$).
\label{fig_ratio_trend}}
\end{figure*}

When solar-like oscillations are detected in a star, it is usually straightforward to estimate the mean large separation of its acoustic modes $\Delta\nu$. We thus chose to show the results of the grid at fixed values of $\Delta\nu$. This time, each evolutionary sequence of our grid is represented as a dot in the $(a_1,a_0)$ plane, provided its large separation matches the chosen value of $\Delta\nu$ at some point along the evolution.

Fig. \ref{fig_ratio_trend} shows the location of the models in the $(a_1,a_0)$ plane for four values of $\Delta\nu$ : 110, 95, 70, and 65 $\mu$Hz. For $\Delta\nu= 110\,\mu$Hz (top left plot), there is a relative degeneracy of the models in the $(a_1,a_0)$ plane. This can be understood because only low-mass unevolved stars reach such a high value of $\Delta\nu$. Higher-mass stars begin the MS with a lower $\Delta\nu$, and this quantity further decreases as the star evolves\footnote{For instance, 1.25-$M_\odot$ stars at solar metallicity reach the ZAMS with $\Delta\nu\sim110\,\mu$Hz, so more massive stars never reach this value.}. For this reason, few of the stars with $\Delta\nu=110\,\mu$Hz have a convective core. And those that have one are still close to the ZAMS, so the $\mu$-gradient has not had time to build up yet and the $r_{010}$ ratio still does not feel it. The diagnostic is thus less efficient for $\Delta\nu\gtrsim110\,\mu$Hz.

For lower values of $\Delta\nu$, different populations are represented: (1) evolved low-mass stars (in the PoMS for the lowest masses) and (2) MS higher-mass stars. In these cases, Fig. \ref{fig_ratio_trend} clearly shows that the location of a model in the $(a_1,a_0)$ plane can be used to estimate:
\begin{itemize}
\item \textbf{the evolutionary state}: as mentioned before, when stars leave the MS, the mean slope $a_1$ of the ratios increases and becomes positive. As a result, PoMS models occupy a place in the $(a_1,a_0)$ plane that is increasingly distinct from that of MS models, as the large separation decreases. This opens the possibility to determine the evolutionary status of a star from its location in the $(a_1,a_0)$ plane. 
%This may seem useless, as usually the post-main-sequence (PoMS) status of solar-like pulsators can be established by the presence of mixed modes in their oscillation spectra, but in fact in lower-mass stars mixed modes do not reach frequencies around $\nu_{\rm max}$ immediately after the turnoff. In this study we indeed find several PoMS stars identified by their location in the $(a_1,a_0)$ plane, and which do not show any detected $l=1$ mixed modes (see Sect. \ref{sect_kepler}).
\item \textbf{the existence and the size of the convective core}: for stars with large separations below $\sim$ 95 $\mu$Hz, models with $\aov=0$, 0.1, 0.2, and 0.3 occupy distinct regions in the $(a_1,a_0)$ plane, which suggests that it should be possible to measure the size of the mixed core by using the location of the star in this plane.
\end{itemize}
We stress that the effects of metallicity on the size of the core are here taken into account in a very conservative way, since the models of Fig. \ref{fig_ratio_trend} include a wide range of metallicities ($-0.4$ to 0.4 dex). In practice, the metallicity of an observed star is usually known with a much better accuracy if spectroscopic measurements are available. We thus conclude that the $r_{010}$ ratios are in principle an efficient tool to measure the size of convective cores, provided the observed star is evolved enough to have developed a glitch in the sound speed at the edge of the core. 

%For even lower values of $\Delta\nu$, a lot of the targets are already in the PoMS. Therefore, mixed modes appear and the $r_{010}$ ratio can no longer be used. Besides, stars that have these lower $\Delta\nu$ are hotter and their analysis is more problematic (see Sect. \ref{sect_analysis}).

\section{Extracting the $r_{010}$ ratios from \kepler\ targets \label{sect_analysis}}

\subsection{Selection of targets}

Based on the tests performed on stellar models in Sect. \ref{sect_test_diagnostic}, we established a set of criteria to select \kepler\ targets for which the $r_{010}$ ratios should provide a good diagnostic for the core structure. We selected stars for which
\begin{itemize}
\item the mean large separation is below $110\,\mu$Hz, so that the diagnostic tool is efficient
\item no mixed modes are contaminating the $r_{010}$ ratios
\item a long enough data set is available, so that a good precision can be attained in the estimates of the parameters $a_i$.  Even with 9 months of \kepler\ data, the $r_{010}$ ratios of a target studied by \cite{silva13} were contaminated by a spurious increase in the low-signal-to-noise part of the spectrum. To avoid these features that might bias our estimates of the $a_i$ parameters, we selected only stars that were observed for at least 9 months.
\item the observed modes are narrow enough: we excluded F stars, whose modes are too wide to unambiguously distinguish the $l=1$ ridge from the $l=0$ and $l=2$ ridges in an \'echelle diagram. We note that Bayesian methods have been proposed to identify the degree of the modes and extract the mode frequencies even in these cases (e.g. \citealt{benomar09a}). However, this type of analysis requires dedicated works, which can be undertaken as an interesting follow-up of this work to explore the sizes of convective cores in higher-mass stars.
%Since the mode width is strongly correlated with the effective temperature (\citealt{appourchaux12}), our sample is biased toward cold stars ($T_{\rm eff}\lesssim6300$ K).
\end{itemize}

We applied these criteria to the solar-like pulsators whose global parameters were determined by \cite{chaplin14} and obtained a list of 24 targets, which are given in Table \ref{tab_param}. 
%For each star, we specified the estimates of $\Delta\nu$ and $\nu_{\rm max}$ obtained by \cite{chaplin14}, and their estimated mass derived from scaling laws. 
Most of these stars were also observed spectroscopically from the ground, which yielded estimates of the effective temperature and of the surface metallicity. When available, these measurements are specified in Table \ref{tab_param}.

\begin{table*}
  \begin{center}
  \caption{Global parameters of the selected targets. \label{tab_global_param} \label{tab_param}}
\begin{tabular}{l c c c c c c}
\hline \hline
\T \B KIC ID & $\Delta\nu$ ($\mu$Hz) & $\nu\ind{max}$ ($\mu$Hz) & $T_{\rm eff}^{\rm photo}$ (K)$^{\rm a}$ & $T_{\rm eff}^{\rm spectro}$ (K)$^{\rm b}$ & $[$Fe/H$]$ (dex)$^{\rm b}$ & $M/M_\odot$  \\
\hline
%\T  9139151 & $117.32  \pm 0.04$ & $2665  \pm 60$ & $6134  \pm  48$ & $6125  \pm  60$ & $ 0.11  \pm 0.06$ & $1.27  \pm0.11$ \\
\T    8394589 & $109.44  \pm 0.04$ & $2373  \pm 39$ & $6251  \pm  54$ & $6114  \pm  60$ & $-0.36  \pm 0.06$ & $1.18  \pm0.08$ \\
    9098294 & $108.92  \pm 0.03$ & $2282  \pm 26$ & $6020  \pm  51$ & $5840  \pm  60$ & $-0.13  \pm 0.06$ & $1.00  \pm0.05$ \\
    9410862 & $107.21  \pm 0.08$ & $2278  \pm 42$ & $6230  \pm  53$ & - & - & $1.17  \pm0.08$ \\
    6225718 & $106.00  \pm 0.03$ & $2316  \pm 38$ & - & $6230  \pm  60$ & $-0.17  \pm 0.06$ & $1.29  \pm0.08$ \\
   10454113 & $105.55  \pm 0.07$ & $2394  \pm 75$ & $6197  \pm  45$ & $6120  \pm  60$ & $-0.06  \pm 0.06$ & $1.41  \pm0.16$ \\
    6106415 & $104.20  \pm 0.02$ & $2224  \pm 25$ & - & $5990  \pm  60$ & $-0.09  \pm 0.06$ & $1.15  \pm0.06$ \\
   10963065 & $103.15  \pm 0.03$ & $2180  \pm 22$ & $6316  \pm  45$ & $6060  \pm  60$ & $-0.20  \pm 0.06$ & $1.15  \pm0.05$ \\
    6116048 & $100.72  \pm 0.02$ & $2081  \pm 22$ & $6072  \pm  49$ & $5935  \pm  60$ & $-0.24  \pm 0.06$ & $1.07  \pm0.05$ \\
    5184732 & $ 95.64  \pm 0.02$ & $2080  \pm 21$ & $5841  \pm 290$ & $5840  \pm  60$ & $ 0.38  \pm 0.06$ & $1.28  \pm0.06$ \\
    3656476 & $ 93.16  \pm 0.02$ & $1910  \pm 10$ & $5684  \pm  56$ & $5710  \pm  60$ & $ 0.34  \pm 0.06$ & $1.06  \pm0.03$ \\
    7296438 & $ 88.68  \pm 0.04$ & $1848  \pm 16$ & $5749  \pm  56$ & - & - & $1.18  \pm0.05$ \\
    4914923 & $ 88.58  \pm 0.02$ & $1800  \pm 15$ & - & $5905  \pm  60$ & $ 0.17  \pm 0.06$ & $1.14  \pm0.05$ \\
   12009504 & $ 88.38  \pm 0.04$ & $1848  \pm 22$ & $6270  \pm  61$ & $6065  \pm  60$ & $-0.09  \pm 0.06$ & $1.30  \pm0.07$ \\
    8938364 & $ 85.59  \pm 0.02$ & $1652  \pm 10$ & $5965  \pm  62$ & $5630  \pm  60$ & $-0.20  \pm 0.06$ & $0.94  \pm0.03$ \\
    7680114 & $ 85.18  \pm 0.02$ & $1697  \pm 10$ & $5800  \pm  56$ & $5855  \pm  60$ & $ 0.11  \pm 0.06$ & $1.11  \pm0.04$ \\
   10516096 & $ 84.43  \pm 0.03$ & $1666  \pm 12$ & $6123  \pm  48$ & $5940  \pm  60$ & $-0.06  \pm 0.06$ & $1.11  \pm0.04$ \\
    7206837 & $ 79.10  \pm 0.07$ & $1653  \pm 23$ & $6392  \pm  59$ & $6304  \pm  60$ & $ 0.14  \pm 0.06$ & $1.54  \pm0.09$ \\
    8176564 & $ 77.86  \pm 0.08$ & $1518  \pm 10$ & $6109  \pm  51$ & - & - & $1.21  \pm0.05$ \\
    8694723 & $ 75.22  \pm 0.04$ & $1431  \pm 11$ & $6351  \pm  62$ & $6120  \pm  60$ & $-0.59  \pm 0.06$ & $1.17  \pm0.05$ \\
   12258514 & $ 74.96  \pm 0.02$ & $1491  \pm 13$ & $5990  \pm  85$ & $5990  \pm  60$ & $ 0.04  \pm 0.06$ & $1.29  \pm0.06$ \\
    6933899 & $ 72.26  \pm 0.02$ & $1377  \pm  8$ & $5841  \pm  56$ & $5860  \pm  60$ & $ 0.02  \pm 0.06$ & $1.14  \pm0.04$ \\
   11244118 & $ 71.50  \pm 0.02$ & $1376  \pm  7$ & $5618  \pm  64$ & $5745  \pm  60$ & $ 0.35  \pm 0.06$ & $1.15  \pm0.04$ \\
    7510397 & $ 62.43  \pm 0.04$ & $1183  \pm 17$ & $6211  \pm  67$ & $6110  \pm  60$ & $-0.23  \pm 0.06$ & $1.39  \pm0.09$ \\
\B  8228742 & $ 62.29  \pm 0.04$ & $1170  \pm  7$ & $6130  \pm  51$ & $6042  \pm  60$ & $-0.14  \pm 0.06$ & $1.33  \pm0.05$ \\
 \hline 
\end{tabular}
\end{center}
{\small \textbf{References:}  $^{\rm a}$\cite{pinsonneault12}, $^{\rm b}$\cite{bruntt12}}
\end{table*}

\subsection{Extraction of the mode frequencies}

The mode frequencies of 13 out of the 24 selected targets were already extracted from \kepler\ observations by \cite{appourchaux12b}. However, this study was performed with nine months of \kepler\ data, whereas at current time almost three years of data are available in the most favorable cases. We thus decided to reanalyze all the targets of the selected sample using the full \kepler\ data sets available (until Q16) to date. For this purpose, we used a maximum likelihood estimation (MLE) method in the same way as previously applied to \corot\ and \kepler\ targets (e.g. \citealt{appourchaux08}, \citealt{deheuvels10a}). For each star, we adjusted Lorentzian profiles to all the modes simultaneously (global fits). We here neglected the rotational splitting of the modes and fitted only one component for each multiplet of degree $l$ and radial order $n$. Since the stars of the sample are expected to be slow rotators, the rotational multiplets should be approximately symmetrical with respect to their $m=0$ component. As a result, we expect negligible bias due to rotation in our estimates of the mode frequencies. We stress that in this work, we were only interested in estimating the $a_i$ parameters of a polynomial fit to the $r_{010}$ ratios of the observed stars. As a result, we did not seek to estimate the frequencies of lower signal-to-noise modes around the edges of the frequency range of observed modes. We obtained estimates of the mode parameters over 9 to 15 overtones for the 24 targets. The results are given in Tables \ref{tab_freq0} to \ref{tab_freq5} in Appendix \ref{app_tabfreq}. Our results are in good agreement with those obtained by \cite{appourchaux12b} for the targets that are among our sample. We indeed found that 31\% (resp. 8\%, 3\%) of the fitted mode frequencies agree within 1 (resp. 2, 3) $\sigma$ with the results of \cite{appourchaux12b}, which is close to what is statistically expected.

We used the estimated mode frequencies to evaluate the global seismic parameters of the selected targets. A linear regression of the frequencies of $l=0$ modes as a function of the radial order $n$ provided an estimate of the mean large separation $\Delta\nu$. The obtained values are given in Table \ref{tab_param}. We then performed a gaussian fit to the mode amplitudes as a function of frequency. The central frequency of the fitted Gaussian provides an estimate of the frequency of maximum power of the oscillations $\nu_{\rm max}$ (see Table \ref{tab_param}). Seismic scaling relations were then used to relate the global seismic parameters $\Delta\nu$ and $\nu_{\rm max}$, and the effective temperature $T_{\rm eff}$ to the stellar mass and radius. The underlying assumption behind seismic scaling relations was already mentioned in Sect. \ref{sect_descript_grid}. Whenever it was available, we used the spectroscopic $T_{\rm eff}$ obtained by \cite{bruntt12}. For the three stars of the sample that were not observed by \cite{bruntt12}, we used a photometric estimate obtained from the recipe proposed by \cite{pinsonneault12}, which was applied to the \textit{griz} photometry available from the \kepler\ input catalogue (KIC). We thus obtained stellar masses ranging from 0.94 to 1.39 $M_\odot$ (see Table \ref{tab_param}). We note that for all the stars for which both spectroscopic and photometric estimates of $T_{\rm eff}$ were available, the agreement on the stellar masses obtained with both sets of $T_{\rm eff}$ is excellent (below 1 $\sigma$ for all stars except one at 1.7 $\sigma$).

%The results of the fits will be discussed in more details in a next version. In particular, \cite{silva13} reported a sudden increase in the $r_{010}$ ratio at high frequency for two \kepler\ MS target that are included in our sample (KIC~6106415 and KIC~12009504). They attributed these features to realization noise owing to the lower SNR of these modes. In our analysis these features have completely disappeared. This can be seen in Fig. \ref{fig_ratio_MLE}a, which shows the $r_{010}$ ratio obtained for KIC~6106415. The star is clearly still in the \textit{linear domain}. Fig. \ref{fig_ratio_MLE}b shows the $r_{010}$ ratio obtained for KIC~12258514, which is evolved enough to have left the linear domain. 

\subsection{Polynomial fit to $r_{010}$ ratios \label{sect_fit_r010}}

\begin{figure*}
\begin{center}
\includegraphics[width=8cm]{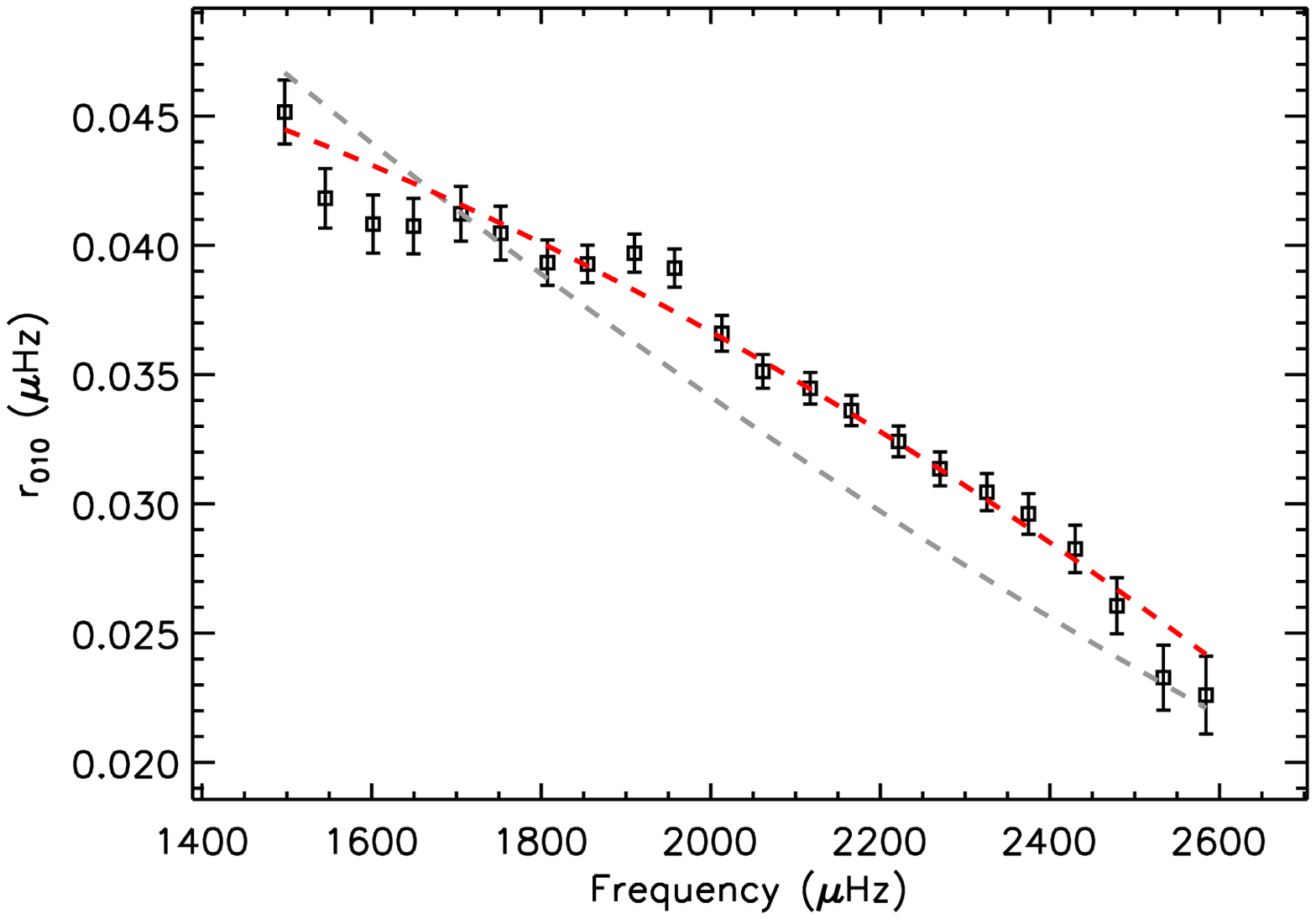}
\includegraphics[width=8cm]{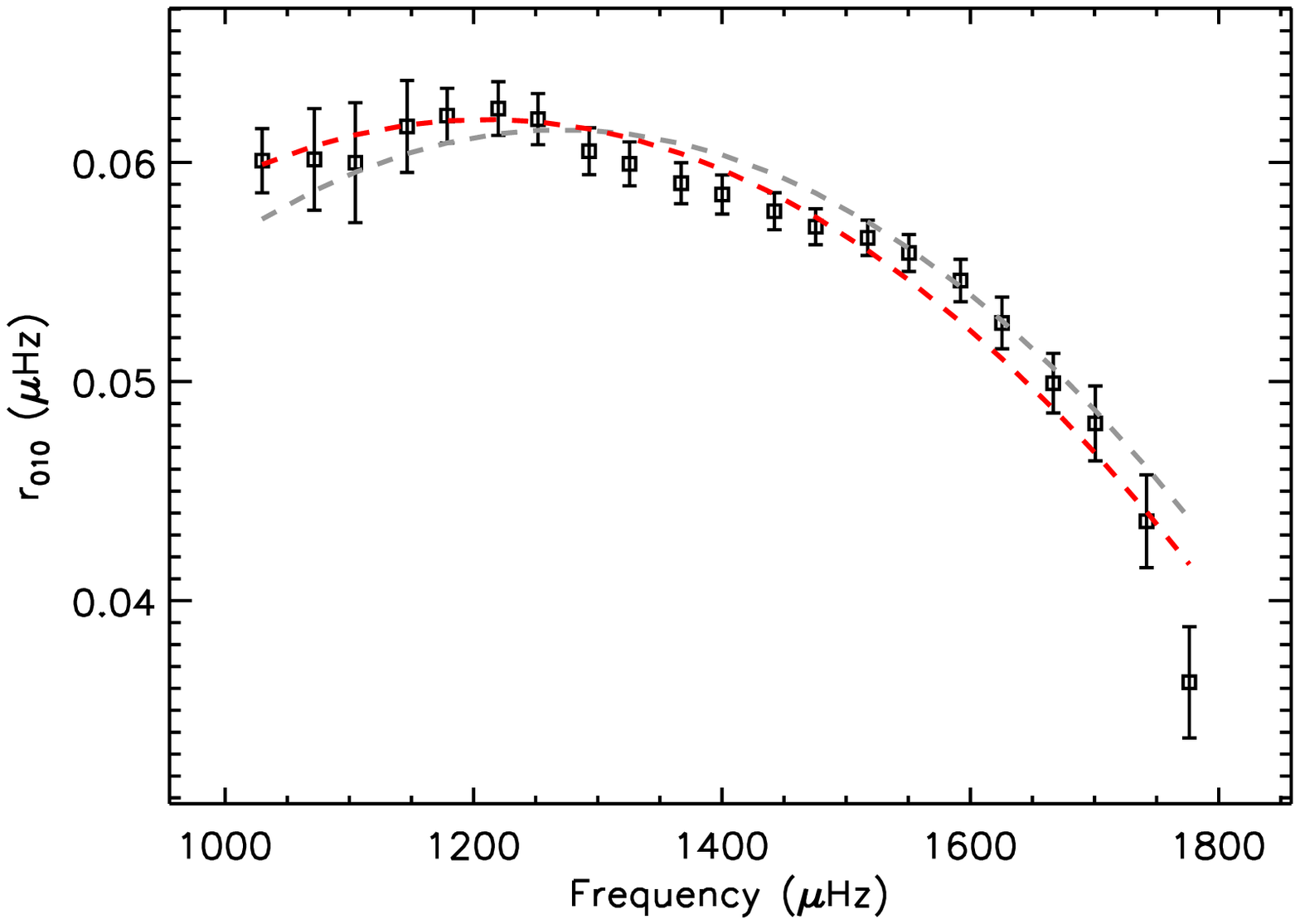}
\end{center}
\caption{Ratios $r_{010}$ computed for KIC~6106415 (left) and KIC~12258514 (right) using the mode frequencies extracted from the \kepler\ oscillation spectra (see text). The colored dashed lines correspond to $2^{\rm nd}$-order polynomial fits to the observed ratios using either the raw covariance matrix (gray lines) or the covariance matrix modified through truncated SVD (see Sect. \ref{sect_fit_r010}). 
\label{fig_ratio_MLE}}
\end{figure*}

We used the fitted mode frequencies listed in Tables \ref{tab_freq0} to \ref{tab_freq5} in Appendix \ref{app_tabfreq} to compute the $r_{010}$ ratios of all the stars of the sample. Two representative examples are shown in Fig. \ref{fig_ratio_MLE}. KIC6106415 (left plot) is still in a phase where the $r_{010}$ ratios are roughly linear in the range of observed frequencies, while the ratios of KIC12258514 have a more parabolic shape. As predicted by stellar models, we found that the observed ratios are well reproduced by 2$^{\rm nd}$ degree polynomials. For several targets, the $r_{010}$ ratios deviate from a mere parabola because of a short-period oscillation around the parabolic general trend. This is expected and corresponds to the signature of the base of the convective envelope. In this work,  the polynomial fit that we applied to the $r_{010}$ ratios filters out this contribution. This is to our advantage here since we are merely interested in probing the core properties in this study. However, we stress that these signatures of the bottom of the convective envelope can potentially yield precious model-independent constraints on the stellar structure (\citealt{mazumdar14}) and deserve further investigation. The dip in the profile of the adiabatic index $\Gamma_1$ corresponding to the region of second ionization of helium can also create a short-period oscillation in seismic indexes, however the $r_{010}$ ratios are almost insensitive to these shallow regions and the amplitude of the corresponding oscillation is expected to be negligible.

To fit polynomials to the observed $r_{010}$ ratios, one needs to take into account the high level of correlation between the data points. Indeed, each mode frequency is used by several data points. The covariance matrix between linear combinations of the mode frequencies (e.g. between the $d_{01}$ and $d_{10}$ separations as defined by Eq. \ref{eq_d01} and \ref{eq_d10}) can easily be computed analytically, but it is much harder for the $r_{010}$ ratios because of the division by the large separations. We therefore resorted to Monte Carlo simulations using the observed mode frequencies and their associated error bars to estimate the covariance matrix $\vect{C}$ for each star. This approach supposes that the errors in the mode frequency estimates are normally distributed, which has been shown to be a valid approximation (\citealt{benomar09a}), except for low signal-to-noise-ratio modes, which we have excluded here\footnote{When fitting the modes following a Bayesian approach coupled with a Markov chain Monte Carlo algorithm, the covariance matrix can be estimated without having to assume normally distributed errors (e.g. \citealt{davies15}).}. The optimal parameters $a_0$, $a_1$, and $a_2$ of the polynomial described in Eq. \ref{eq_poly} were then obtained by a least-square minimization of the residuals weighted by the coefficients of the inverse $\vect{W}$ of the covariance matrix, as described in Appendix \ref{app_polyfit}. This type of fitting is now applied routinely to fit stellar models constrained by combinations of mode frequencies (e.g. \citealt{silva13}, \citealt{lebreton14}). However, when applied directly to our simple case of a polynomial fit of the $r_{010}$, we obtained poor fits to the observed ratios (see gray dashed lines in Fig. \ref{fig_ratio_MLE}).

After careful inspection of the results, we found that the covariance matrix $\vect{C}$ is in fact ill-conditioned, with a conditioning of the order of $10^5$ or $10^6$. As a result, the covariance matrixes are nearly non-invertible, which explains the poor agreement obtained by direct fitting. This property is not specific to our particular case, and we expect any covariance matrix built with combinations of frequencies to show similar behavior as the number of points increases. The conditioning of matrix $\vect{C}$ increases as the number of modes involved in the combinations of frequencies increases, which explains why the problem is so obvious for the $r_{010}$ ratios, but with a large enough number of points, it also arises for three-point separations. To remedy this problem, we applied truncated SVD to the covariance matrix as explained in Appendix \ref{app_polyfit}. We found that suppressing the 5 smallest eigenvalues of matrix $\vect{C}$ is generally enough to obtain satisfactory fits to the observed $r_{010}$ ratios (red dashed lines in Fig. \ref{fig_ratio_MLE}).

\section{Measuring the size of mixed cores in \kepler\ targets \label{sect_grid}}

Since the $r_{010}$ ratios have been shown to efficiently cancel out the contribution from the outer layers (\citealt{roxburgh03}), the observed ratios could be directly compared to those of models. We thus confronted the observed $r_{010}$ ratios to those of two grids of models: the one computed with \cesam, which was described in Sect. \ref{sect_test_diagnostic}, and a second equivalent grid that was built with the evolutionary code \mesa\ (\citealt{paxton11}, \citealt{paxton13}), which is described below. Obviously, these grids are too coarse to provide in themselves statistically reliable estimates of the stellar parameters, and in particular of the amount of core overshooting. However, based on the tests performed in Sect. \ref{sect_test_diagnostic}, these grids can be used to identify stars with a convective core and  obtain a rough estimate of the extension of the mixed core in these stars. As a second step presented in Sect. \ref{sect_calibrate}, these estimates were refined using a more sophisticated optimization procedure.

%\subsection{Using grids of models \label{sect_grid}}

\begin{figure*}
\begin{center}
\includegraphics[width=8cm]{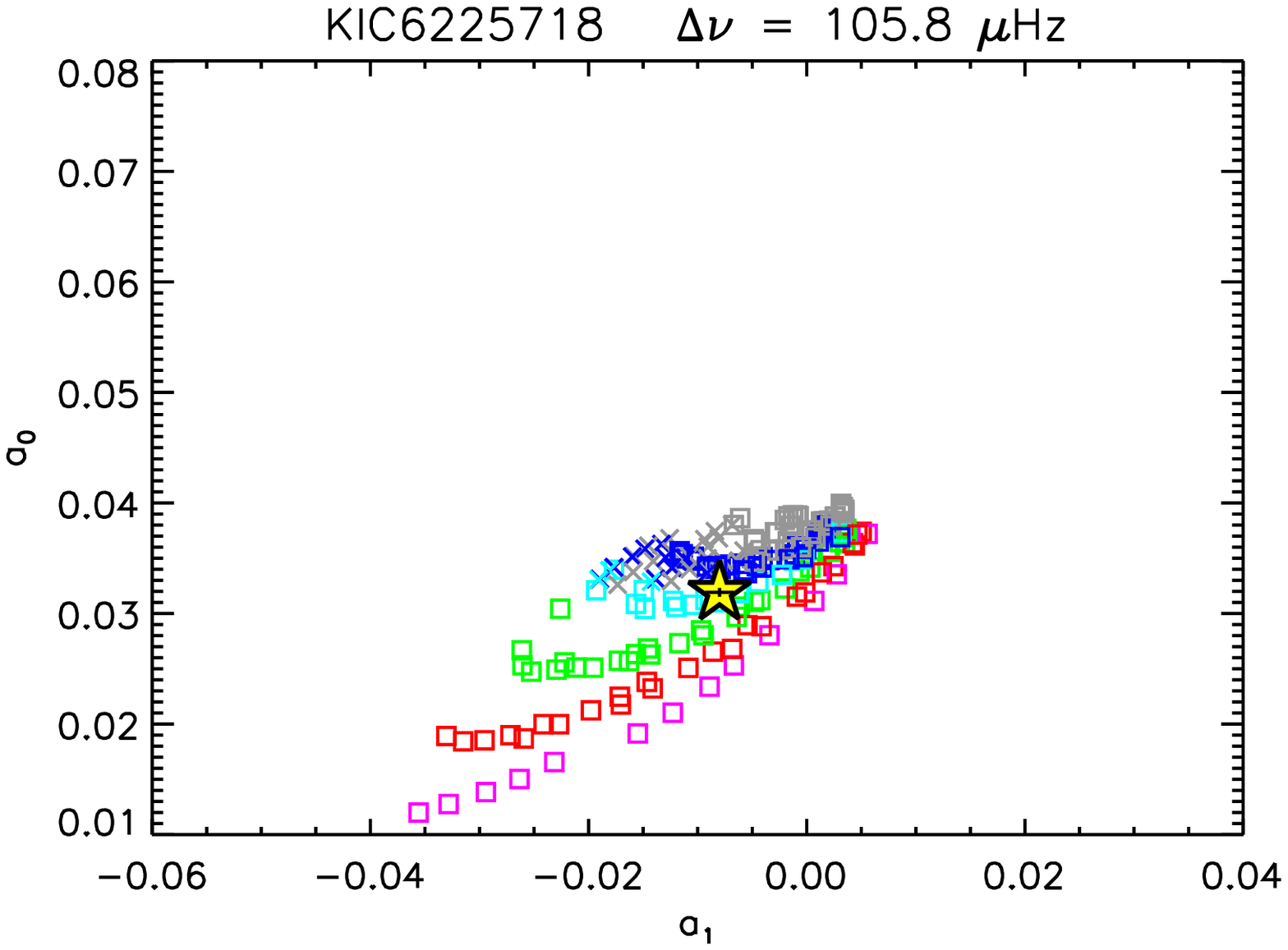}
\includegraphics[width=8cm]{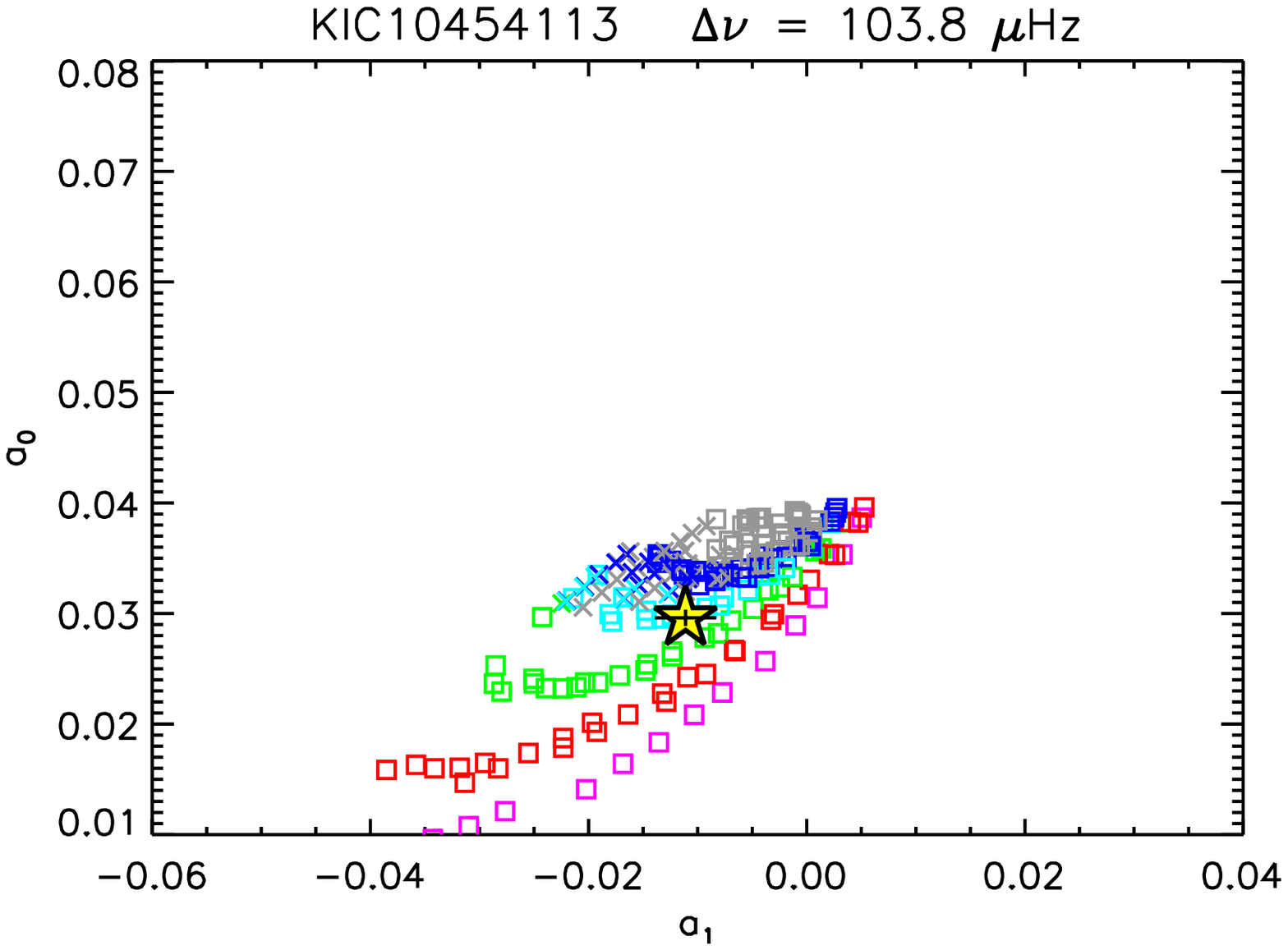}
\includegraphics[width=8cm]{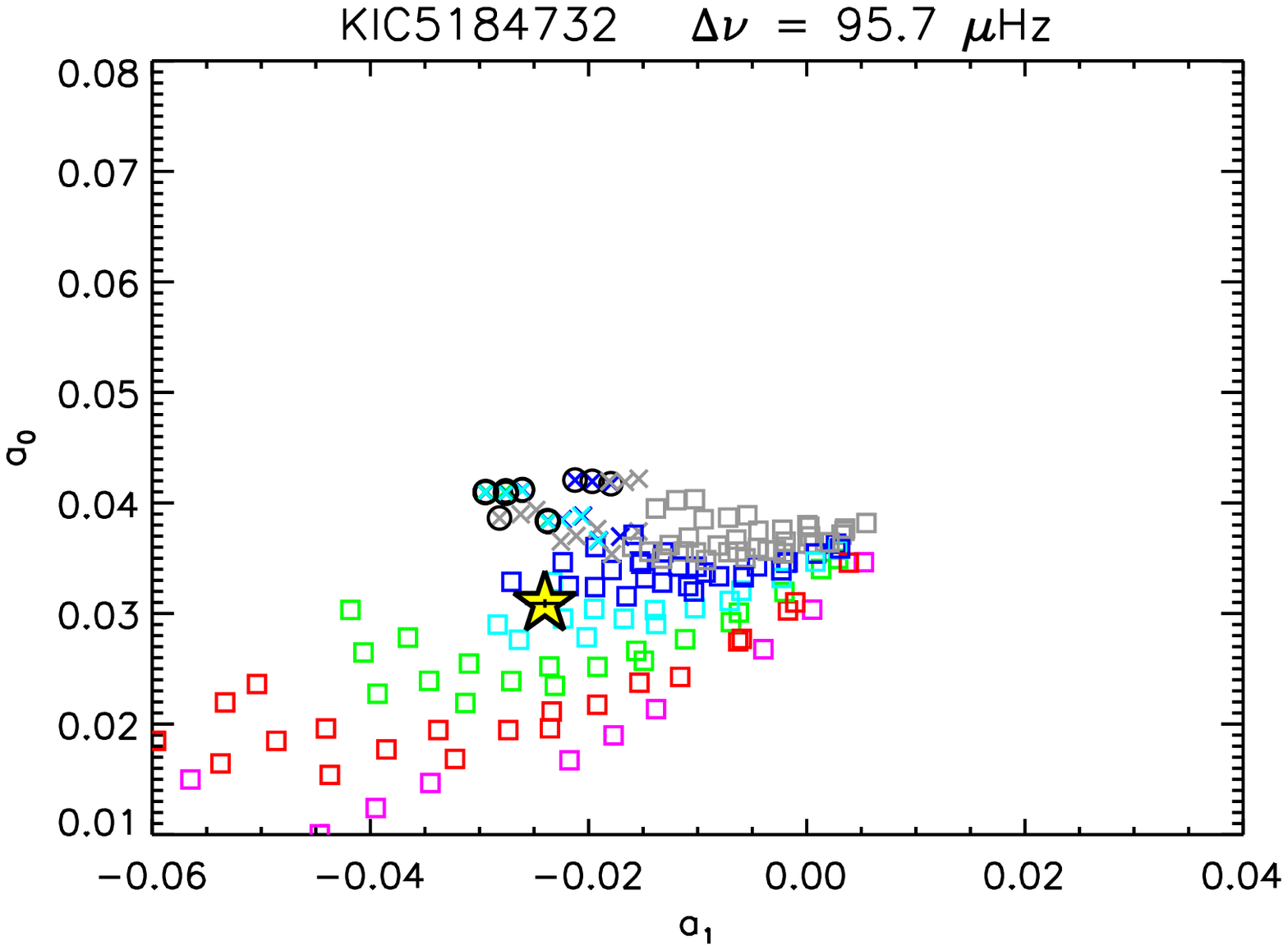}
\includegraphics[width=8cm]{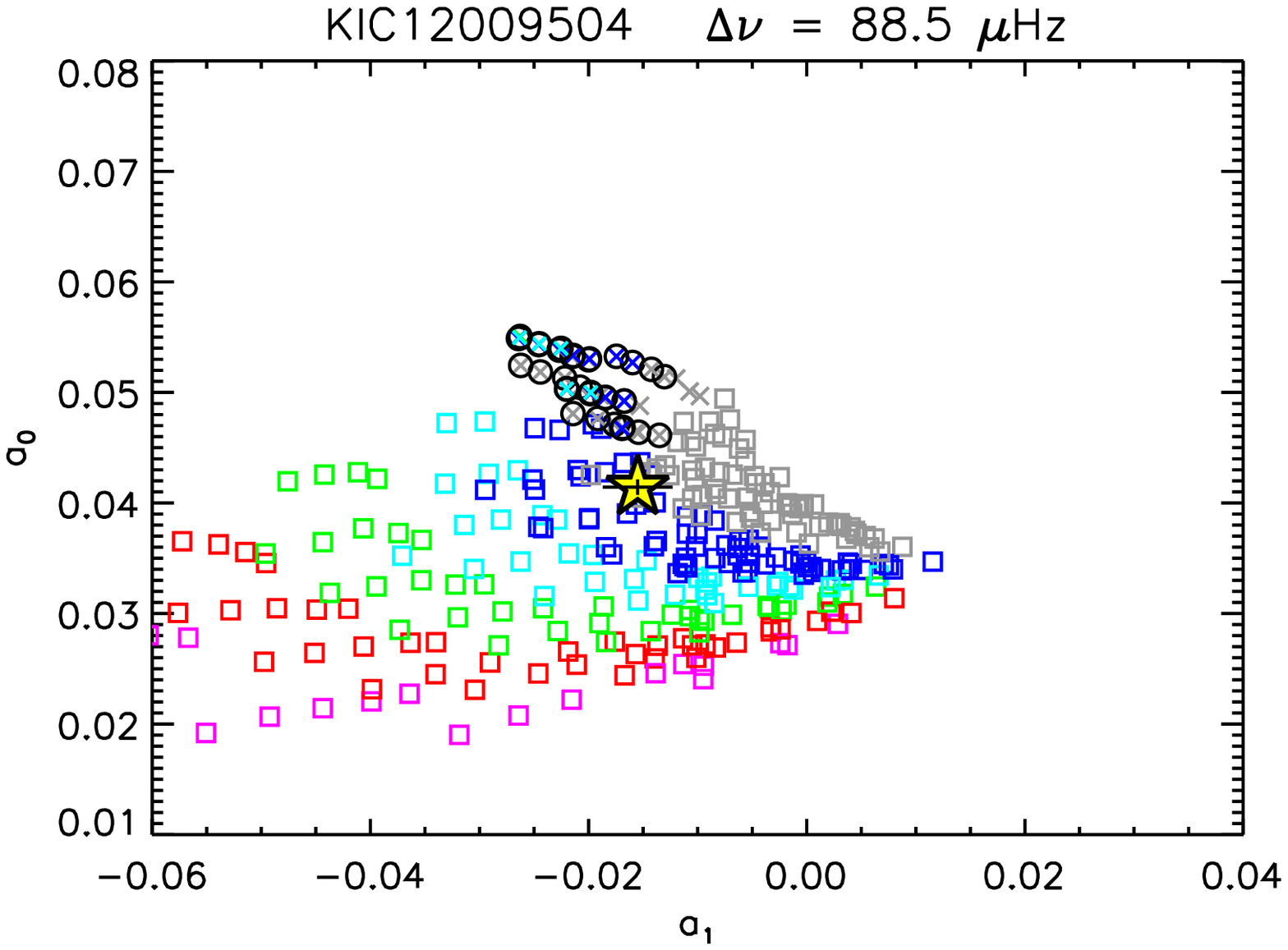}
\includegraphics[width=8cm]{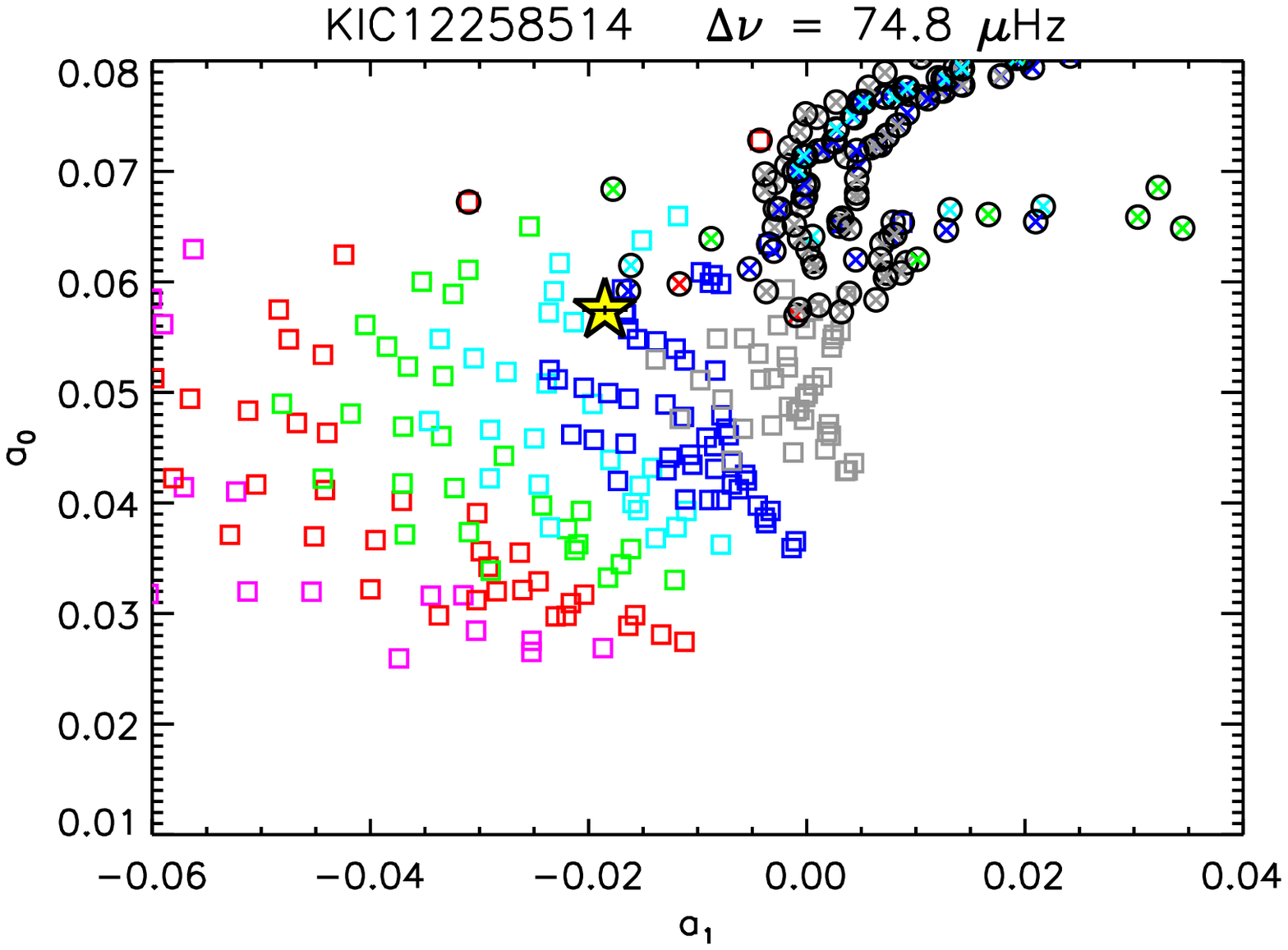}
\includegraphics[width=8cm]{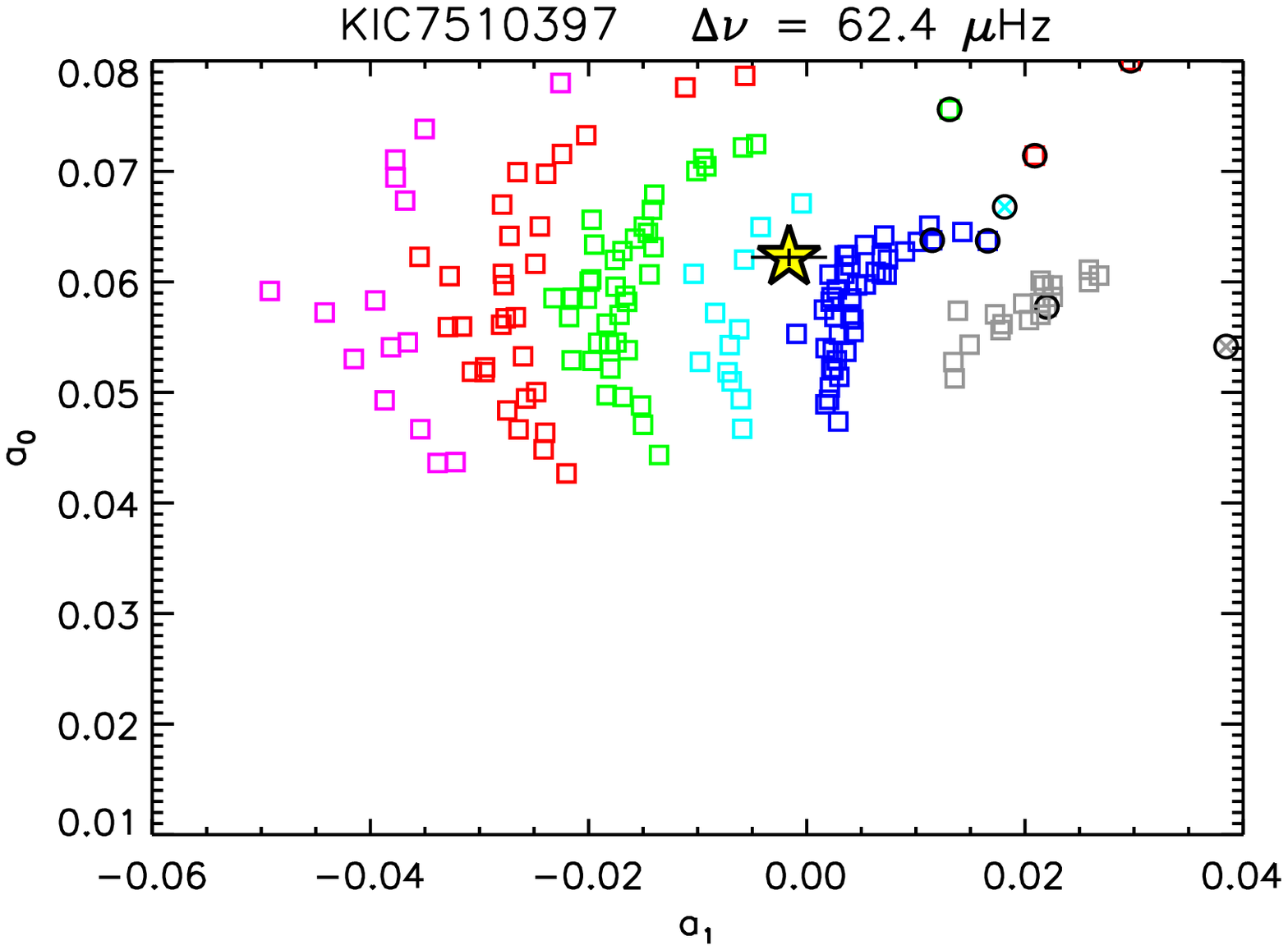}
\includegraphics[width=8cm]{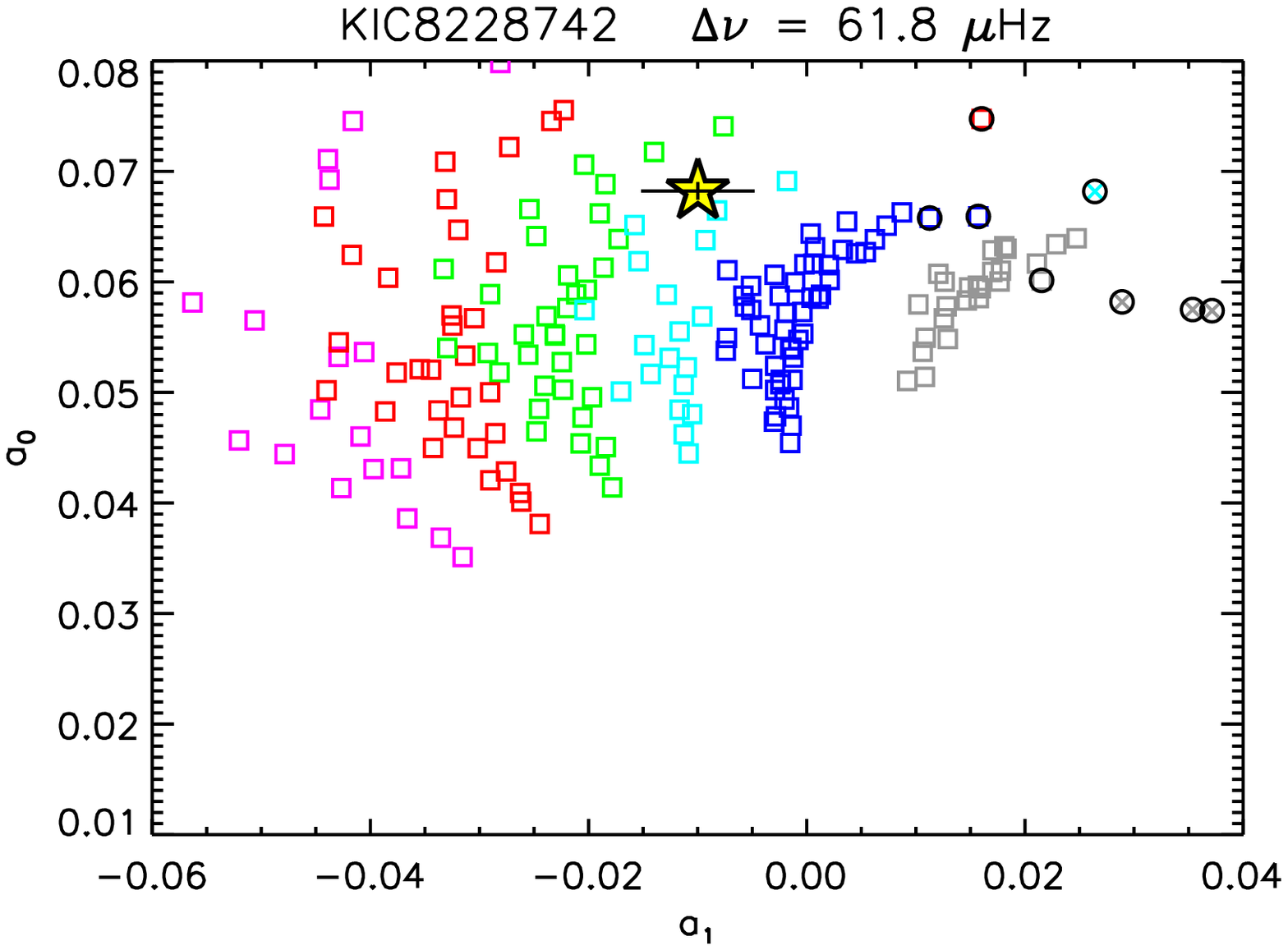}
\includegraphics[width=8cm]{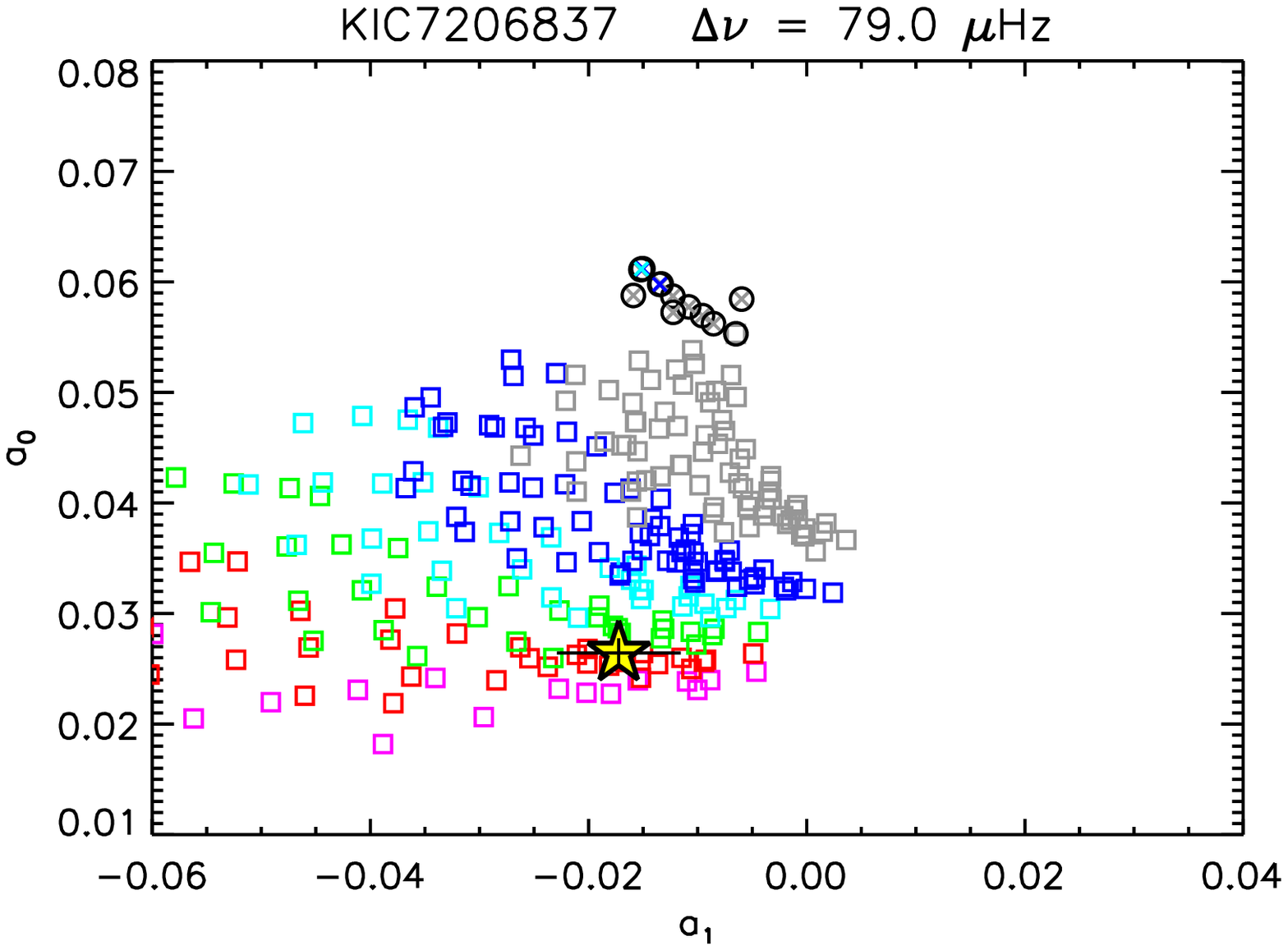}
\end{center}
\caption{Location in the $(a_1,a_0)$ plane (star symbols and black error bars) of the stars of the sample that were found to be on the MS with a convective core in this study. Models that reproduce the observed large separation, the spectroscopic estimate of metallicity, and the stellar mass derived from scaling laws within 3 $\sigma$ errors are overplotted. The symbols have the same meaning as in Fig. \ref{fig_ratio_trend}. 
\label{fig_ratio_kepler1}}
\end{figure*}

\begin{figure*}
\begin{center}
\includegraphics[width=8cm]{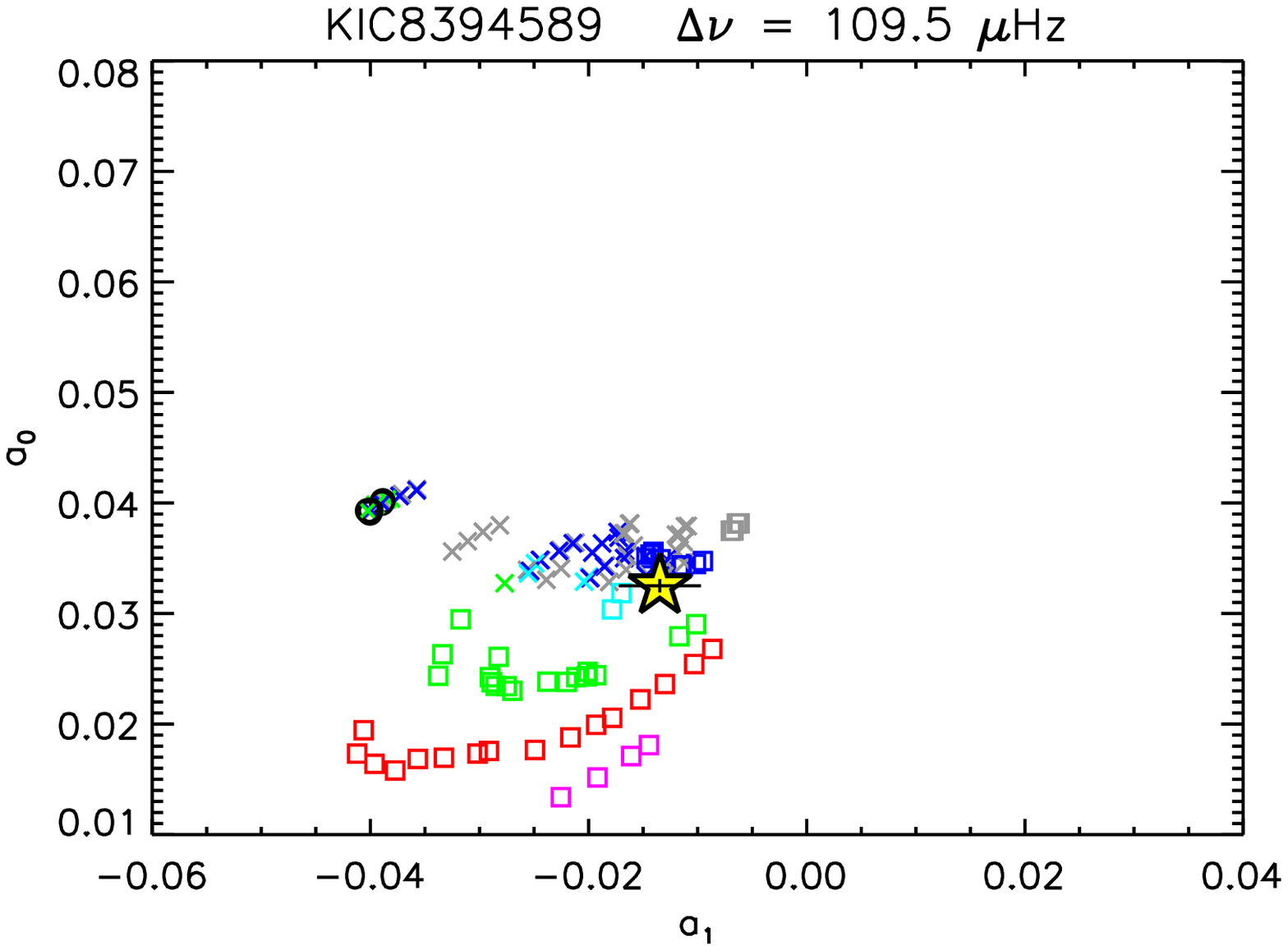}
\includegraphics[width=8cm]{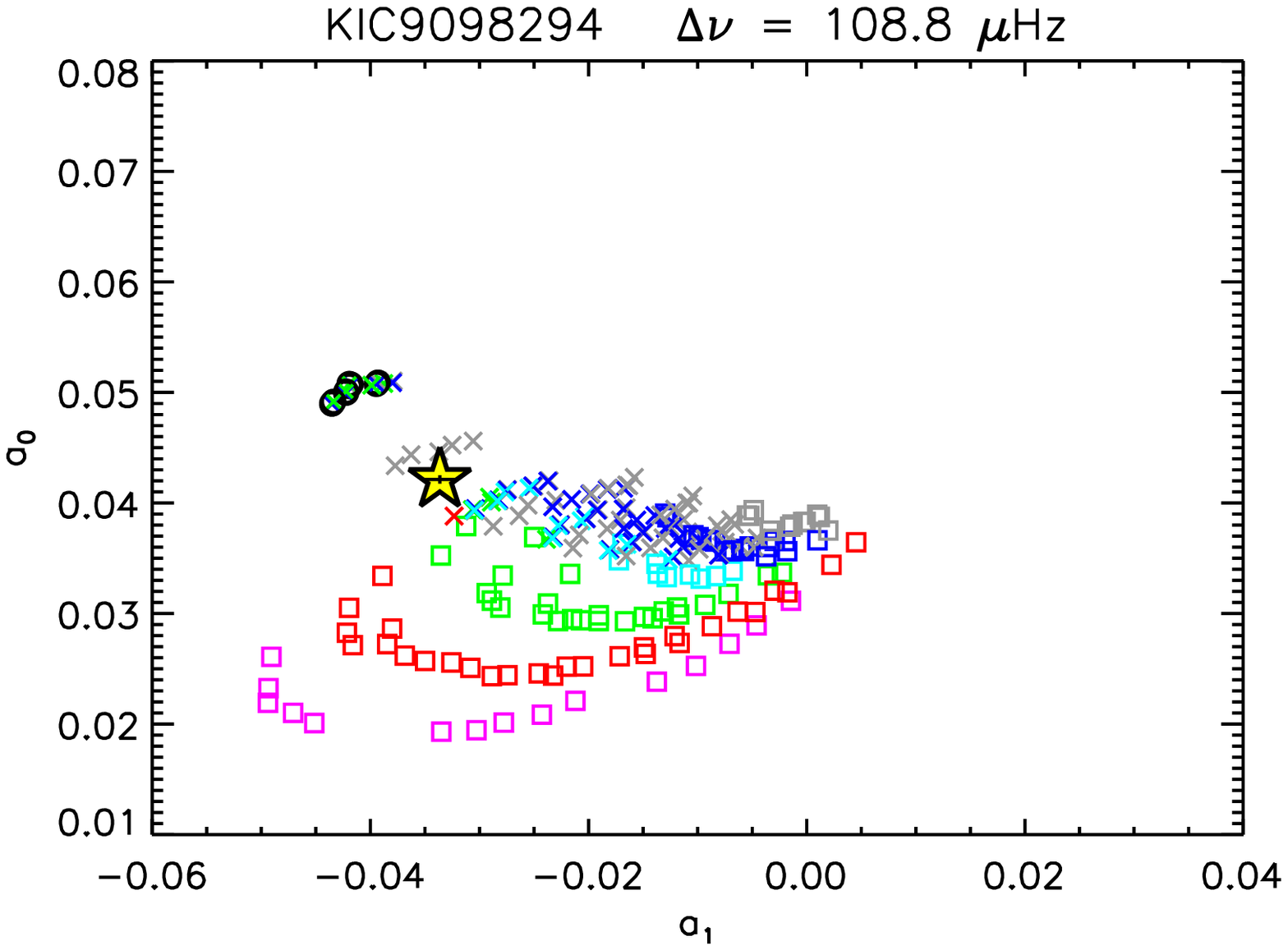}
\includegraphics[width=8cm]{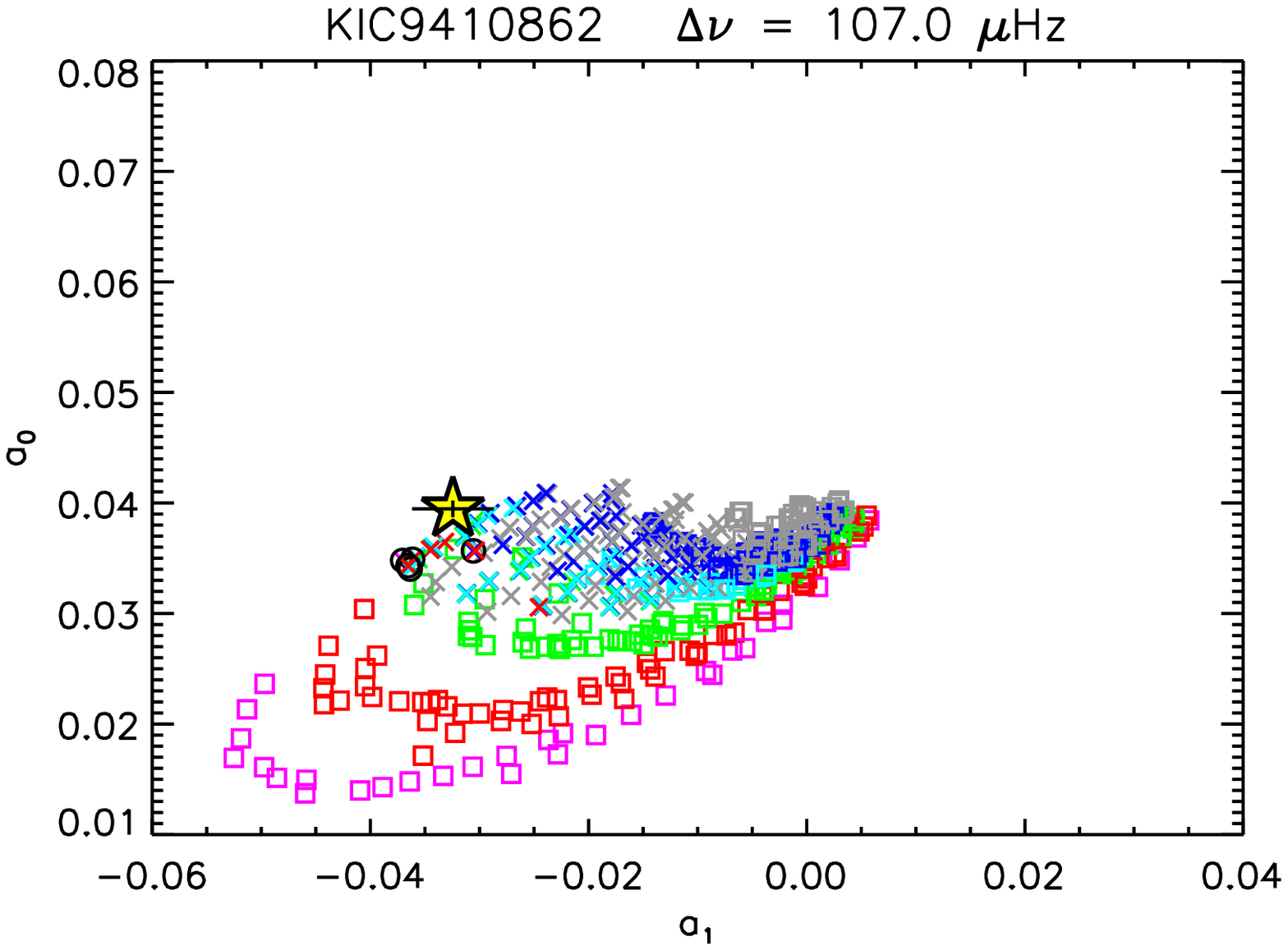}
\includegraphics[width=8cm]{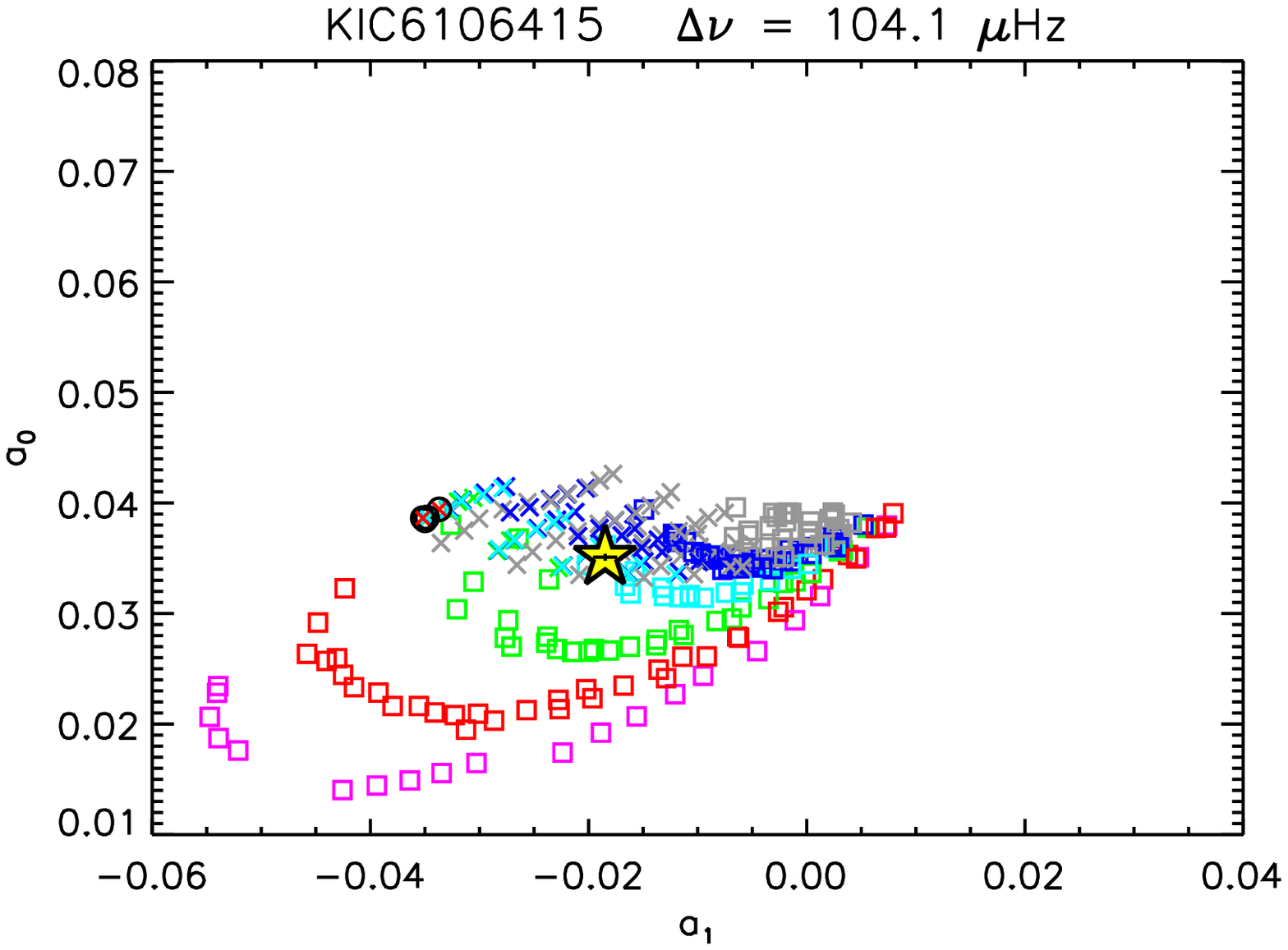}
\includegraphics[width=8cm]{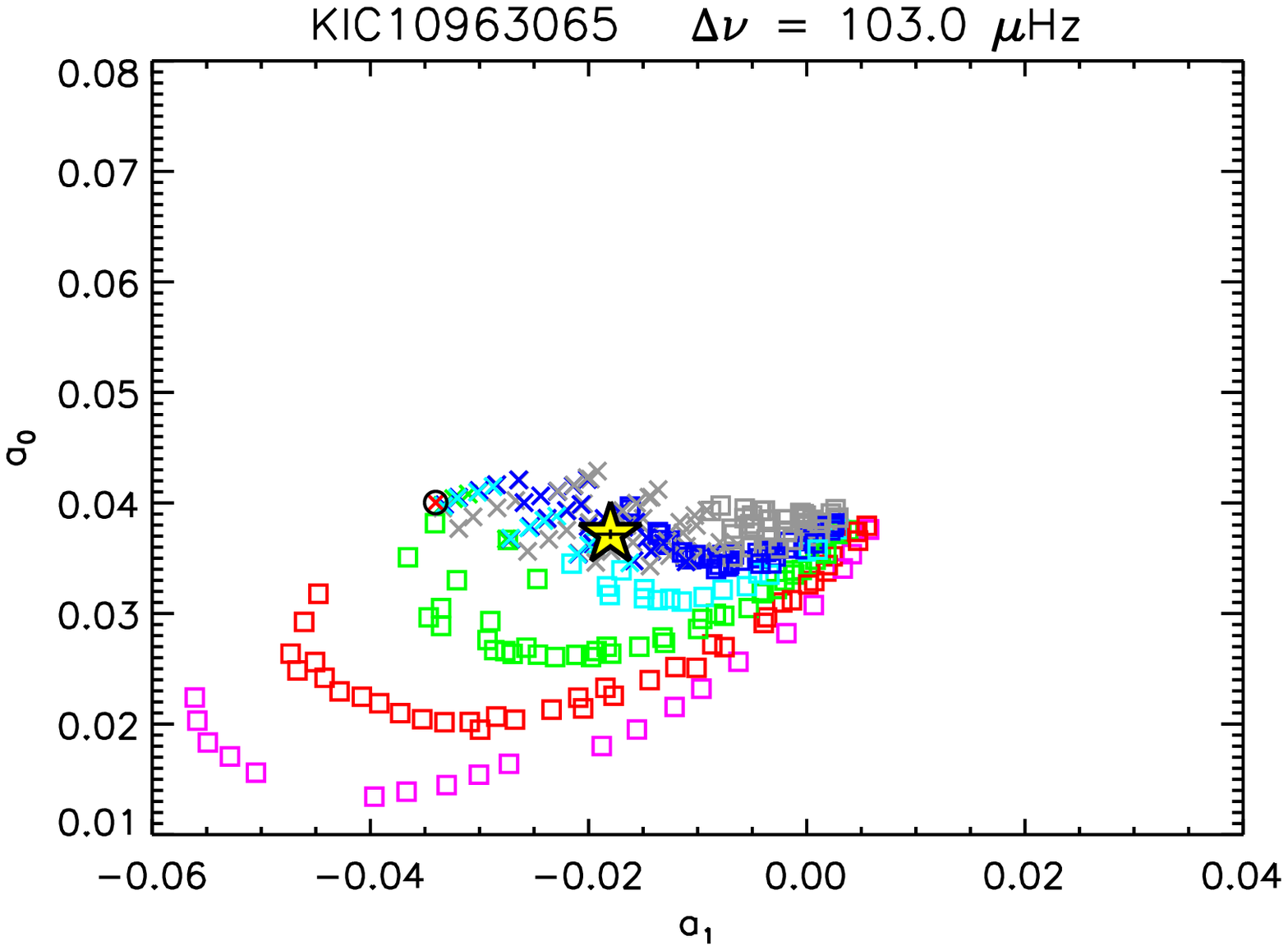}
\includegraphics[width=8cm]{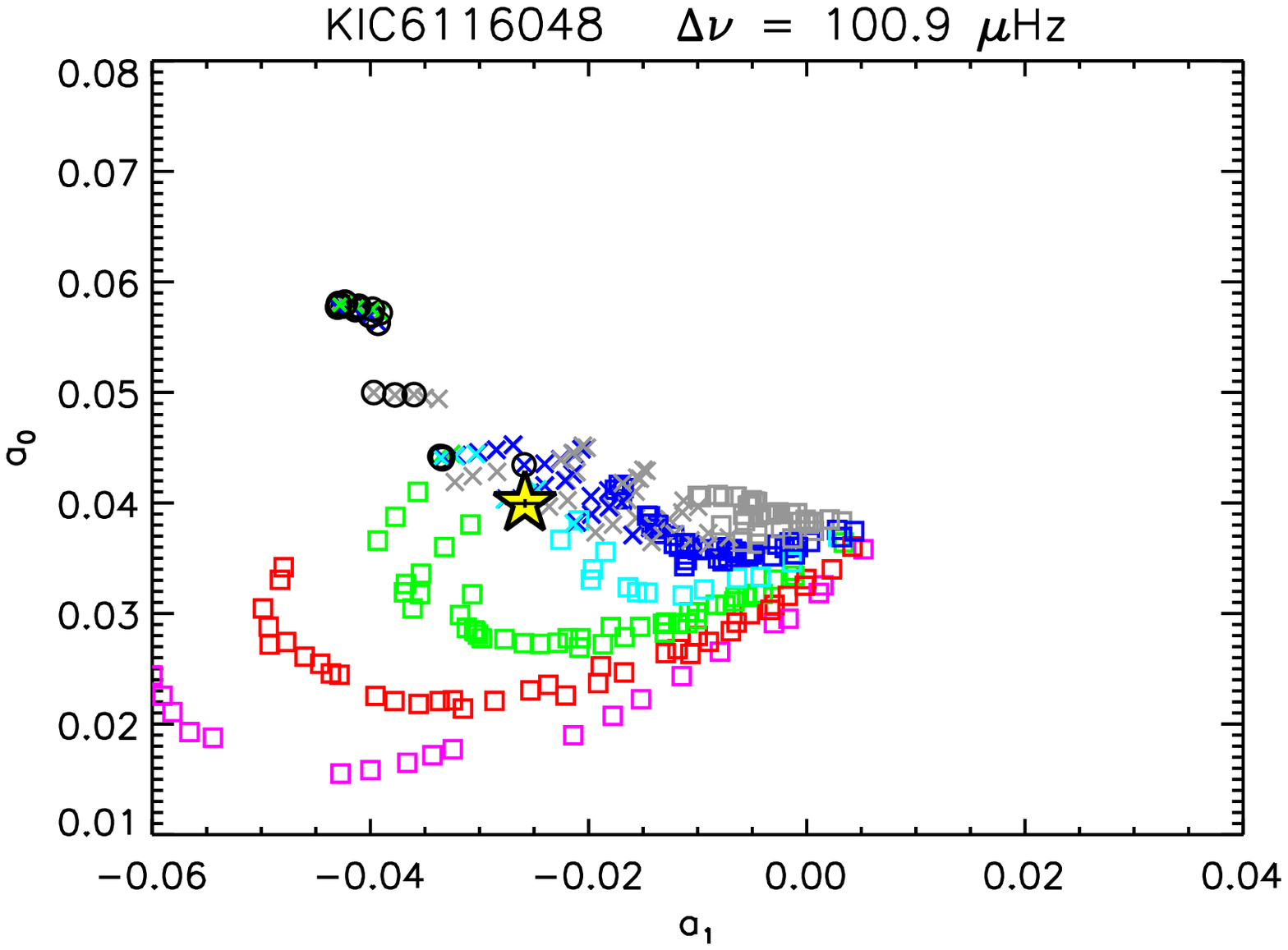}
\end{center}
\caption{Location in the $(a_1,a_0)$ plane of the MS stars for which the presence of a convective core is uncertain. The symbols have the same meaning as in Fig. \ref{fig_ratio_kepler1}.
\label{fig_ratio_kepler2}}
\end{figure*}

\begin{figure*}
\begin{center}
\includegraphics[width=8cm]{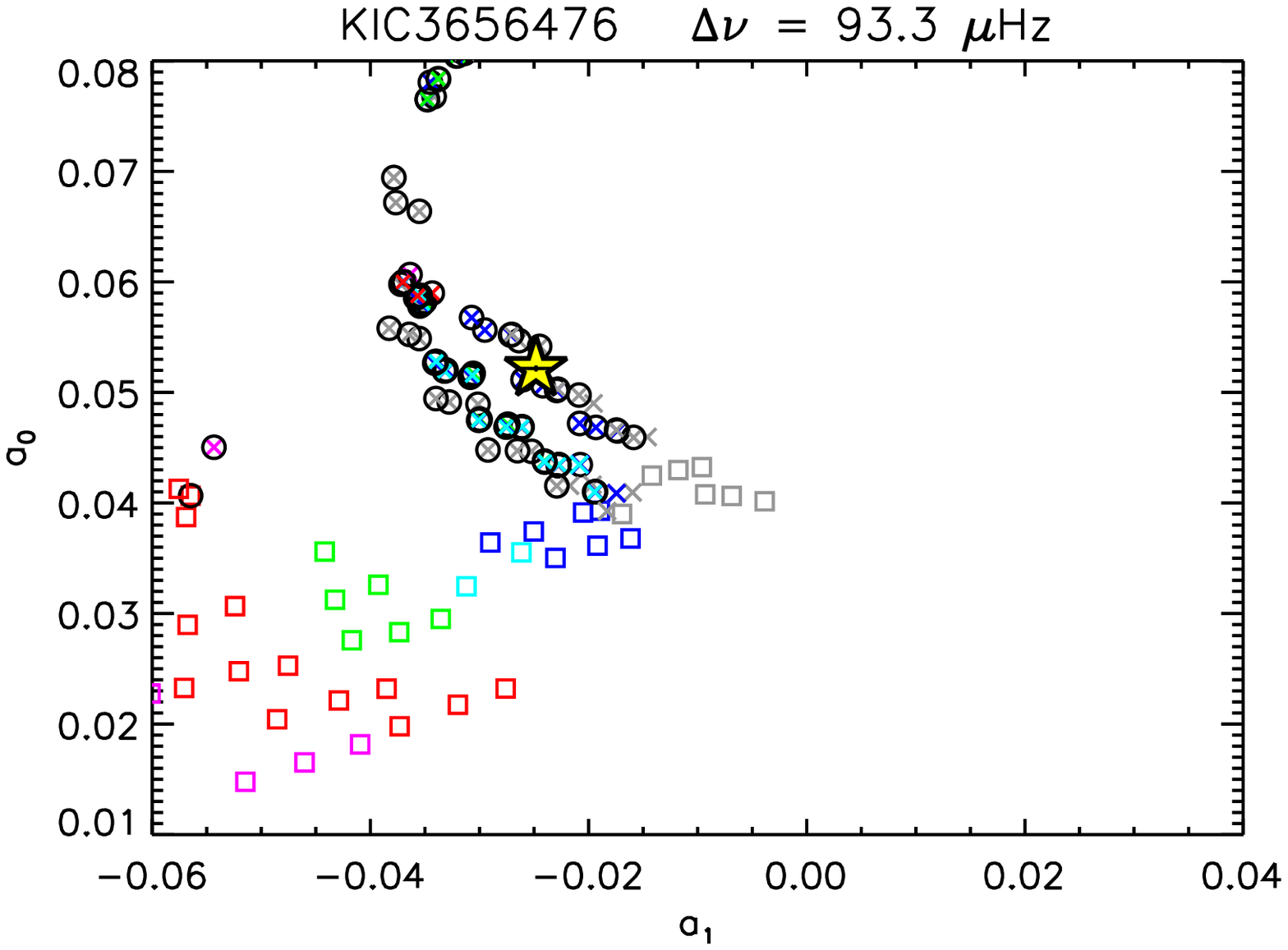}
\includegraphics[width=8cm]{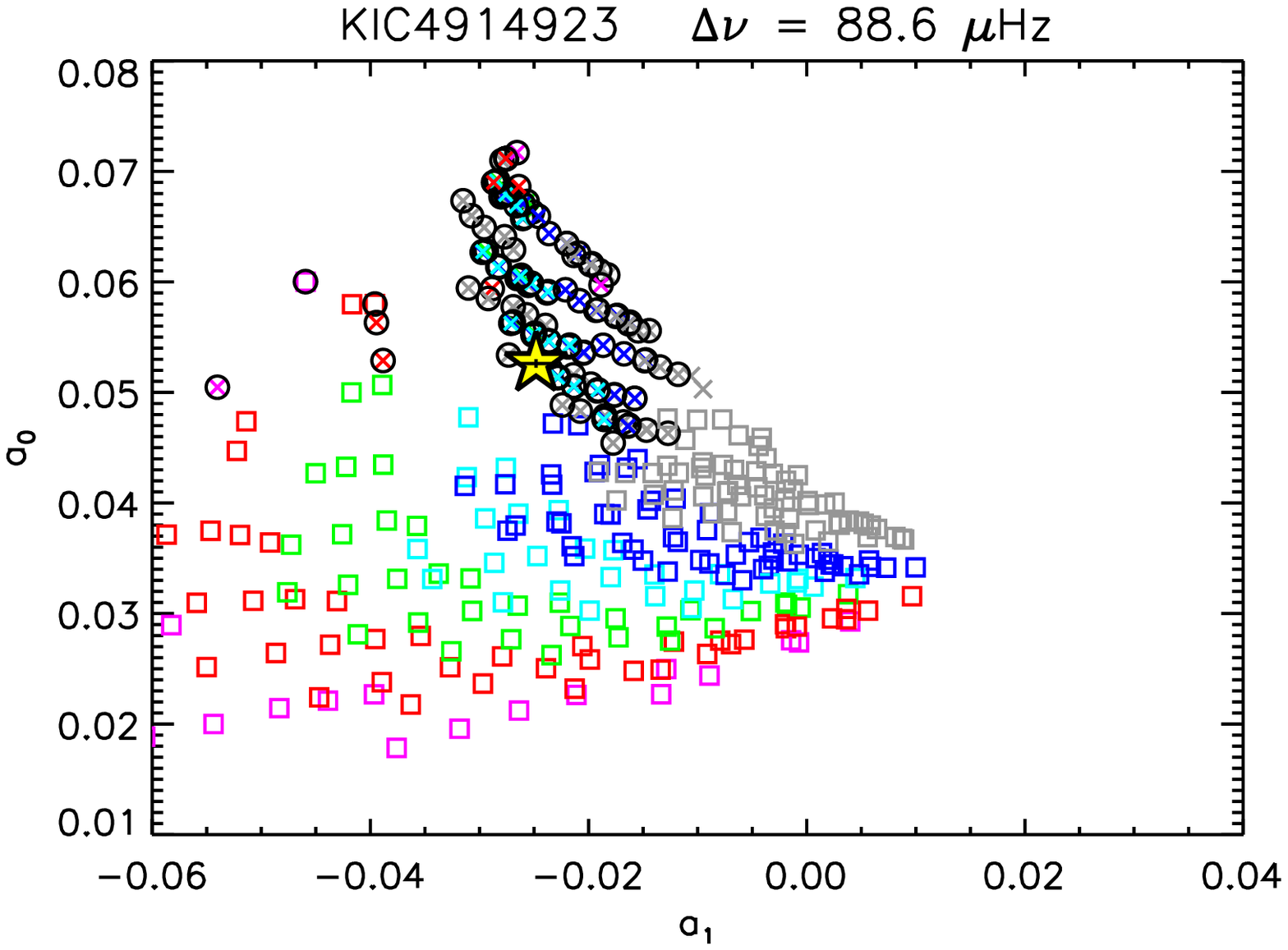}
\includegraphics[width=8cm]{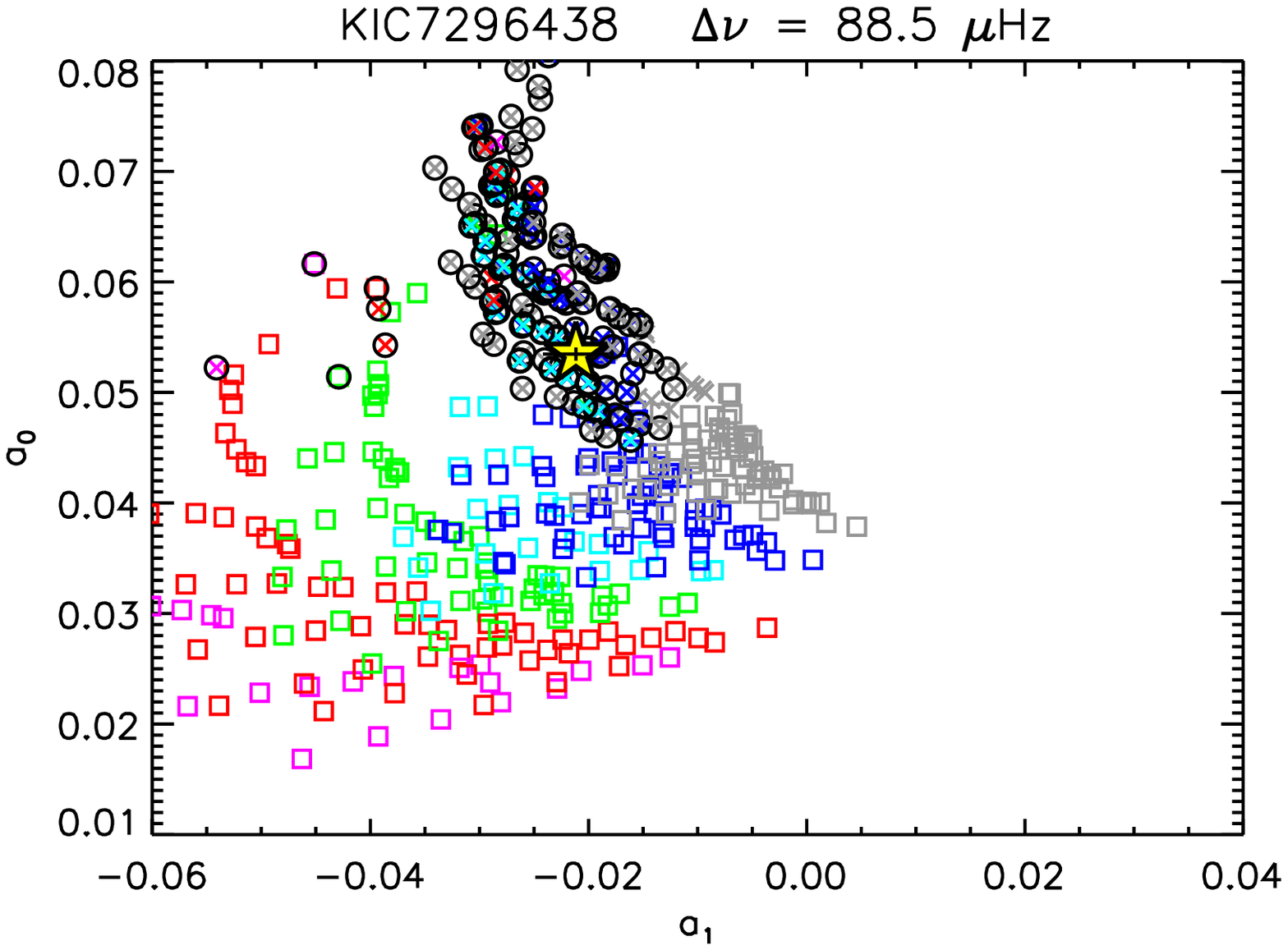}
\includegraphics[width=8cm]{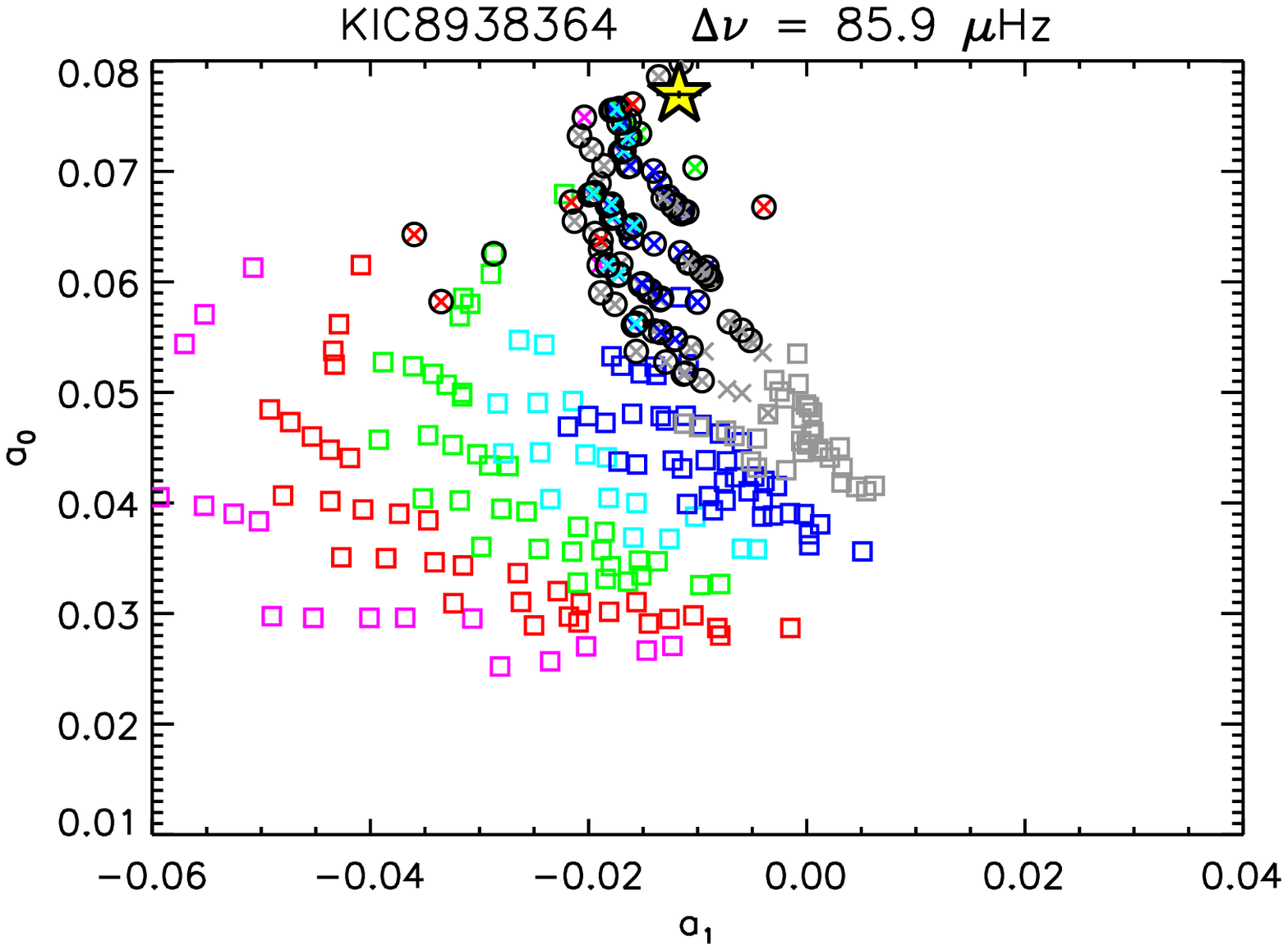}
\includegraphics[width=8cm]{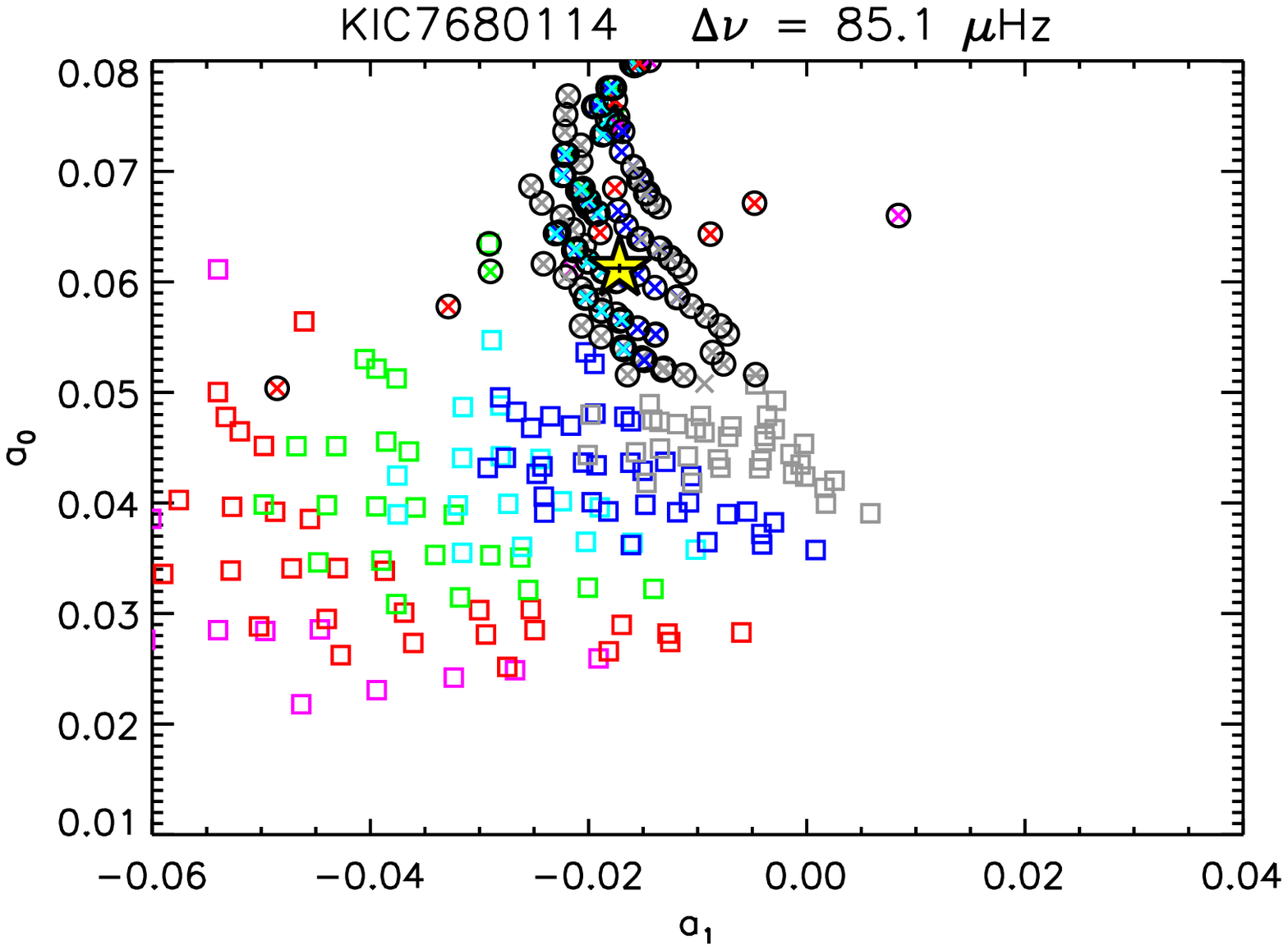}
\includegraphics[width=8cm]{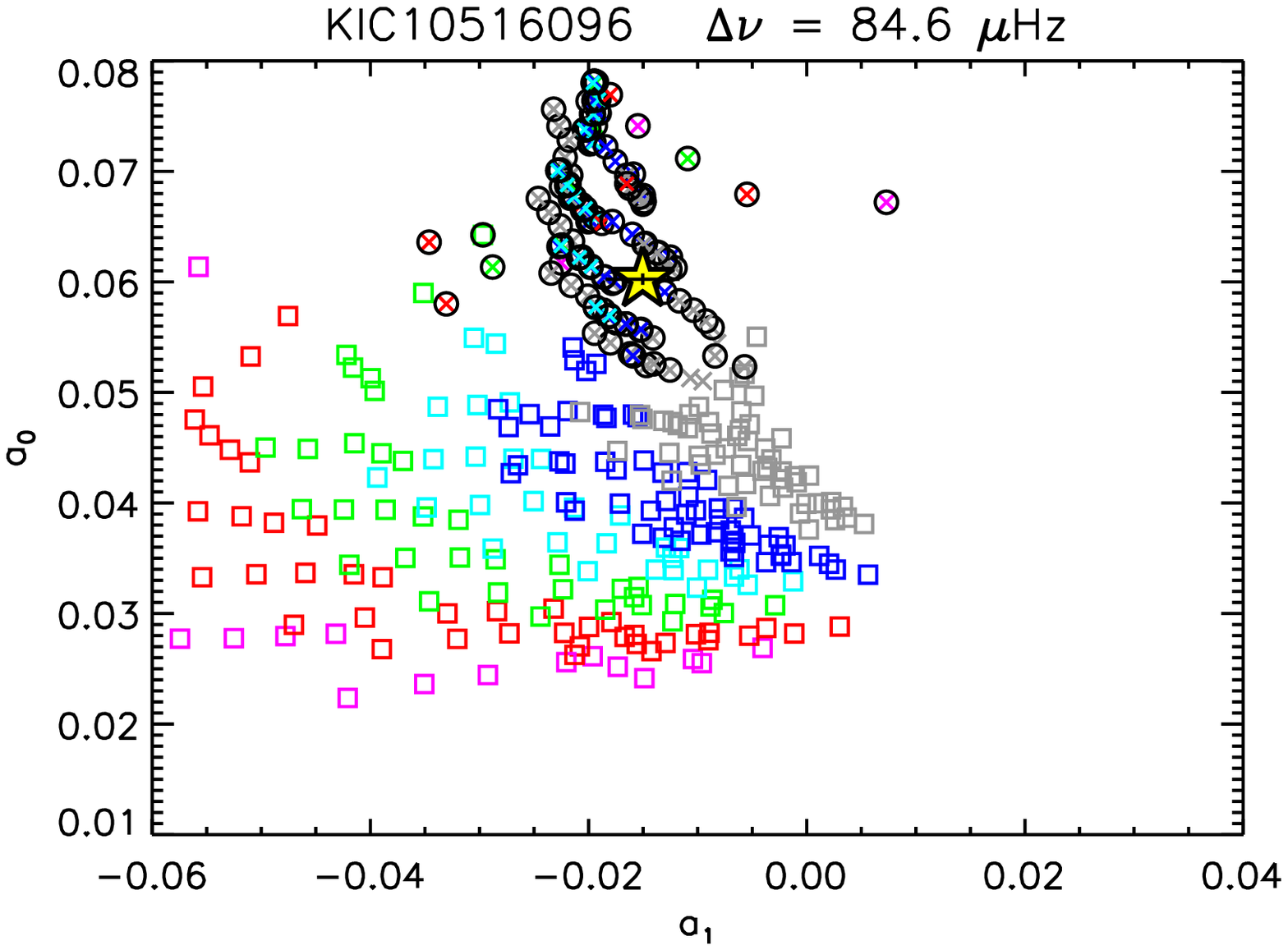}
\end{center}
\caption{Location in the $(a_1,a_0)$ plane of the first six PoMS stars of the sample. The symbols have the same meaning as in Fig. \ref{fig_ratio_kepler1}.
\label{fig_ratio_kepler3}}
\end{figure*}

\begin{figure*}
\begin{center}
\includegraphics[width=8cm]{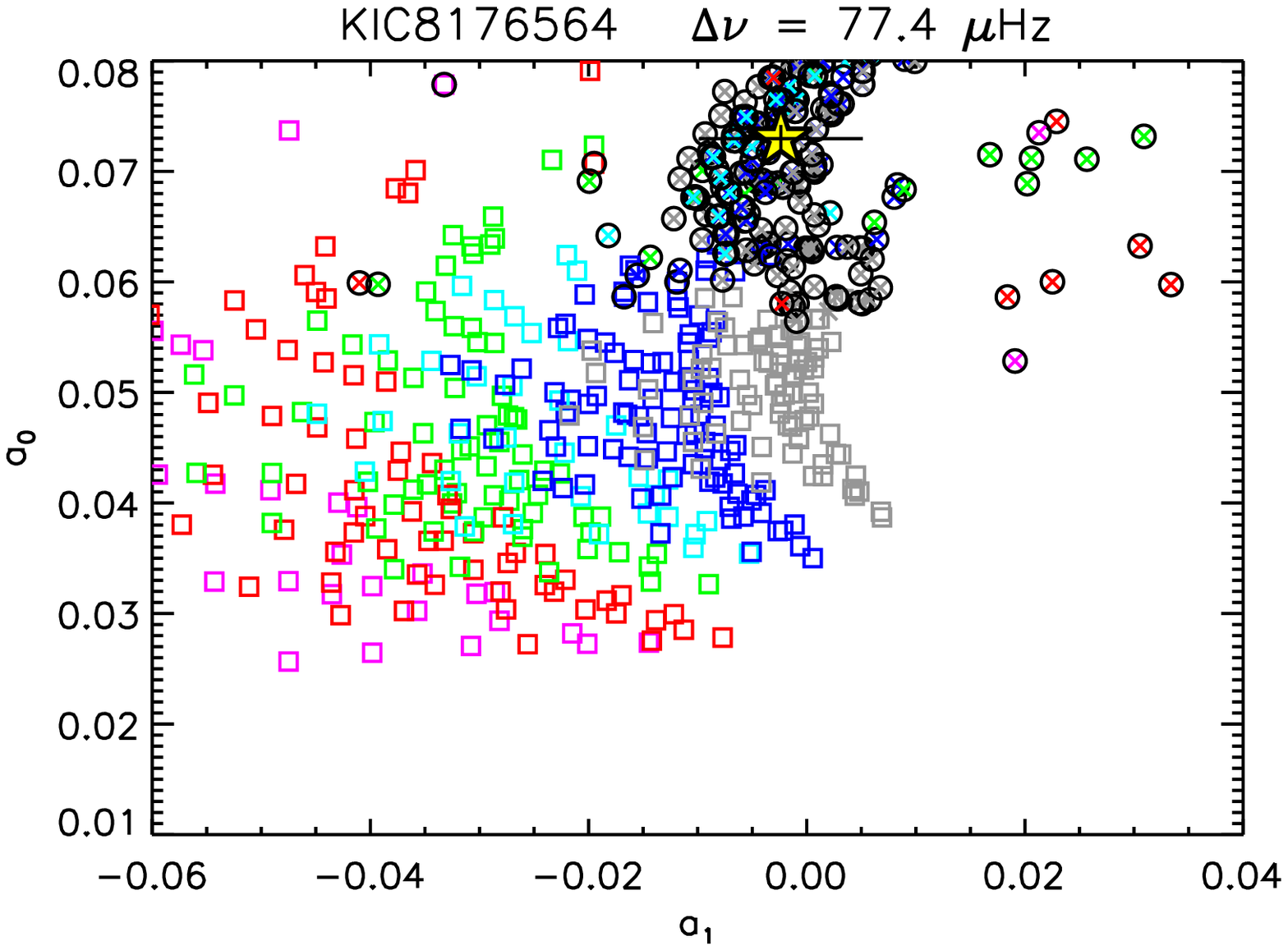}
\includegraphics[width=8cm]{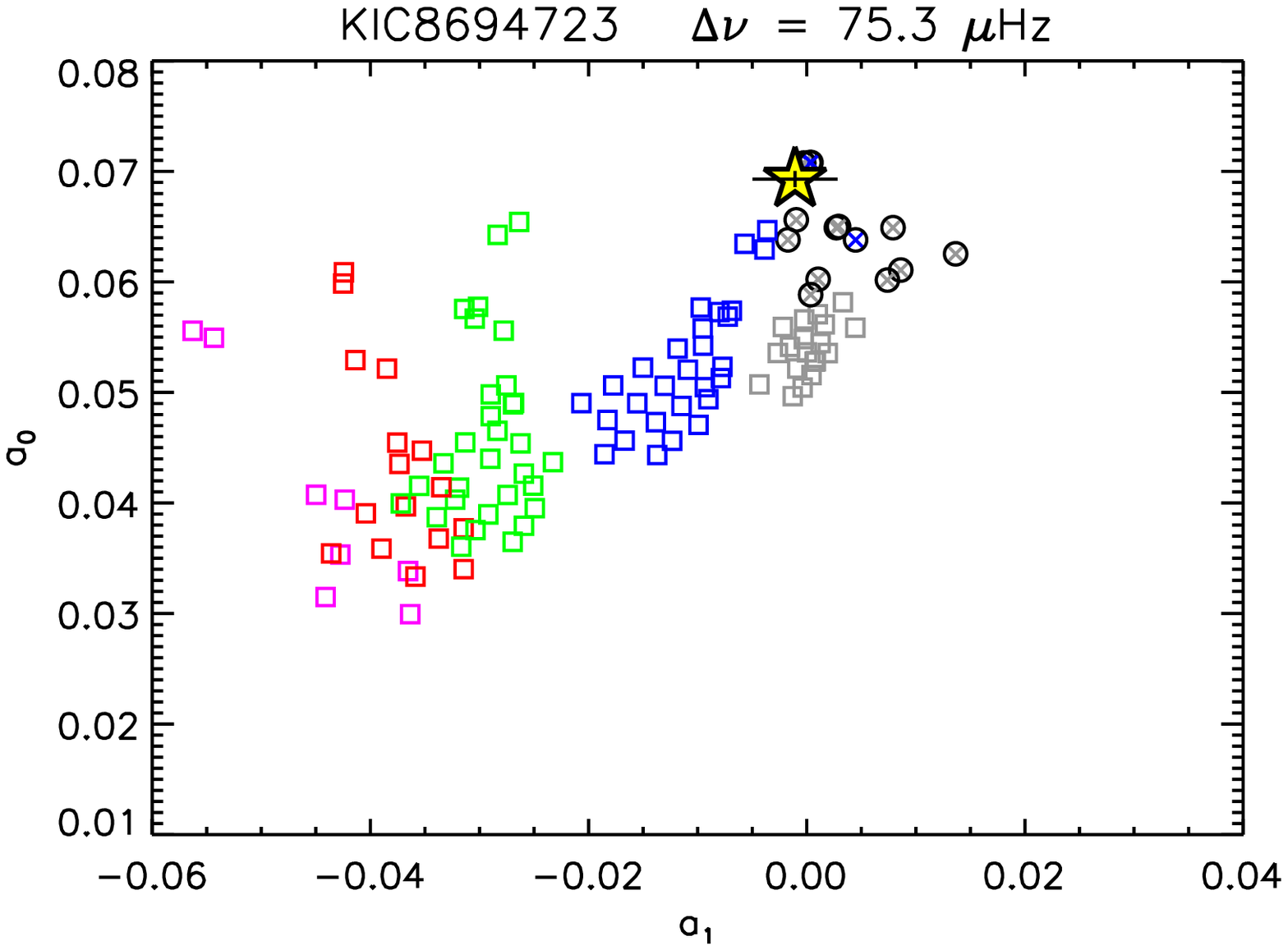}
\includegraphics[width=8cm]{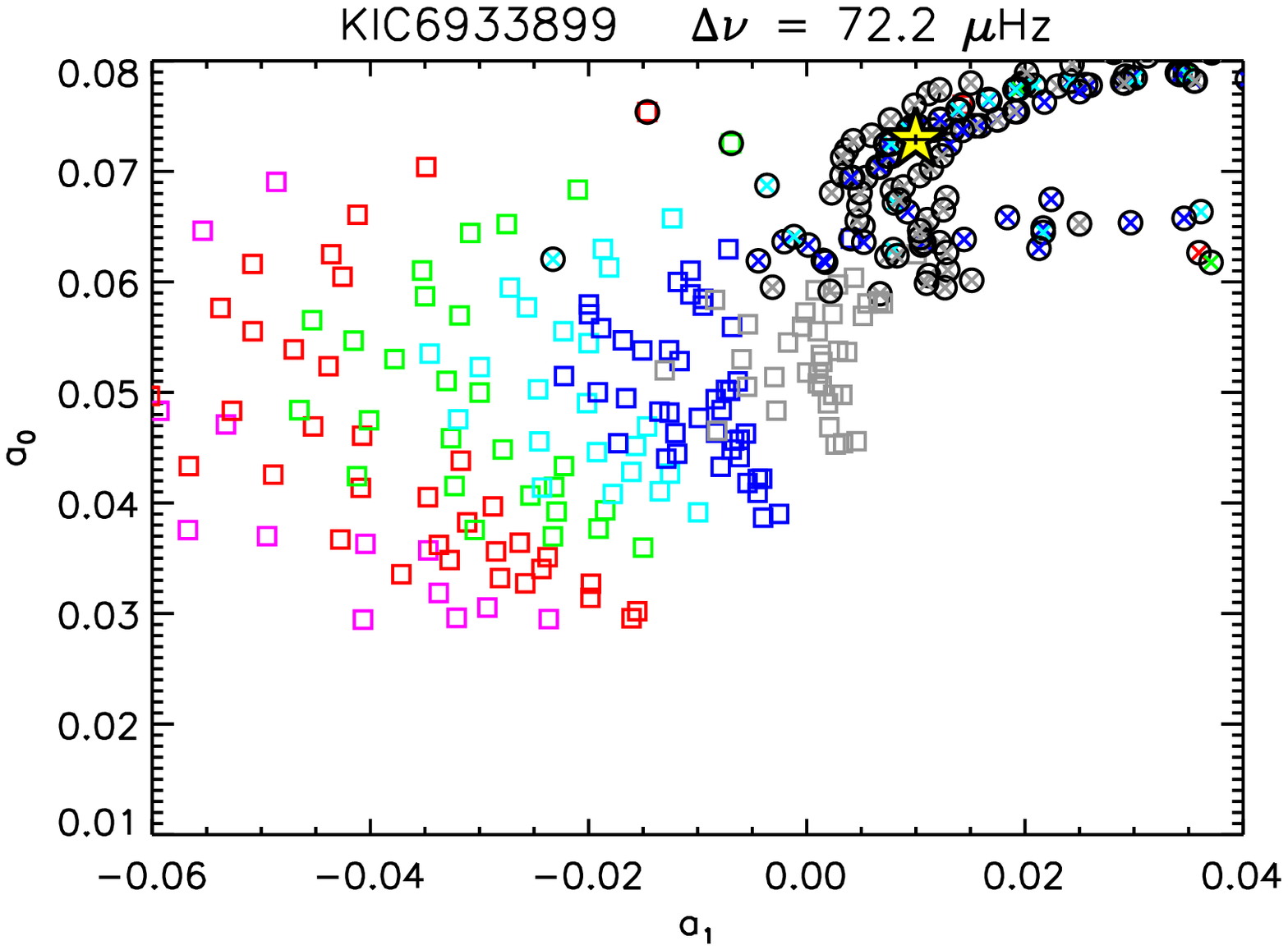}
\includegraphics[width=8cm]{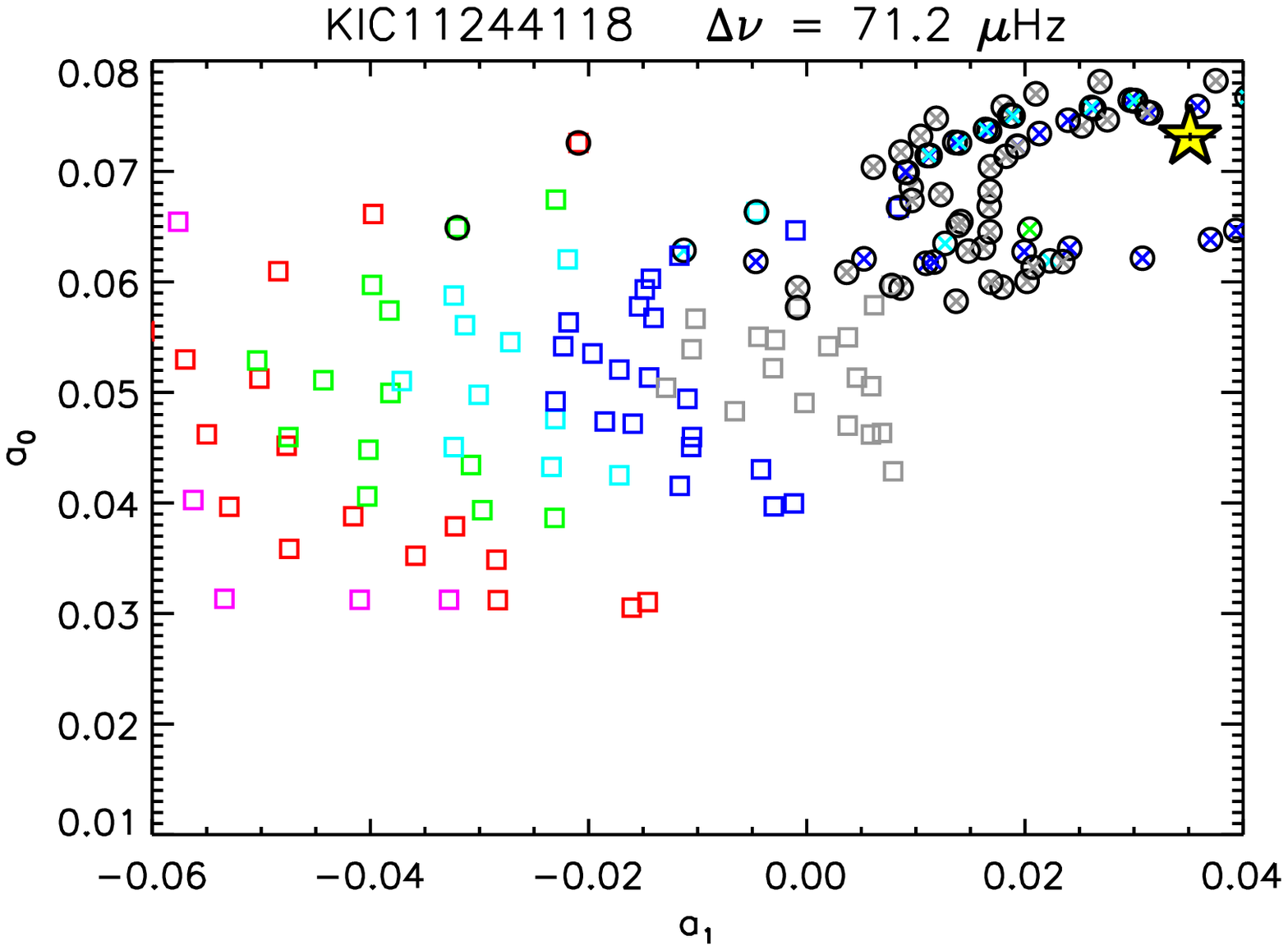}
\end{center}
\caption{Location in the $(a_1,a_0)$ plane of the last four PoMS stars of the sample. The symbols have the same meaning as in Fig. \ref{fig_ratio_kepler1}.
\label{fig_ratio_kepler4}}
\end{figure*}

\subsection{\cesam\ models \label{sect_cesam}}

For each star of the sample, we selected among the grid described in Sect. \ref{sect_descript_grid} the models that have a surface metallicity within 3 $\sigma $ of the spectroscopic $[$Fe/H$]$ (all metallicities were included in the cases where no spectroscopic measurement was available), and a stellar mass within 3 $\sigma$ of the estimate obtained from scaling laws (see Table \ref{tab_param}). Among the selected evolutionary sequences, we retained only the models whose mean large separations bracket the observed $\Delta\nu$. We note that for both models and observations, the mean value of $\Delta\nu$ was estimated using only the modes below $\nu_{\rm max}$ so that the corresponding large separations are only slightly affected by near-surface effects. 

For the selected models, we fitted polynomials to the $r_{010}$ ratio as defined by Eq. \ref{eq_poly}. For this purpose, we used the same modes and the same values of $\beta$, $\gamma_1$, and $\gamma_2$ (see Eq. \ref{eq_poly}) as those found from the observations, so that the parameters $a_i$ of the models can be directly compared to the observed ones. Since the models that we retained do not exactly match the observed large separation, we performed an interpolation to obtain the parameters $a_i$ that correspond exactly to the observed $\Delta\nu$. This process was repeated for all the stars of the sample. Fig. \ref{fig_ratio_kepler1} through \ref{fig_ratio_kepler4} show the location of the selected models and the observations in the $(a_1,a_0)$ plane. 
%For evolved stars, the $a_2$ parameter is not negligible and should also be used to compare observations and models. We ignored it in this section, it is used when performing optimizations (see Sect. \ref{sect_optim}).

The first comforting observation is that all the observed stars occupy a place in the $(a_1,a_0)$ plane that is populated by models. This shows that in all cases, there exist models that simultaneously reproduce the observed trend of the $r_{010}$ ratio and the other global observational constraints. 

Secondly, as anticipated in the previous section, the evolutionary status of the observed stars can be unambiguously established in most cases using the diagnostic from the $r_{010}$ ratios. For 13 stars of the sample, the profile of the $r_{010}$ ratio is only compatible with MS models, the PoMS models lying at least several $\sigma$ away in the $(a_1,a_0)$ plane (see Fig. \ref{fig_ratio_kepler1} and \ref{fig_ratio_kepler2}). Conversely, 10 stars are clearly in the PoMS phase judging by their location in the $(a_1,a_0)$ plane (see Fig. \ref{fig_ratio_kepler3} and \ref{fig_ratio_kepler4}). We stress that it was not obvious at first sight that these 10 stars are in the subgiant phase. Indeed, the PoMS status of solar-like pulsators is generally established by the presence of mixed modes in their oscillation spectrum. However, at the beginning of the subgiant phase, the lowest order g modes have not yet reached the frequency range of observed modes and such a diagnostic cannot be applied. It is the case for these 10 stars of the sample, and we here showed that the general trend of the $r_{010}$ ratios is a powerful diagnostic for the evolutionary status in this case. The evolutionary status remains ambiguous only for one star of the sample, KIC9410862, which is either at the end of the main sequence or at the beginning of the subgiant phase (Fig. \ref{fig_ratio_kepler2}). 

Among the 13 MS targets, eight have values of the parameters $a_0$ and $a_1$ that can be reproduced only by models that have a convective core. These stars are listed in Table \ref{tab_convcore} and their locations in the $(a_1,a_0)$ plane are shown in Fig. \ref{fig_ratio_kepler1}. As predicted in Sect. \ref{sect_test_diagnostic}, we were able to use the position in the $(a_1,a_0)$ plane of the stars that have a convective core to obtain an estimate of the amount of core overshooting. 
%This is all the more true for stars that have a large separation below about 95 $\mu$Hz. 
Interestingly, the eight stars draw a quite consistent picture of the extension of convective cores in low-mass stars.
\begin{itemize}
\item {All the targets require an extended core compared to the classical Schwarzschild criterion}. Indeed, all the stars that have a convective core lie several $\sigma$ away from models computed without overshooting.
\item {None of the targets were found to be consistent with a core overshooting above $\aov=0.2$}.
\item The only target which is consistent with a core overshooting around $\aov=0.2$ (KIC7206837) corresponds to the highest-mass star of the sample ($1.54\pm0.09\,M_\odot$ according to seismic scaling relations). This raises the question of a potential mass-dependence of the amount of core overshooting as implemented in the evolution code \cesam, which is addressed in more details in Sect. \ref{sect_calibrate}.
\end{itemize} 

We stress that seismology provides information about the size of the mixed core at the current age of the star. The amounts of overshooting that are quoted above are those required so that the evolution code \cesam\ produces cores with an appropriate size. One should be careful that the values that were obtained for $\aov$ hold only for the prescription of core overshooting that is implemented in \cesam\ and they should not be directly applied to other codes. We discuss this point in details in Sect. \ref{sect_calibrate}.

\begin{figure}
\begin{center}
\includegraphics[width=9cm]{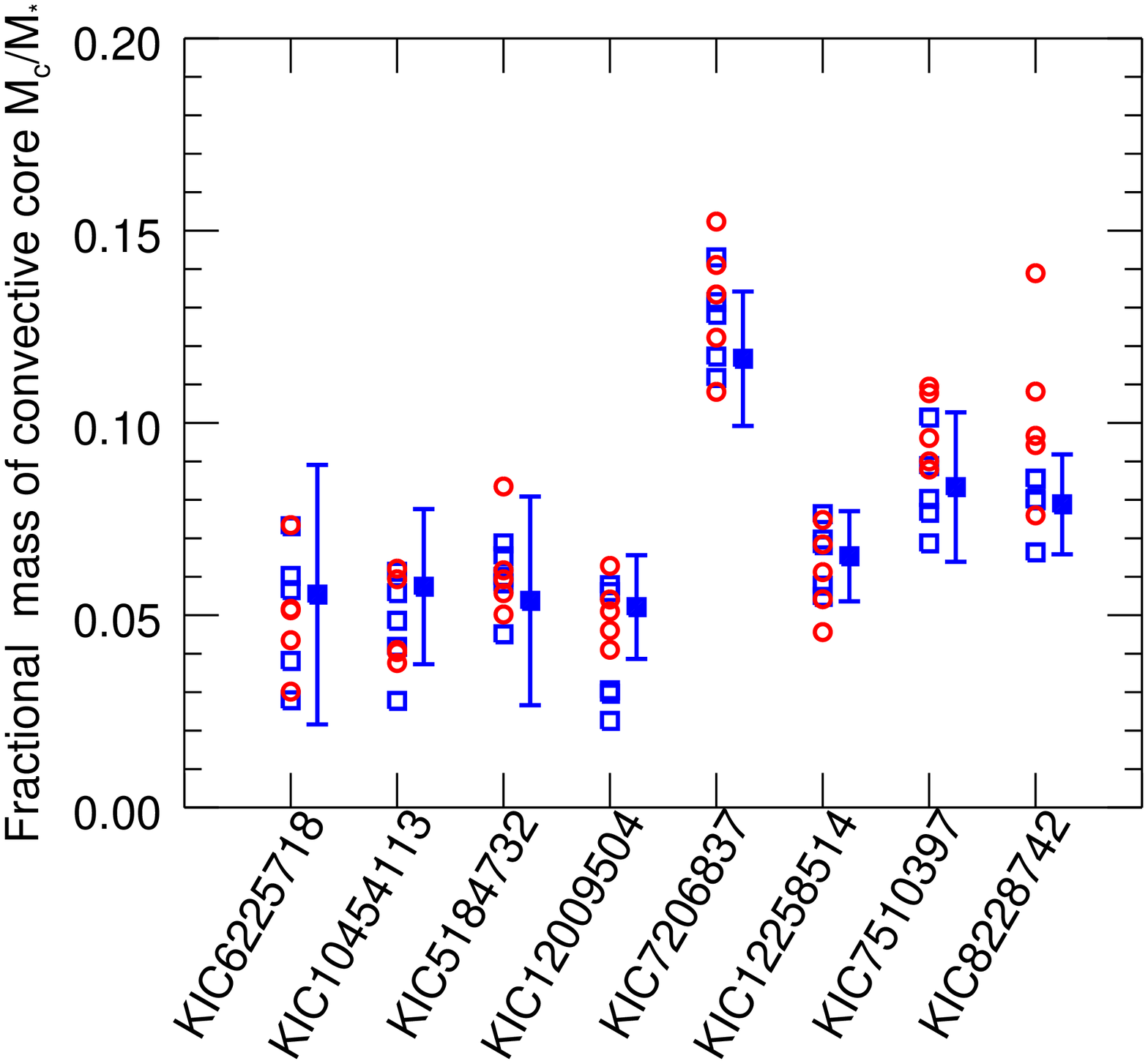}
\end{center}
\caption{Fractional mass of the convective core for the eight stars that were found to have a convective core in this study. For each star, the open symbols correspond to the core size of the five models of the two grids (blue squares for \cesam\ models, red circles for \textsc{mesa} models) that yield the lowest values of $\chi^2$ as defined by Eq. \ref{eq_chi2}. The filled squares give the core sizes obtained from a Levenberg-Marquardt optimization and the evolution code \cesam\ (see Sect. \ref{sect_optim}).
\label{fig_compare_cc}}
\end{figure}

A more relevant result to quote is the extent of the mixed core obtained from seismic constraints. To determine this for each of the stars for which a convective core was detected, we selected a subset of optimal models from the grid of models, defined as those that minimize the quantity
\begin{equation}
\chi^2 = \sum_{i=1}^N \frac{(\mathcal{O}_i^{\rm mod}-\mathcal{O}_i^{\rm obs})^2}{\sigma_i^2}
\label{eq_chi2}
\end{equation}
where the $\mathcal{O}_i^{\rm obs}$ correspond to the $N$ observables used to constrain the models, namely the effective temperature $T_{\rm eff}$, the surface metallicity $(Z/X)$ (if available), the asteroseismic $\log g$, and the parameters $a_0$ and $a_1$ of the $2^{\rm nd}$ order polynomial fit of the observed $r_{010}$ ratio. The $\sigma_i$ are the measurement errors, and the $\mathcal{O}_i^{\rm mod}$ are the values corresponding to the observables computed from the models. We note that the observables can be regarded as independent (since we fitted a sum of orthogonal polynomials to the observed ratios) so that Eq. \ref{eq_chi2} holds. For each star, the fractional mass of the convective core $M_{\rm c}/M_\star$ for the five best models is shown in Fig. \ref{fig_compare_cc} (blue squares for \cesam\ models). We note that the spreads in $M_{\rm c}/M_\star$ observed in Fig. \ref{fig_compare_cc} cannot be interpreted as uncertainties on this quantity. Indeed, to estimate proper uncertainties one should have chosen the set of optimal models based on the variations of the $\chi^2$ function compared to the lowest value of $\chi^2$ in the grid ($\Delta\chi^2=1$, 4, and 9 provide 1, 2, and 3~$\sigma$ errors, respectively) but the grid computed here is too coarse to make such an approach possible\footnote{Proper uncertainties on the core sizes are obtained from optimizations in Sect. \ref{sect_optim}}.

\subsection{\textsc{MESA} models \label{sect_mesa}}

\begin{figure*}
\begin{center}
\includegraphics[width=8cm]{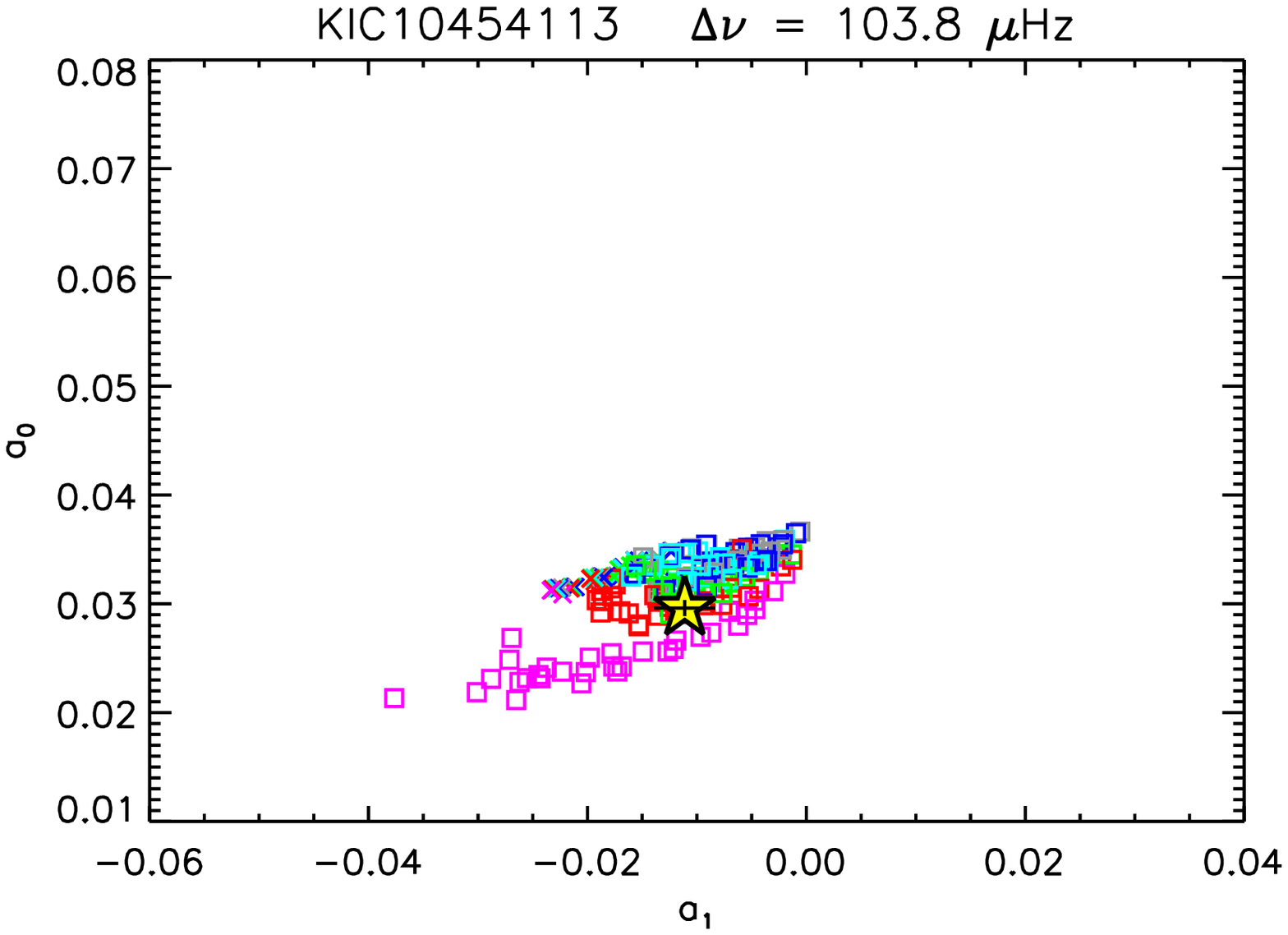}
\includegraphics[width=8cm]{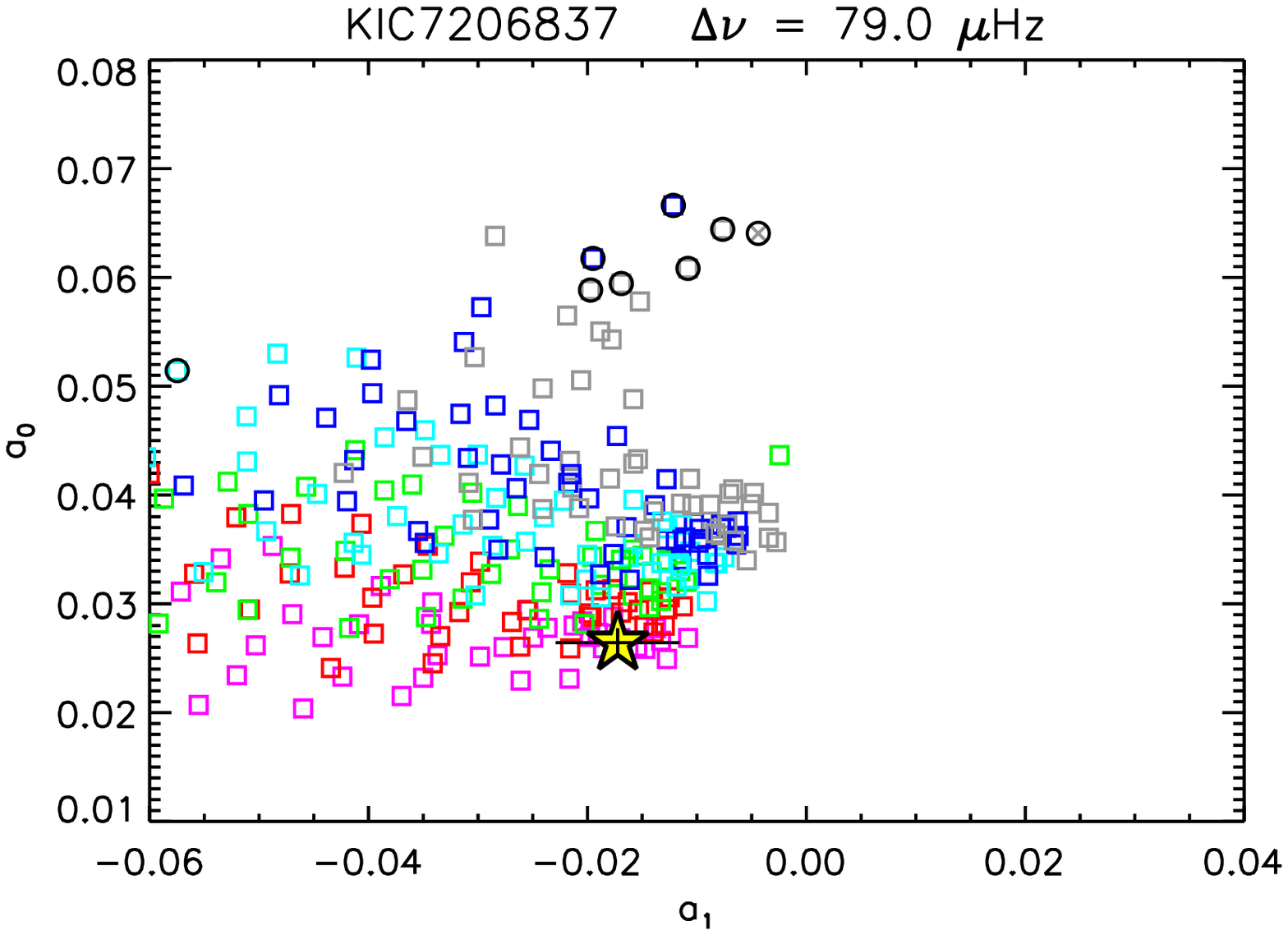}
\includegraphics[width=8cm]{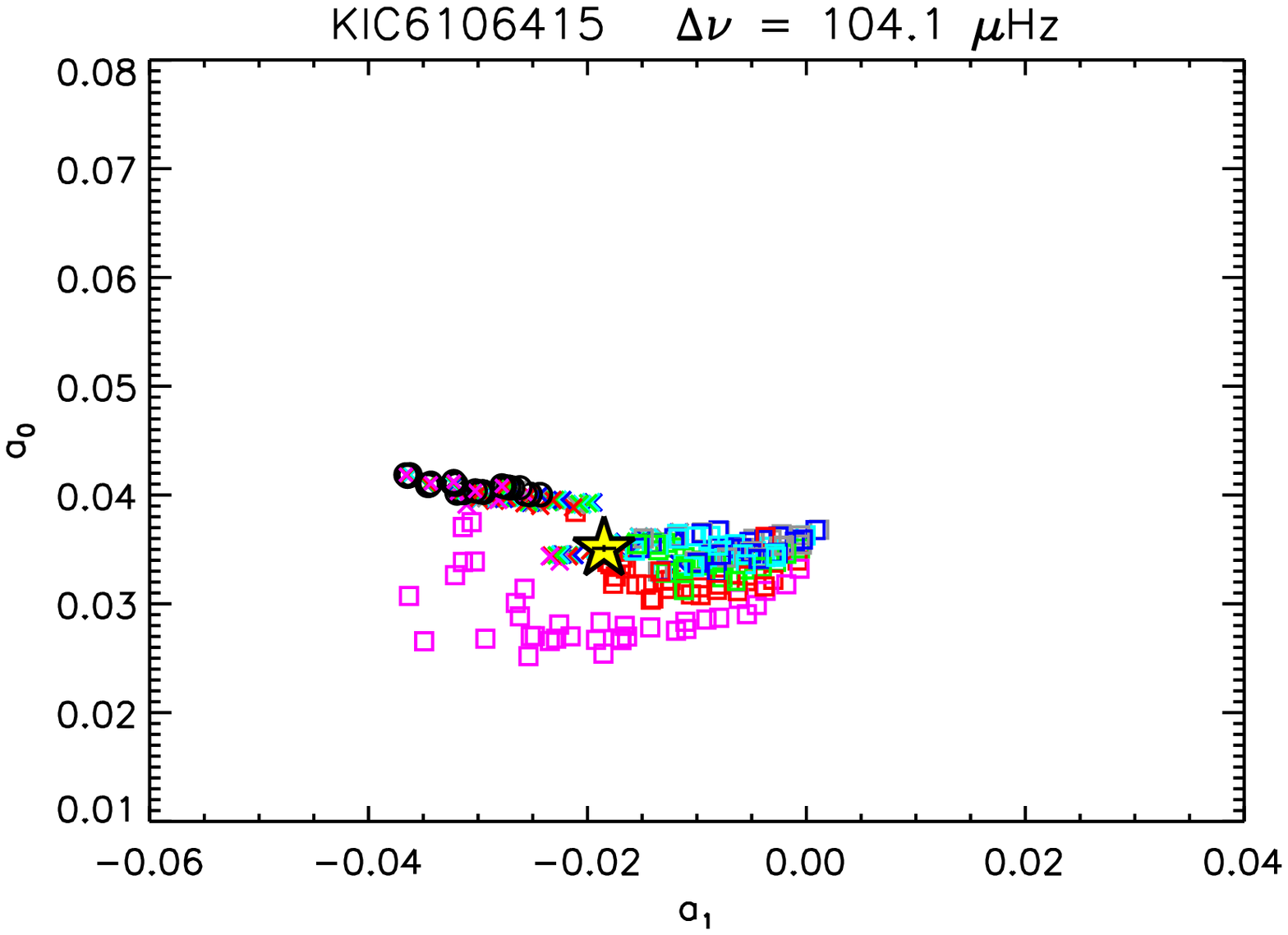}
\includegraphics[width=8cm]{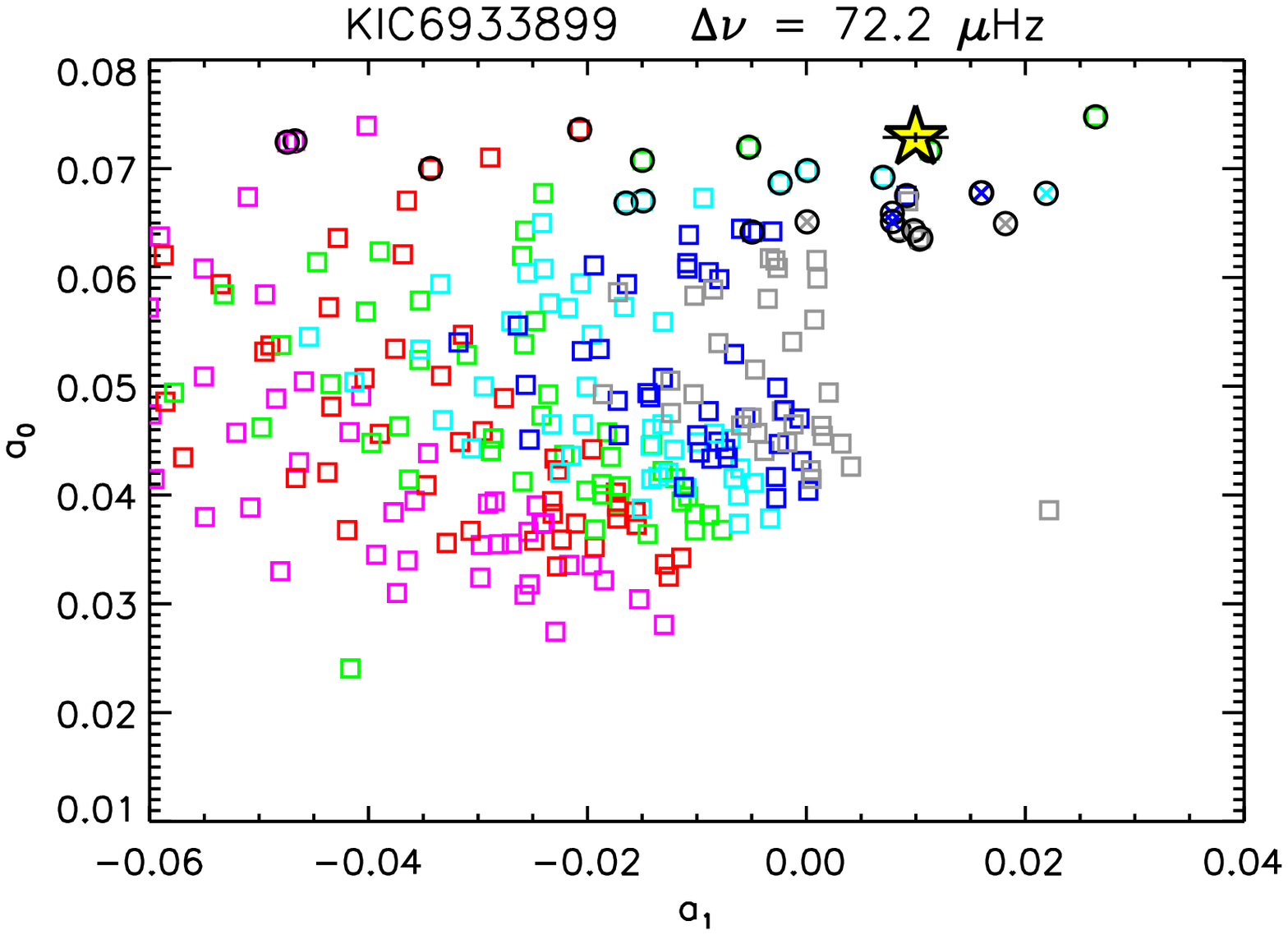}
\end{center}
\caption{Location in the $(a_1,a_0)$ plane of four stars of the sample compared to the location of models computed with the evolution code \mesa. Symbols are the same as in Fig. \ref{fig_ratio_kepler1}, except for the colors, which indicate diffusive overshooting parameters of: $f=0.004$ (gray), 0.010 (blue), 0.016 (cyan), 0.022 (green), 0.028 (red), 0.035 (magenta).
\label{fig_ratio_kepler_isa}}
\end{figure*}

As mentioned above, we have also computed a second grid of models with the evolution code \mesa\ (\citealt{paxton11}, \citealt{paxton13}). 

The \mesa\ models were computed using the OPAL 2005 equation of state from the tables of \cite{rogers02}, which are completed at lower temperature by the tables of \cite{saumon95}. \mesa\ opacity tables are constructed by combining radiative opacities with the electron conduction opacities from \cite{cassisi07}. Radiative opacities are taken from \cite{ferguson05} for $2.7<\log T<4.5$ and OPAL opacities \citet{iglesias93,iglesias96} for $3.75<\log T<8.7$. The low temperature opacities of \cite{ferguson05} include the effects of molecules and grains on the radiative opacity. The nuclear reaction rates module from \mesa\ contains the rates computed by \cite{caughlan88} and \cite{angulo99} (NACRE), with preference given to the NACRE rates when available. The atmosphere was described as Hopf's gray law. We used the solar mixture from \cite{grevesse93}. Convection was treated using the classical mixing-length theory (MLT, \citealt{bohm58}) with a fixed mixing length parameter $\alpha_{\rm MLT}=1.9$, which corresponds to a solar calibration (\citealt{paxton11}). 

Core overshooting is included and described as a diffusive process, following \cite{herwig00}. For this purpose, an extra diffusion is added at the edge of the core, with a coefficient
\begin{equation}
D_{\rm ov}(r) = D_0 \exp \left[ -\frac{2(r-r_{\rm s})}{f H_P} \right]
\end{equation}
where $D_0$ is the MLT-derived diffusion coefficient near the Schwarzschild boundary, $H_P$ is the pressure scale height at this location, and $f$ is the adjustable overshooting parameter. To avoid unrealistically large extensions of convective cores, the current version of \mesa\ uses a modified value $\widetilde{H}_P$ for the pressure scale height, defined as
\begin{equation}
\widetilde{H}_P = r_{\rm s}/\alpha_{\rm MLT}
\label{eq_hptilde}
\end{equation}
in the case where the mixing length $\ell_{\rm MLT} = \alpha_{\rm MLT} H_P$ becomes larger than the Schwarzschild limit $r_{\rm s}$ of the core. This prescription is different from the one adopted in the \cesam\ code. When using the same prescription for core overshooting (instantaneous or diffusive) and the same overshooting parameter at the boundary of small convective cores, the approach followed by \mesa\ is expected to yield core extensions that are smaller by a factor $\alpha_{\rm MLT}$ compared to the extensions produced with the \cesam\ approach in the saturated regime.

Gravitational settling and chemical diffusion are taken into account by solving the equations of \cite{burgers69} using the method and diffusion coefficients of \cite{thoul94}.

For each star of the sample, we performed the same model selection as was done with \cesam\ models, and for each selected model we fitted $2^{\rm nd}$-order polynomials to the $r_{010}$ ratios in the same way as described in Sect. \ref{sect_cesam}. This allowed us to compare the location of the observed stars in the $(a_1,a_0)$ plane to that of \mesa\ models. Fig. \ref{fig_ratio_kepler_isa} shows the  results obtained for four stars of the sample, which are representative of the different cases identifid in Sect. \ref{sect_cesam}: KIC8228742 and KIC7206837 are in the MS and have a convective core, KIC6106415 is in the MS but has no convective core, and KIC6933899 is in the PoMS.

The \mesa\ grid agrees with the \cesam\ grid on the evolutionary status of all the stars of the sample. The star KIC94110862, whose evolutionary status was uncertain based on \cesam\ models, was found to be more consistent with models shortly after the end of the MS using the \mesa\ grid. Additionally, the eight stars identified as having a convective core with the \cesam\ grid were also found to have one with the \mesa\ grid. The locations of two of these stars in the $(a_1,a_0)$ plane are shown in the upper panels of Fig. \ref{fig_ratio_kepler_isa}. It is clear that the extension can be estimated from the $a_0$ and $a_1$ parameters, as was claimed in Sect. \ref{sect_cesam}. Interestingly, all the conclusions reached with \cesam\ models about the amount of overshooting that is required are confirmed. The eight stars with convective cores all require an extended core with overshooting parameters ranging from 0.010 to 0.035, and the star that requires the largest amount of overshooting corresponds to the highest-mass stars of the sample (KIC7206837) as was found in Sect. \ref{sect_cesam}.

Obviously, the overshooting parameters obtained from the \mesa\ models are not directly comparable to those found from the \cesam\ grid because a diffusive overshooting was chosen in \mesa\ models. A more detailed comparison is provided in Sect. \ref{sect_cesam_vs_mesa}, but we can already compare directly the absolute sizes of the extended cores found with both evolution codes. For all the stars that have a convective core, we selected the five models of the \mesa\ grid that minimize the $\chi^2$ function as defined by Eq. \ref{eq_chi2}. The fractional mass of the mixed core $M_{\rm c}/M_\star$ in these models is shown in Fig. \ref{fig_compare_cc}. Interestingly, there is a quite good agreement on the size of the extended cores obtained with both evolution codes, in spite of the different prescriptions for core overshooting. This is further indication that the seismic diagnostic based on $r_{010}$ ratios can provide a measurement of the size of the mixed core mostly independently of the input physics, as was already suggested by \cite{silva11}.

\subsection{Instantaneous vs diffusive mixing beyond convective cores \label{sect_diff_vs_step}}

\begin{figure}
\begin{center}
\includegraphics[width=9cm]{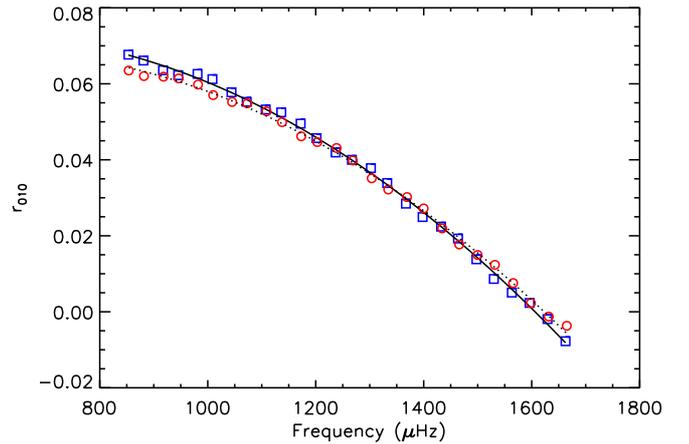}
\end{center}
\caption{Profile of $r_{010}$ for two \mesa\ models: one with a diffusive overshooting ($f=0.02$) and a 1.3-$M_\odot$ mass (blue squares) and the other with a step overshooting ($\aov=0.22$) and a 1.31-$M_\odot$ mass (red circles). Both models have the same mean large separation.
\label{fig_r010_diff_vs_step}}
\end{figure}

In this study, we have chosen to adopt two different prescriptions for core overshooting, an instantaneous overshooting (\cesam\ models) and a diffusive overshooting (\mesa\ models), with the aim to confront the two most frequently used prescriptions to \kepler\ data. It is interesting to address the question whether we can distinguish between these two types of mixing beyond convective cores using $r_{010}$ ratios. The \mesa\ code offers the possibility to test this since both treatments have been implemented. We computed a 1.3-$M_\odot$ \mesa\ model including diffusive overshooting with a parameter $f=0.020$, which we evolved until $X_{\rm c}$ has dropped to 0.2 (chosen arbitrarily). We also computed a \mesa\ model including a step overshooting with $\aov=0.22$ and a slightly higher mass (1.31 $M_\odot$) evolved until it has the same large separation as the diffusive-overshooting model. We found that both models are undistinguishable from an observational point of view (within typical observational errors), and they also share a very similar behavior of the $r_{010}$ ratios, as is shown in Fig. \ref{fig_r010_diff_vs_step}. This shows that the seismic diagnostic based on the $r_{010}$ ratios is unfortunately not capable of distinguishing between the two scenarios regarding the nature of the extra mixing beyond the core. 

%\begin{table}
%  \begin{center}
%  \caption{ \label{tab_diff_vs_step}}
%\begin{tabular}{l c c }
%\hline \hline
%\T  $T_{\rm eff}$ (K) & 5967 & 6002 \\
%$L/L_\odot$ & 3.677 & 3.763 \\
%$\log g$ (dex) & 4.042 & 4.045 \\
%$(Z/X)_{\rm surf}$ & & \\
%\hline 
%\end{tabular}
%\end{center}
%\end{table}

\section{Toward a calibration of core overshooting for low-mass stars \label{sect_calibrate}}

%\section{What can we learn about the extra mixing beyond convective cores?  \label{sect_calibrate}}

%\subsection{Instantaneous vs diffusive mixing}
%
%In Sect. \ref{sect_grid}, we have modeled the extension of convective cores either as an instantaneous mixing (\cesam\ models) or as a diffusive mixing (\mesa\ models). An interesting question is whether or not we can distinguish between these two approaches. At first sight, 

%\subsection{Efficiency of extra-mixing beyond cores in low-mass stars}

In Sect. \ref{sect_grid}, we were able to measure the sizes of mixed cores in eight low-mass stars using seismology. The question is then how these results can be used to estimate the efficiency of the extra-mixing beyond convective cores. Answering this question is not straightforward. One could consider simply comparing the convective core masses obtained in Sect. \ref{sect_grid} to the convective core masses that would be obtained with identical stellar parameters but no mixing beyond the core. This is however inapplicable in practice because increasing the size of the convective core at the beginning of the main sequence has large subsequent effects on its composition and evolution. For stars in the mass range that we considered here, the main effect is that the abundance of $^3$He in the core increases, which increases its luminosity, and thus also its size because the Schwarzschild radius increases. For instance, extending the convective core of a 1.3-$M_\odot$ star over 10\% of the Schwarzschild radius in fact results in an increase of the core radius of as much as 50\% during the main sequence. Another consequence is that the lifetime of small convective cores can be dramatically extended (see \citealt{roxburgh85}, \citealt{deheuvels10a}). For instance, a stellar model of KIC62245718 computed with the same stellar parameters as those found in Sect. \ref{sect_grid} but without including any extra-mixing beyond the core has lost its convective core at current age.

It therefore seems that the problem of the efficiency of convective core extensions cannot be studied independently from the evolution of the star, even though asteroseismology only tells us about the size of the mixed core at current age. We thus chose to estimate the efficiency of the extra-mixing beyond the core by adjusting the overshooting parameter ($\aov$ for instantaneous mixing or $f$ for diffusive mixing) considered constant throughout the evolution, so that stellar models have the right convective core size at current age. As mentioned in the introduction, so far we have had to model convective core extensions using such simplistic parametric models because we lack observational constraints that would justify using more complex models. Our aim in this section is to search for correlations between the efficiency of overshooting and properties of stellar interiors, which might eventually give us better insight on the physical processes that are responsible for core extensions, and lead us to prefer more realistic modelings of this phenomenon. On the shorter term, this type of study can enable us to propose a calibration of the overshooting parameter, which can later be used in 1D stellar models.

%We start by pointing out differences in the prescriptions used to extend small convective cores in stellar models (Sect. \ref{sect_ov_lowmass}). We then show how \kepler\ data can be quantitatively used to obtain a calibration of core overshooting for low-mass stars with the \cesam\ code (Sect. \ref{sect_optim}) and we address the question whether it can be adapted to \mesa\ in Sect. \ref{sect_cesam_vs_mesa}.

%This opens the interesting opportunity to try to calibrate core overshooting in the mass range covered by the stars of our sample. This question is directly related to the consistency in the way core overshooting is implemented in evolution codes. So far, few comparisons have been made between the sizes of convective cores predicted by the available stellar evolution codes when overshooting is included. For 2-$M_\odot$ stars, \cite{lebreton08b} have compared the size of the convective core along the evolution for three different codes (\cesam, \textsc{CLES}, and \textsc{ASTEC}). For a core overshooting of $\aov=0.15$, they found a reasonable agreement between the three codes. The situation is however less clear for lower-mass stars, which have smaller convective cores.

\subsection{Calibration of core overshooting in \cesam \label{sect_optim}}

To calibrate core overshooting in \cesam\ models, we needed to obtain more quantitative estimates of the amounts of core overshooting that are required for the stars of the sample.

\subsubsection{Stars with a convective core \label{sect_convcore}}

We performed optimizations for the eight stars that were found to have a convective core in Sect. \ref{sect_grid}. For this purpose, we used the Levenberg-Marquardt algorithm, which is an appealing alternative to grid-search minimization when the number of free parameters is large. This algorithm combines the low sensitivity to initial guesses of the gradient search method and the rapidity of convergence of the Newton-Raphson method. Its use has first been suggested for the purpose of stellar modeling by \cite{miglio05}. The main drawback of such an optimization technique is the risk to converge toward a secondary minimum of the cost function if the initial guesses are to far from the optimum set of parameters. In our particular case, this risk is minimized since we used the best models of the grid computed in Sect. \ref{sect_cesam} as initial guesses. 

To find optimal models, we minimized the quantity $\chi^2$ as defined in Eq. \ref{eq_chi2}. %The observables that were considered in the $\chi^2$ computation in Sect. \ref{sect_mesa} do not include the observed large separation because the models of the grid were already selected to match it. 
We used the same observables as those listed in Sect. \ref{sect_mesa}, to which we added the frequency of the lowest-order observed radial mode. This observable is preferred to the observed mean large separation because of its lower dependence on the structure of the outer layers. We note that the $a_2$ parameter of the $2^{\rm nd}$ order polynomial fit of the observed $r_{010}$ ratio was here included as a constraint. This parameter becomes constraining for evolved stars, for which the observed $r_{010}$ ratios depart from a simple linear relation (see Fig. \ref{fig_ratio_parabola}). To reproduce these observables five parameters were left free: the stellar mass, age, initial helium abundance $Y_{\rm i}$, initial metallicity $(Z/X)_{\rm i}$, and the parameter of core overshooting $\alpha_{\rm ov}$. We imposed a lower limit of 0.24 for $Y_{\rm i}$ in order to exclude models with initial helium abundances significantly below the standard big bang nucleosynthesis (SBBN) values of $Y_0=0.248\pm0.007$ (\citealt{steigman10}). To limit the number of free parameters, we kept the mixing length fixed to $\alpha_{\rm CGM}=0.64$, which was obtained from a solar calibration. As a consequence, the fit that we performed has two degrees of freedom and a reduced value $\chi^2_{\rm red}$ was thus obtained by dividing the regular $\chi^2$ by two. For each star, two types of optimizations were performed, one where the effects of microscopic diffusion are neglected, and another that includes these effects following the formalism of \citealt{burgers69}. This procedure enabled us to test the influence of microscopic diffusion on the amount of core overshooting that is required. As mentioned above, diffusion increases the abundance of heavy elements in the core and thus the opacity, which results in an increase in the size of the convective core. We therefore expected to require less core overshooting when microscopic diffusion is included. Since \cesam\ does not include the computation of radiative accelerations of chemical elements, their effect was neglected in this study. Since radiative levitation acts against gravitational settling in the interior of stars with masses above about 1.2 $M_\odot$, our models including microscopic diffusion likely overestimate the sinking of heavy elements in this mass range. We thus expect our models computed with microscopic diffusion and stellar masses above $1.2\,M_\odot$ to provide us with an upper limit to the effects of diffusion, in particular on the sizes of convective cores.

%\modif{In this section, we have neglected the effects of microscopic diffusion, despite its expected impact on the size of the mixed core. The reason for doing so is that the efficiency of atomic diffusion in stars hotter than the Sun is still a matter of debate. For stellar models with thin convective envelopes, atomic diffusion alone is known to produce an important depletion of helium and heavier elements at the surface, which is at odds with observations. The effects of atomic diffusion on our results are assessed in Sect. \ref{sect_sensitivity}}.
%To prevent the quick draining of heavy elements in models with thin convective envelopes, we added an extra mixing diffusion coefficient taken equal to the kinematic radiative viscosity, as prescribed by \citealt{morel02}.

%There are two major advantages of having computed a grid of models prior to the optimization process: first it provided good initial guesses, which ensured an easy convergence of the algorithm, and secondly it made it less likely to reach a secondary minimum of the $\chi^2$ function.

\begin{figure*}
\begin{center}
\includegraphics[width=9cm]{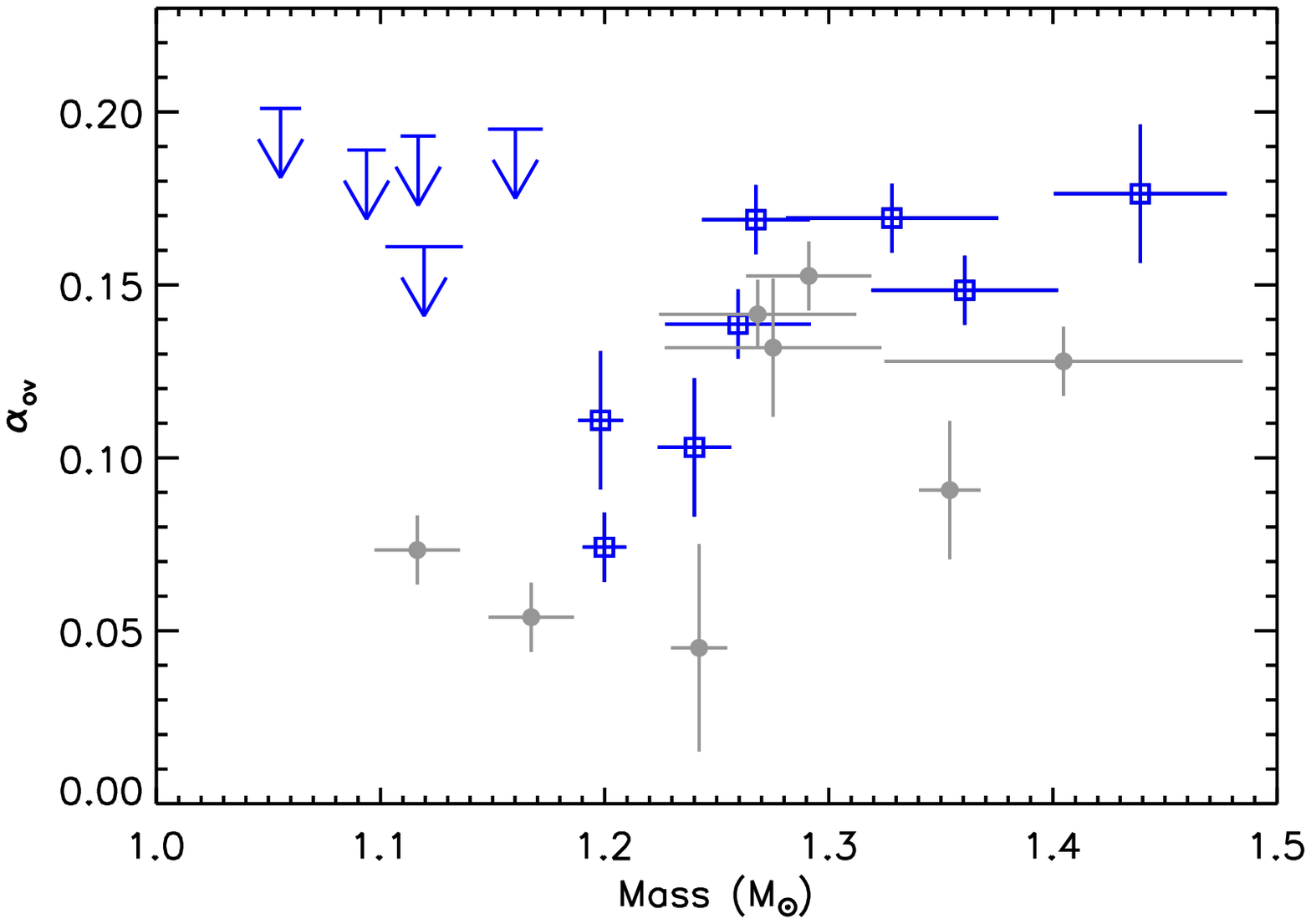}
\includegraphics[width=9cm]{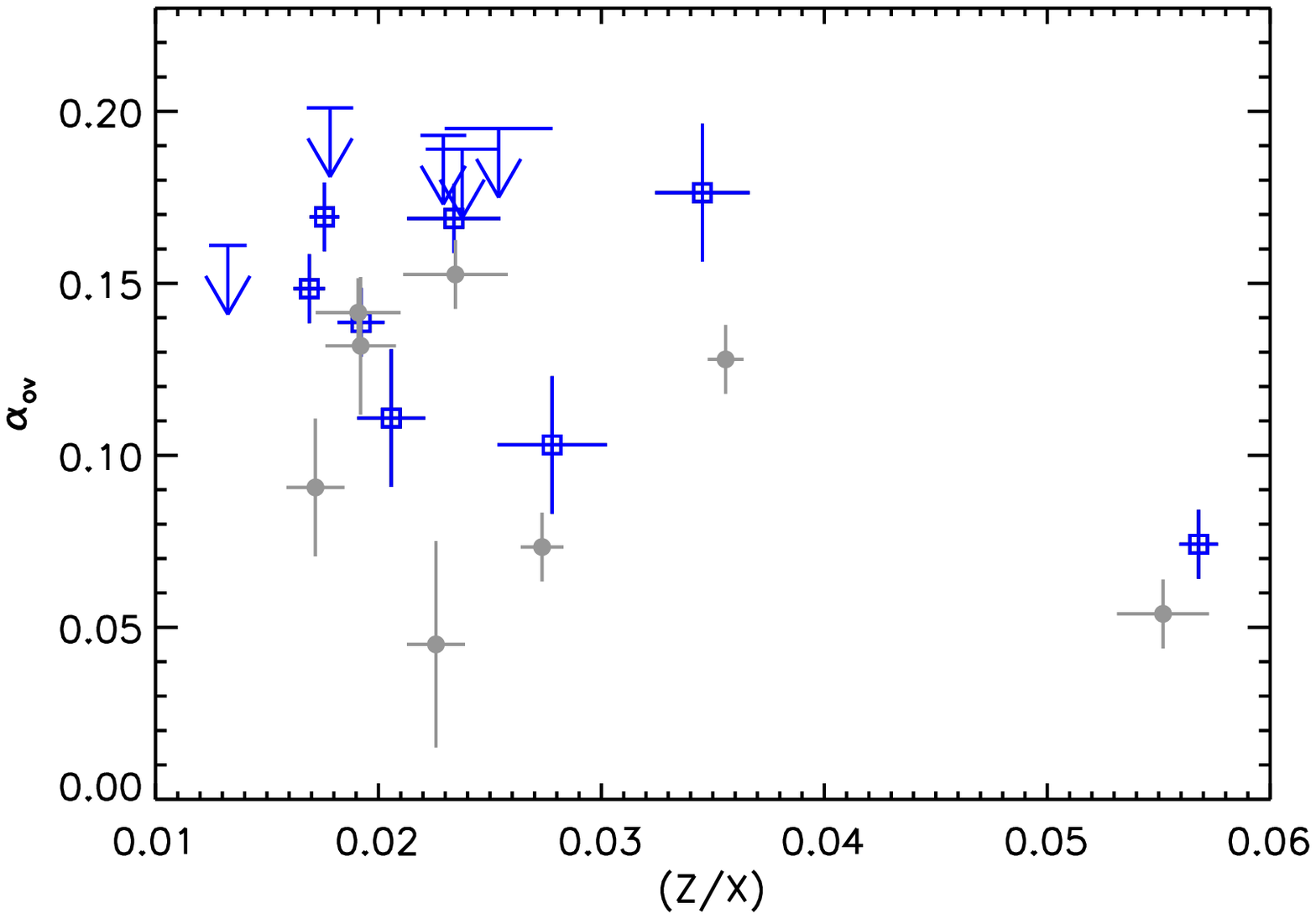}
\end{center}\caption{Amount of core overshooting found for the stars of the sample that have a convective core as a function of the fitted stellar mass (\textbf{left}) and as a function of the fitted initial metallicity (\textbf{right}). Blue squares indicate models computed without microscopic diffusion and grey circles, models where microscopic diffusion is included following \cite{burgers69}. The vertical arrows indicate upper limits of $\aov$ (see Sect. \ref{sect_noconvcore}).
\label{fig_mass_ov}}
\end{figure*}

The parameters of the best-fit models are given in Table \ref{tab_convcore}. The quoted error bars were obtained as the diagonal coefficients of the inverse of the Hessian matrix. The results confirm that the amount of core overshooting can be well constrained by using the parameters $a_i$. The values obtained for $\aov$ range from 0.07 to 0.18 in the case without diffusion, which is in good agreement with the results of the grids of models (Sect. \ref{sect_cesam}). As foreseen, the models that include microscopic diffusion require lower amounts of core overshooting to reproduce the seismic observations, with values ranging from 0.05 to 0.15. However, our results show that the effects of diffusion cannot account in themselves for the entire extension of convective cores since core overshooting was required for all eight stars of the sample. We note that for several stars of the sample, the fitted value of the initial helium abundance $Y_{\rm i}$ coincides with the lower limit of 0.24 that we have imposed to avoid sub-SBBN helium abundances. Similar results have been found in several studies where seismic modelings were performed (e.g. \citealt{metcalfe14}, \citealt{silva15}). This is potentially the consequence of the well-known correlation between stellar mass and helium abundance (\citealt{lebreton12}). For these stars, we have performed additional fits imposing a higher limit to $Y_{\rm i}$ (0.26) and found results that agree within 1-$\sigma$ errors with the values quoted in Table \ref{tab_convcore} (in particular, we found very little difference in the sizes of convective cores, which is our main interest here). The optimizations also provided estimates of the stellar mass, which are given in Table \ref{tab_convcore}. The agreement with estimates from scaling laws is quite good (below 1.3~$\sigma$ for all the stars). We note that KIC12009504 was already modeled by \cite{silva13} who already found that this stars possesses a convective core that extends beyond the Schwarzschild limit. Our results for this star are in good agreement with those of \cite{silva13}. 

The values of $\chi^2_{\rm red}$ for some of our fits are significantly larger than 1, which in principle indicates either disagreements between models and observations, or underestimated error bars for the observables. Table \ref{tab_convcore} gives the level of agreement with observations for each fitted parameter normalized by observational 1-$\sigma$ errors. It shows that a very good level of agreement is reached for the $a_0$ and $a_1$ parameters, as was expected based on the results of Sect. \ref{sect_grid}. On the contrary,  disagreements above the 3-$\sigma$ level arise for the $a_2$ parameter. This occurs mainly for stars where the $r_{010}$ ratios vary nearly linearly with frequency, so that the $a_2$ coefficient is small. In this case, the observational estimate of $a_2$ can be altered by the short-period oscillation that arises because of the glitch at the base of the convection zone (see Sect. \ref{sect_fit_r010}). Disagreements above the 2-$\sigma$ level also arise for the effective temperature and the surface metallicity. We note that we have used the error bars of \cite{bruntt12} for these quantities, which have been deemed somewhat underestimated in previous studies (\citealt{silva13}). This might at least partly explain this disagreement. Also, we note that the agreement with the observed surface metallicities improves when including microscopic diffusion in models.

Using our optimizations, we could also obtain estimates of the total size of the mixed core in the eight stars. Since the size of the core is not a fitted parameter, the optimization algorithm does not directly provide error bars on these obtained values. However, they can be deduced from the relation
\begin{equation}
\sigma_{M_{\rm c}} = \sqrt{\sum_{j=1}^{P} \sigma_j^2\left(\frac{\partial M_{\rm c}}{\partial b_j}\right)^2}
\end{equation}
where the $b_j$ terms correspond to the $P$ free parameters and the derivatives $\left(\partial M_{\rm c}/\partial b_j\right)$ can be evaluated with the models used to compute the Hessian matrix. The fractional masses of the convective cores for the eight stars are plotted along with their error bars in Fig. \ref{fig_compare_cc}. 

The refined estimates of the amount of core overshooting in the eight stars that have a convective core enabled us to test correlations between the overshooting parameter $\aov$ and other stellar parameters. Fig. \ref{fig_mass_ov}a shows the obtained values of $\aov$ as a function of the stellar mass. We observe that there seems to be a tendency of core overshooting to increase with stellar mass in this mass range. This tendency is less clear for the models where microscopic diffusion was included (grey circles), but we still found in this case that the three less massive stars of the sample require less core overshooting that the five more massive ones. Clearly more data points are required to be more conclusive, but if such a tendency is confirmed, then an empirical law could be derived and implemented in the \cesam\ code in order to better model the extent of mixed cores for stars in this mass range. We also note that we have found no apparent dependency of the amount of core overshooting required with stellar metallicity (see Fig. \ref{fig_mass_ov}b).

\afterpage{
\begin{landscape}
\begin{table}
  \begin{center}
  \caption{Fitted parameters and characteristics of the best-fit models obtained for the stars of the sample that have a convective core. \label{tab_convcore}}
\begin{tabular}{| l c | c c c c c | c | c c c c c c | c |}
\hline 
\T & & \multicolumn{5}{ c |}{Free parameters} & Mixed core size & \multicolumn{6}{ c |}{Level of agreement normalized} & \\
\B & & & & & &  &   & \multicolumn{6}{ c |}{by observational 1-$\sigma$ errors} & \\
\hline
\T\B KIC ID & diffusion & $M$ $(M_\odot)$ &  $(Z/X)_0$ & $Y_0$ & Age (Myr) & $\aov$ & $M_{\rm c}/M_\star$ & $T_{\rm eff}$ &  $\log g$ &  $(Z/X)_{\rm s}$  & $a_0$ & $a_1$ & $a_2$ & $
\chi^2_{\rm red}$\\
\hline
%\T   9139151 & core? \\
%   8394589 & core? \\
\T  6225718 & no  & $1.26 \pm 0.03$ & $0.019 \pm 0.001$ & $0.24 \pm 0.01$ & $ 1603 \pm  225$ & $0.14 \pm 0.01$ & $0.06 \pm 0.03$ & $ 2.1$ & $ 0.5$ & $ 2.9$ & $ 0.2$ & $ 0.3$ & $ 2.4$ & $ 9.4$ \\
\B & B69 & $1.27 \pm 0.04$ & $0.019 \pm 0.002$ & $0.24 \pm 0.03$ & $ 1311 \pm  209$ & $0.14 \pm 0.01$ & $0.05 \pm 0.03$ & $ 2.2$ & $ 0.6$ & $ 2.0$ & $ 0.0$ & $ 0.0$ & $ 2.7$ & $ 8.3$ \\
\T 10454113 & no  & $1.27 \pm 0.02$ & $0.023 \pm 0.002$ & $0.24 \pm 0.02$ & $ 1761 \pm  145$ & $0.17 \pm 0.01$ & $0.06 \pm 0.02$ & $ 1.3$ & $ 0.6$ & $ 2.4$ & $ 0.2$ & $ 0.2$ & $ 1.3$ & $ 4.9$ \\
\B & B69 & $1.29 \pm 0.03$ & $0.023 \pm 0.002$ & $0.24 \pm 0.02$ & $ 1413 \pm  423$ & $0.15 \pm 0.01$ & $0.06 \pm 0.03$ & $ 2.6$ & $ 0.7$ & $ 1.7$ & $ 0.5$ & $ 1.1$ & $ 0.9$ & $ 6.2$ \\
\T  5184732 & no  & $1.20 \pm 0.01$ & $0.057 \pm 0.001$ & $0.31 \pm 0.01$ & $ 3957 \pm  416$ & $0.07 \pm 0.01$ & $0.05 \pm 0.01$ & $ 0.2$ & $ 0.2$ & $ 1.4$ & $ 0.1$ & $ 0.2$ & $ 0.9$ & $ 1.7$ \\
\B & B69 & $1.17 \pm 0.02$ & $0.055 \pm 0.002$ & $0.32 \pm 0.01$ & $ 3770 \pm  351$ & $0.05 \pm 0.01$ & $0.05 \pm 0.03$ & $ 0.2$ & $ 0.0$ & $ 0.5$ & $ 0.1$ & $ 0.1$ & $ 0.9$ & $ 0.6$ \\
\T 12009504 & no  & $1.20 \pm 0.01$ & $0.021 \pm 0.002$ & $0.25 \pm 0.02$ & $ 4275 \pm  939$ & $0.11 \pm 0.02$ & $0.06 \pm 0.02$ & $ 2.0$ & $ 0.2$ & $ 2.0$ & $ 0.1$ & $ 1.9$ & $ 0.6$ & $ 6.0$ \\
\B & B69 & $1.24 \pm 0.01$ & $0.023 \pm 0.001$ & $0.24 \pm 0.01$ & $ 3977 \pm  365$ & $0.05 \pm 0.03$ & $0.03 \pm 0.06$ & $ 1.1$ & $ 0.4$ & $ 1.4$ & $ 0.1$ & $ 1.1$ & $ 3.8$ & $ 9.4$ \\
\T  7206837 & no  & $1.44 \pm 0.04$ & $0.035 \pm 0.002$ & $0.25 \pm 0.02$ & $ 2250 \pm  147$ & $0.18 \pm 0.02$ & $0.12 \pm 0.02$ & $ 0.9$ & $ 0.5$ & $ 1.9$ & $ 1.1$ & $ 1.9$ & $ 1.0$ & $ 5.2$ \\
\B & B69 & $1.40 \pm 0.08$ & $0.036 \pm 0.001$ & $0.26 \pm 0.03$ & $ 2179 \pm  276$ & $0.13 \pm 0.01$ & $0.10 \pm 0.02$ & $ 0.1$ & $ 0.3$ & $ 0.1$ & $ 0.0$ & $ 0.2$ & $ 0.0$ & $ 0.1$ \\
\T 12258514 & no  & $1.24 \pm 0.02$ & $0.028 \pm 0.002$ & $0.28 \pm 0.02$ & $ 4472 \pm  138$ & $0.10 \pm 0.02$ & $0.07 \pm 0.01$ & $ 0.1$ & $ 0.5$ & $ 2.0$ & $ 0.5$ & $ 0.5$ & $ 3.3$ & $ 7.9$ \\
\B & B69 & $1.12 \pm 0.02$ & $0.027 \pm 0.001$ & $0.35 \pm 0.01$ & $ 3640 \pm  132$ & $0.07 \pm 0.01$ & $0.07 \pm 0.01$ & $ 2.1$ & $ 0.1$ & $ 0.6$ & $ 0.1$ & $ 0.2$ & $ 2.4$ & $ 5.4$ \\
\T  7510397 & no  & $1.36 \pm 0.04$ & $0.017 \pm 0.001$ & $0.24 \pm 0.02$ & $ 3385 \pm  129$ & $0.15 \pm 0.01$ & $0.08 \pm 0.02$ & $ 2.5$ & $ 0.8$ & $ 3.0$ & $ 0.6$ & $ 0.6$ & $ 1.1$ & $ 8.8$ \\
\B & B69 & $1.35 \pm 0.01$ & $0.017 \pm 0.001$ & $0.24 \pm 0.01$ & $ 3397 \pm   31$ & $0.09 \pm 0.02$ & $0.07 \pm 0.01$ & $ 1.6$ & $ 0.7$ & $ 1.2$ & $ 0.2$ & $ 0.4$ & $ 0.2$ & $ 2.4$ \\
\T  8228742 & no  & $1.33 \pm 0.05$ & $0.018 \pm 0.001$ & $0.24 \pm 0.03$ & $ 3968 \pm   88$ & $0.17 \pm 0.01$ & $0.08 \pm 0.01$ & $ 1.3$ & $ 0.2$ & $ 1.7$ & $ 0.3$ & $ 0.3$ & $ 0.6$ & $ 2.5$ \\
\B & B69 & $1.28 \pm 0.05$ & $0.019 \pm 0.002$ & $0.27 \pm 0.03$ & $ 3716 \pm   80$ & $0.13 \pm 0.02$ & $0.08 \pm 0.03$ & $ 1.2$ & $ 0.0$ & $ 1.0$ & $ 0.4$ & $ 0.5$ & $ 0.5$ & $ 1.6$ \\
\hline 
\end{tabular}
\end{center}
\end{table}
\end{landscape}
}

\subsubsection{Stars without a convective core \label{sect_noconvcore}}

\begin{figure}
\begin{center}
\includegraphics[width=9cm]{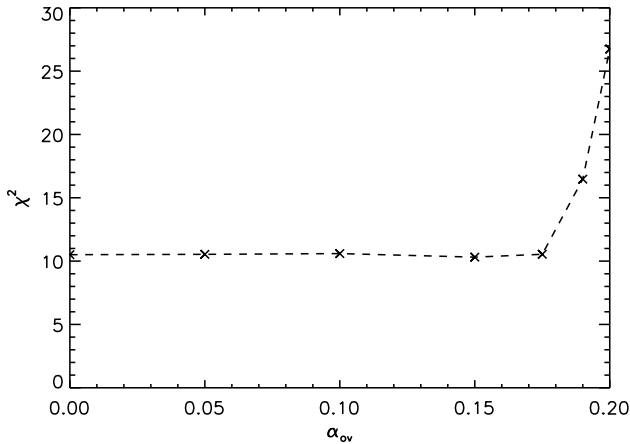}
\end{center}
\caption{Value of the $\chi^2$ of the best-fit model as a function of the (fixed) amount of core overshooting for the star KIC~10516096.
\label{fig_ov_chi2}}
\end{figure}

Information about core overshooting can also be drawn from stars that have no convective core but lie just below the mass limit for having one. Indeed, above a certain amount of core overshooting, the models all develop a convective core and the profile of the $r_{010}$ becomes at odds with the observations. So these targets can be used to obtain an upper limit to the amount of core overshooting. For these targets, we performed optimizations using the Levenberg-Marquardt algorithm as before, except that we fixed the parameter of core overshooting to predefined values ranging from 0 to 0.3. The result of this procedure is shown as an example for the case of KIC~10516096. For $\aov=0$, the fit converges toward a PoMS model with a mass of 1.12 $M_\odot$, an age of about 6.4 Gyr, a metallicity of $(Z/X)=0.0229$ and no convective core. For $0\leqslant\aov\leqslant0.15$, the fits converge toward roughly the same model. The only difference between the best-fit models is that the initial convective core survives longer for higher values of $\aov$ (about 1 Gyr for $\aov=0.15$ compared to 30 Myr for $\aov=0$). However, even with $\aov=0.15$ the convective core vanishes long before the end of the MS and its effect on the core structure has been washed out by the age of 6.4 Gyr. On the contrary, for $\aov=0.2$ this model keeps a convective core until the end of the MS. As a result, the duration of the MS is extended and by the time the model reaches the observed large separation, it is still in the MS with a convective core and the $r_{010}$ ratio of this model is in poor agreement with the observations. Therefore, to decrease the $\chi^2$, the fit converges toward a model with higher metallicity ($Z/X = 0.0281$) for which the convective core vanishes before the end of the MS. However, this latter model is in less good agreement with the observations as can be seen in Fig. \ref{fig_ov_chi2}. We thus obtained an upper limit of the overshooting parameter of about 0.19 for this star (value of $\aov$ above which the obtained $\chi^2$ is larger than $\min(\chi^2)+9$). Similar results were found for one other PoMS star (KIC~6933899) and three MS stars (KIC~6106415, KIC~6116048, and KIC~8394589). For all these stars, the agreement deteriorates for an upper limit $0.16<\alpha_{\rm lim}<0.20$. These constraints were added as vertical arrows in Fig. \ref{fig_mass_ov}. Unfortunately, they are too loose to confirm the tendency of $\aov$ to increase with mass that was found in Sect. \ref{sect_convcore}.

%\subsection{Effects of microscopic diffusion (in progress)}
%
%For low-mass stars, radiative forces are negligible and atomic diffusion causes elements heavier than hydrogen to sink. As a result, the increasing metallicity in the core generates an increase in the opacity, and thus in the radiative temperature gradient, which favors the onset of convection. Models with microscopic diffusion are thus expected to have bigger convective cores than models computed without diffusion. Problems in this range of mass:
%\begin{itemize}
%\item radiative forces need to be taken into account for $M\gtrsim1.3\,M_\odot$
%\item above a certain (about $1.3\,M_\odot$), the convective envelopes that acts as a reservoir of heavy elements for lower mass stars becomes so thin that it is depleted from elements heavier than H on a timescale of the order of 100 Myr. Observations are at odds with the severe depletion of heavy elements predicted by pure diffusion for the surface of these stars. Other processes such as thermohaline convection should play a role.
%\end{itemize}
%
%Results: see Table \ref{tab_convcore}. Diffusion decreases the fitted mass, especially for the lowest-mass stars. It also decreases the amount of core overshooting needed by about 0.03 $H_p$. Core overshooting still needed. Apparent trend in mass not affected.

\begin{table*}
  \begin{center}
  \caption{Characteristics of the best-fit models obtained for KIC5184732 when modifying the chosen input physics. The first column gives the alternate choices adopted in each new optimization. As mentioned in the text, the reference models have \textsc{OPAL05} equation of state, NACRE+LUNA reaction rates, no microscopic diffusion, and AGSS09 solar mixture. \label{tab_sensitivity}}
\begin{tabular}{l c c c c c c c c}
\hline \hline
\T \B Tested input physics & $M$ $(M_\odot)$ &  $(Z/X)_0$ & $Y_0$ & Age (Myr) & $\aov$ & $M_{\rm c}/M_\star$ & $\chi^2_{\rm red}$\\
\hline
%\T   9139151 & core? \\
%   8394589 & core? \\
\T  \textbf{Reference} & $1.20 \pm 0.01$ & $0.057 \pm 0.001$ & $0.31 \pm 0.01$ & $ 3957 \pm  416$ & $0.07 \pm 0.01$ & $0.05 \pm 0.01$ & $ 1.7$ \\
\T \textbf{Equation of State} & & & & & & & \\
\B \textsc{OPAL01} & $1.21 \pm 0.03$ & $0.056 \pm 0.005$ & $0.30 \pm 0.02$ & $ 3879 \pm 124$ & $0.08 \pm 0.02$ & $0.05 \pm 0.02$ & $ 1.6$ \\
\T \textbf{Nuclear reaction rates} & & & & & & & \\
\B  NACRE  & $1.21 \pm 0.02$ & $0.059 \pm 0.002$ & $0.32 \pm 0.01$ & $ 3408 \pm 555$ & $0.03 \pm 0.01$ & $0.06 \pm 0.03$ & $ 2.2$ \\
% \T \textbf{Microscopic diffusion} & & & & & & & \\
%\B  \cite{burgers69} & $1.17 \pm 0.02$ & $0.055 \pm 0.002$ & $0.32 \pm 0.01$ & $ 3770 \pm  351$ & $0.05 \pm 0.01$ & $0.05 \pm 0.03$ & $ 0.6$ \\
 \T \textbf{Solar mixture} & & & & & & & \\
 \B GN93 & $1.20 \pm 0.02$ & $0.055 \pm 0.004$ & $0.30 \pm 0.01$ & $ 3867 \pm 318$ & $0.07 \pm 0.01$ & $0.06 \pm 0.02 $ & $3.8$ \\
%mod18 & diff & $1.28 \pm 0.05$ & $0.019 \pm 0.002$ & $0.27 \pm 0.03$ & $ 3716 \pm   80$ & $0.13 \pm 0.02$ & $0.08 \pm 0.03$ & $ 1.6$ \\

\hline 
\end{tabular}
\end{center}
\end{table*}

\subsubsection{Sensitivity to input physics \label{sect_sensitivity}}

We here briefly address the question of the sensitivity of our results to some of the choices of the model input physics. We focused on one star (KIC5184732) chosen arbitrarily among the stars which was found to have a convective core and was modeled in Sect. \ref{sect_convcore}. We performed additional optimizations of this target modifying each time one assumption on the model input physics. We note that the influence of microscopic diffusion, in particular on the size of the mixed core, was already addressed in Sect. \ref{sect_convcore}. We did not expect the measurement of the mixed core size to be modified because we have confirmed in this study that its inference is mostly independent of the model physics. However, the amount of overshooting required to produce the appropriate core size at current age does depend on the input physics.

%In principle, the asteroseismic measurement of the extent of the mixed core in terms of fractional mass of the star should be mostly independent of the input physics. However, the amount of overshooting required to produce the appropriate core size at current age for the stars that we studied depends on the size of the convective core that would be produced without extra mixing, and thus it depends on the input physics. To study how much our estimate of the For this purpose, we focused on one star (KIC5184732) chosen arbitrarily among the stars which was found to have a convective core and was modeled in Sect. \ref{sect_optim}.

\paragraph{Equation of state}

Our reference \cesam\ models were computed using the OPAL05 equation of state (\citealt{rogers02}). To estimate uncertainties linked to the choice of EoS, we performed a new optimization for the target KIC5184732 using the OPAL01 EoS instead. As can be seen in Table \ref{tab_sensitivity}, the fitted parameters all lie within 1-$\sigma$ errors of the results obtained with the OPAL05 EoS.

\paragraph{Nuclear reaction rates}

We also calculated models of KIC5184732 conserving the NACRE nuclear reaction rate for the $^{14}$N$({\rm p},\gamma)^{15}$O reaction instead of the revised rate obtained from the LUNA facility (\citealt{formicola04}), which was used in Sect. \ref{sect_optim}. Table \ref{tab_sensitivity} gives the obtained fitted parameters. The amount of overshooting that is required to produce a mixed core with the appropriate size is significantly reduced. This is understandable since the NACRE cross section for the $^{14}$N$({\rm p},\gamma)^{15}$O reaction was about 30 \% higher than the revised LUNA rate. As a consequence, models computed with this previous cross section have a larger luminosity in the core, and thus a larger mixed core. The other fitted parameters are little modified compared to the reference fit. In particular, the size of the mixed core is unchanged, within statistical errors.

\paragraph{Solar mixture}

We adopted the solar mixture of AGSS09 for which $(Z/X_\odot) = 0.0181$ in our reference models in Sect. \ref{sect_optim}. We here explored the impact of considering instead the solar mixture of \cite{grevesse93} (GN93), for which $(Z/X_\odot) = 0.0244$. As can be seen from Table \ref{tab_sensitivity}, this new optimization converged toward a solution with roughly the same abundance of heavy elements as in the reference fit using AGSS09. As a result, the fitted parameters are very similar to the reference case. 

\subsection{Applicability to \mesa\ models \label{sect_cesam_vs_mesa}}

We now address the question whether the prescription obtained for the \cesam\ code in Sect. \ref{sect_optim} can be applied to \mesa\ models. We found in Sect. \ref{sect_diff_vs_step} that an instantaneous overshooting with $\aov$ is roughly equivalent to a diffusive overshooting with $f\sim\aov/10$, as was already pointed out in several studies before (e.g. \citealt{noels10}). At first sight this correspondence leads to believe that the \mesa\ models require core extensions larger than the \cesam\ models. For instance, for the target KIC7206837, an instantaneous overshooting with $\aov=0.18$  was found necessary with \cesam\ (see Table \ref{tab_convcore}), while \mesa\ models required a diffusive overshoot parameter of $f=0.035$, which would translate into  $\aov\approx0.35$ according to the established correspondence. However, we have mentioned in Sect. \ref{sect_mesa} that when using the same prescription for core overshooting (instantaneous or diffusive) and the same overshooting parameter, \mesa\ yields core extensions that are smaller by a factor $\alpha_{\rm MLT}$ compared to the extensions produced with \cesam. Since \mesa\ models were computed with $\alpha_{\rm MLT}=1.9$, the overshooting parameters obtained with \mesa\ should be divided by a factor 1.9 to be compared to the \cesam\ overshoot parameters. By doing this, we find that the diffusive overshooting parameter of $f=0.035$ obtained with \mesa\ for KIC7206837 is equivalent to an instantaneous overshooting with $\aov=0.35/1.9\approx0.18$ using the \cesam\ formalism, which is in agreement with the value of $\aov$ obtained with \cesam\ for this star.

\begin{figure}
\begin{center}
\includegraphics[width=9cm]{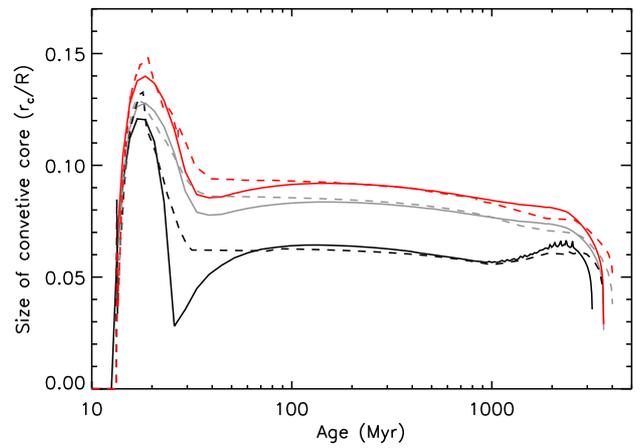}
\end{center}
\caption{Variations in the size of the convective core with age for a 1.3 $M_\odot$ model without overshooting (black lines) and with $\aov=0.1$ (red lines). The gray lines indicate the Schwarzschild limit for the case with overshooting. \cesam\ models are shown as solid lines, while \mesa\ models are represented by dashed lines.
\label{fig_cc_cesam_mesa}}
\end{figure}

To push further the comparison between \cesam\ and \mesa\ in terms of convective core size, we checked that if the exact same formalism is used for core overshooting (and therefore the same prescription for small convective cores), the two codes provide similar sizes for the extended convective cores. For this purpose, we evolved a 1.3-$M_\odot$ model with both \cesam\ and \mesa, either without or with overshooting. In the latter case we used an instantaneous overshooting with $\aov=0.1$ in both codes, and redefined in \mesa\ the overshooting distance for small convective cores using Eq. \ref{eq_dov} instead of Eq. \ref{eq_hptilde}. As shown by Fig. \ref{fig_cc_cesam_mesa}, the variations in the core size with age are very similar for \cesam\ (solid lines) and for \mesa\ (dashed lines), both in the case without overshooting (black lines) and in the case with overshooting (red lines). The only slight differences occur right after the exhaustion of the initial $^{12}$C in the core, whose burning outside of equilibrium creates the sharp peak in the core size between 15 and 25 Myr, and at the end of the main sequence, whose duration is slightly different in the two codes because of small differences in the input physics.

We thus conclude that the prescription for the overshooting parameter as a function of stellar mass obtained with \cesam\ models should also be applicable to \mesa\ models, provided the exact same formalism is considered for core overshooting. Consistency tests such as the one presented above should be performed before applying this prescription to other stellar evolution codes.

\section{Conclusion}

The main result of this paper is the detection of a convective core in eight main-sequence solar-like pulsators observed with the \kepler\ space mission, and the asteroseismic measurement of the extent of the core in these stars. 

For this purpose, we tested the seismic diagnostic for the size of the core based on the $r_{010}$ ratios, which had been successfully applied to isolated targets before (e.g. \citealt{silva13}) but whose general validity had not been addressed. By computing a grid of stellar models with varying mass, age, helium abundance, metallicity, and core overshooting, we established that the slope and mean value of the $r_{010}$ ratios can be used to estimate (1) whether the star has left the main sequence or not, (2) whether it has a convective core or not, and (3) the extent of the convective core if the star possesses one. The efficiency of this diagnostic stems from the presence of a sharp $\mu$-gradient at the boundary of the mixed core, which adds an oscillatory component to the $r_{010}$ ratios. Since unevolved stars have not yet built up such a $\mu$-gradient, the diagnostic is ineffective on these targets. 

Based on this, we selected a subset of 24 G and late-F solar-like pulsators among \kepler\ targets, avoiding too unevolved stars. We extracted the oscillation mode frequencies of these stars using the complete \kepler\ data set (nearly four years) and fitted second-order polynomials to the observed $r_{010}$ ratios. At this occasion, we realized that the covariance matrix of the observables is very ill-conditioned, which in some cases leads the fit astray. We therefore resorted to truncated SVD to solve the problem. This issue should be kept in mind, as it can be suspected to occur in any seismic modeling where combinations of mode frequencies are used as observables, as is now frequently done (e.g. \citealt{lebreton14}).

By confronting the slope and mean value of the $r_{010}$ ratios of the 24 selected targets to those of a grid of models computed with the \cesam\ code, we were able to establish that
\begin{itemize}
\item 10 of these targets are in the post main sequence and therefore do not possess convective cores,
\item 13 targets are in the main sequence (the evolutionary status of the remaining target is uncertain) and among them eight stars have a convective core,
\item the convective cores of these eight targets extend beyond the classical Schwarzschild boundary.
\end{itemize}
Interestingly, identical conclusions were reached using a similar grid of models computed with the \mesa\ code. We were able to obtain measurements of the extent of the convective cores of the eight targets that possess one, with a good agreement between the values obtained with \cesam\ and \mesa. We also produced precise estimates of the stellar parameters of these eight stars that we obtained through seismic modelings. Consequently, these stars are ideal targets to test and potentially calibrate theoretical models of physical processes that could be responsible for the extension of convective cores, such as core overshooting itself or rotational mixing. Before realistic models of these processes are available, the results obtained in this paper can be used to calibrate the simple parametric models of convective core extensions that are included in most 1D stellar evolution codes.

We addressed this question using the code \cesam, in which cores are extended over a fraction $\aov$ of either the pressure scale height $H_P$, or the radius of the core in the sense of the Schwarzschild limit if it is smaller than $H_P$. We were able to efficiently constrain $\aov$ for the eight stars, obtaining values ranging from 0.07 to 0.18. We showed that microscopic diffusion is responsible for only a small fraction of the core extension. Interestingly, we observed a tendency of $\aov$ to increase with stellar mass, which opens the possibility to derive an empirical law for $\aov(M)$ in the mass range of observed targets ($1.1\leqslant M/M_\odot\leqslant 1.5$), and thus to a calibration of what is usually referred to as core overshooting, but in fact encompasses the effects of all non-standard processes that extend convective cores. One must be careful that such a calibration necessarily depends on the prescription chosen to model the extension of convective cores in 1D stellar models. We can also suspect that it depends on the evolution code itself. We have however verified in this study that the sizes of convective cores produced by the code \mesa\ are very similar to those produced by the code \cesam, provided the same prescription for core overshooting is adopted.

This study thus constitutes a first step towards the calibration of the extension of convective cores in low-mass stars. Constraints on the extent of the convective cores of more stars will be required to confirm and enrich our results. In that respect, the \plato\ mission (\citealt{rauer14}), which was recently selected by ESA, will be particularly helpful. Reciprocally, obtaining a calibration of the distance over which convective cores extend will reduce the uncertainties on stellar ages, which will be useful to stellar physics in general, and in particular to the \plato\ mission, for which the precise determination of stellar ages is crucial.

\begin{acknowledgements}
The authors wish to thank the anonymous referee for suggestions that helped clarify the paper. This work was performed using HPC resources from CALMIP (Grant 2015-P1435). We acknowledge support from the Centre National d'\'Etudes Spatiales (CNES, France). IMB and MSC are supported by Funda\c{c}\~ao para a Ci\^encia e a Tecnologia (FCT) through the Investigador FCT contract of reference IF/00894/2012 and POPH/FSE (EC) by FEDER funding through the program COMPETE. Funds for this work were provided also by the FCT research grant UID/FIS/04434/2013 and by EC, under FP7, through the project FP7-SPACE-2012-312844. Funding for the Stellar Astrophysics Centre is provided by The Danish National Research Foundation (Grant agreement no.: DNRF106). The research is supported by the ASTERISK project (ASTERoseismic Investigations with SONG and Kepler) funded by the European Research Council (Grant agreement no.: 267864). V.S.A acknowledges support from VILLUM FONDEN (research grant 10118).

\end{acknowledgements}

\bibliographystyle{aa.bst} % style aa.bst
\bibliography{biblio} % your references Yourfile.bib

\begin{appendix}

\section{Mode parameters of the selected targets {\label{app_tabfreq}}}

Tables \ref{tab_freq0} through \ref{tab_freq5} give the fitted frequencies of the oscillation modes in the 24 selected stars among \kepler\ targets (see Sect. \ref{sect_analysis}).

\begin{table*}
 \begin{center}
 \caption{Mode parameters of KIC8394589, KIC9098294, KIC9410862, and KIC6225718. \label{tab_freq0}}
 \subfloat[KIC8394589]{
 \begin{tabular}{l c}
 \hline \hline
 \T \B $l$ & $\nu_{n,l}$ ($\mu$Hz) \\
 \hline
\T    0 & $1677.88 \pm  0.30$ \\
   0 & $1787.14 \pm  0.36$ \\
   0 & $1893.93 \pm  0.32$ \\
   0 & $2001.74 \pm  0.17$ \\
   0 & $2109.86 \pm  0.09$ \\
   0 & $2219.20 \pm  0.08$ \\
   0 & $2328.74 \pm  0.12$ \\
   0 & $2438.21 \pm  0.13$ \\
   0 & $2547.51 \pm  0.20$ \\
   0 & $2656.90 \pm  0.37$ \\
   1 & $1836.34 \pm  0.40$ \\
   1 & $1944.02 \pm  0.23$ \\
   1 & $2051.60 \pm  0.17$ \\
   1 & $2160.78 \pm  0.08$ \\
   1 & $2270.52 \pm  0.10$ \\
   1 & $2380.08 \pm  0.10$ \\
   1 & $2489.64 \pm  0.16$ \\
   1 & $2599.10 \pm  0.23$ \\
   1 & $2708.88 \pm  0.30$ \\
   1 & $2820.47 \pm  0.56$ \\
   2 & $1886.27 \pm  0.64$ \\
   2 & $1993.13 \pm  0.33$ \\
   2 & $2101.48 \pm  0.26$ \\
   2 & $2210.36 \pm  0.29$ \\
   2 & $2320.83 \pm  0.23$ \\
   2 & $2429.72 \pm  0.49$ \\
   2 & $2540.04 \pm  0.45$ \\
\B    2 & $2648.84 \pm  0.77$ \\
 \hline
 \end{tabular}}
 \quad
 \subfloat[KIC9098294]{
 \begin{tabular}{l c}
 \hline \hline
 \T \B $l$ & $\nu_{n,l}$ ($\mu$Hz) \\
 \hline
\T    0 & $1685.86 \pm  0.06$ \\
   0 & $1793.57 \pm  0.21$ \\
   0 & $1901.48 \pm  0.07$ \\
   0 & $2008.61 \pm  0.11$ \\
   0 & $2117.15 \pm  0.07$ \\
   0 & $2225.88 \pm  0.08$ \\
   0 & $2335.30 \pm  0.12$ \\
   0 & $2443.70 \pm  0.11$ \\
   0 & $2552.86 \pm  0.12$ \\
   1 & $1626.16 \pm  0.06$ \\
   1 & $1734.69 \pm  0.05$ \\
   1 & $1842.58 \pm  0.12$ \\
   1 & $1949.98 \pm  0.04$ \\
   1 & $2058.70 \pm  0.12$ \\
   1 & $2167.76 \pm  0.11$ \\
   1 & $2276.98 \pm  0.09$ \\
   1 & $2386.35 \pm  0.12$ \\
   1 & $2495.71 \pm  0.16$ \\
   1 & $2605.07 \pm  0.18$ \\
   2 & $1893.80 \pm  0.11$ \\
   2 & $2002.60 \pm  0.28$ \\
   2 & $2110.57 \pm  0.25$ \\
   2 & $2220.22 \pm  0.13$ \\
   2 & $2329.51 \pm  0.26$ \\
   2 & $2438.96 \pm  0.23$ \\
\B    2 & $2549.36 \pm  0.25$ \\
 \hline
 \end{tabular}}
 \quad
 \subfloat[KIC9410862]{
 \begin{tabular}{l c}
 \hline \hline
 \T \B $l$ & $\nu_{n,l}$ ($\mu$Hz) \\
 \hline
\T    0 & $1652.49 \pm  0.20$ \\
   0 & $1758.46 \pm  0.14$ \\
   0 & $1863.76 \pm  0.24$ \\
   0 & $1969.83 \pm  0.20$ \\
   0 & $2077.60 \pm  0.15$ \\
   0 & $2184.90 \pm  0.10$ \\
   0 & $2291.83 \pm  0.18$ \\
   0 & $2399.43 \pm  0.24$ \\
   0 & $2506.47 \pm  0.56$ \\
   1 & $1699.03 \pm  0.30$ \\
   1 & $1806.64 \pm  0.15$ \\
   1 & $1911.90 \pm  0.11$ \\
   1 & $2019.10 \pm  0.14$ \\
   1 & $2127.06 \pm  0.13$ \\
   1 & $2234.81 \pm  0.11$ \\
   1 & $2342.33 \pm  0.13$ \\
   1 & $2449.40 \pm  0.26$ \\
   1 & $2558.77 \pm  0.44$ \\
   2 & $1751.55 \pm  0.25$ \\
   2 & $1856.44 \pm  0.22$ \\
   2 & $1962.25 \pm  0.33$ \\
   2 & $2070.13 \pm  0.40$ \\
   2 & $2178.25 \pm  0.18$ \\
   2 & $2286.00 \pm  0.26$ \\
\B    2 & $2393.10 \pm  0.52$ \\
 \hline
 \end{tabular}}
 \quad
 \subfloat[KIC6225718]{
 \begin{tabular}{l c}
 \hline \hline
 \T \B $l$ & $\nu_{n,l}$ ($\mu$Hz) \\
 \hline
\T    0 & $1407.05 \pm  0.14$ \\
   0 & $1510.12 \pm  0.24$ \\
   0 & $1614.89 \pm  0.18$ \\
   0 & $1720.58 \pm  0.12$ \\
   0 & $1825.47 \pm  0.11$ \\
   0 & $1929.19 \pm  0.11$ \\
   0 & $2032.73 \pm  0.09$ \\
   0 & $2137.47 \pm  0.08$ \\
   0 & $2243.38 \pm  0.07$ \\
   0 & $2349.66 \pm  0.07$ \\
   0 & $2455.61 \pm  0.09$ \\
   0 & $2561.13 \pm  0.13$ \\
   0 & $2666.54 \pm  0.21$ \\
   0 & $2772.79 \pm  0.30$ \\
   1 & $1454.19 \pm  0.12$ \\
   1 & $1558.48 \pm  0.19$ \\
   1 & $1664.17 \pm  0.17$ \\
   1 & $1769.93 \pm  0.12$ \\
   1 & $1873.82 \pm  0.14$ \\
   1 & $1977.34 \pm  0.12$ \\
   1 & $2081.48 \pm  0.08$ \\
   1 & $2186.85 \pm  0.09$ \\
   1 & $2293.00 \pm  0.08$ \\
   1 & $2399.40 \pm  0.08$ \\
   1 & $2505.35 \pm  0.11$ \\
   1 & $2611.44 \pm  0.13$ \\
   1 & $2717.51 \pm  0.17$ \\
   1 & $2824.08 \pm  0.27$ \\
   2 & $1498.60 \pm  0.79$ \\
   2 & $1606.16 \pm  0.35$ \\
   2 & $1710.67 \pm  0.34$ \\
   2 & $1816.21 \pm  0.26$ \\
   2 & $1920.10 \pm  0.23$ \\
   2 & $2024.09 \pm  0.35$ \\
   2 & $2128.69 \pm  0.14$ \\
   2 & $2234.81 \pm  0.15$ \\
   2 & $2340.91 \pm  0.14$ \\
   2 & $2447.01 \pm  0.18$ \\
   2 & $2553.09 \pm  0.22$ \\
   2 & $2658.63 \pm  0.30$ \\
\B    2 & $2765.31 \pm  0.46$ \\
 \hline
 \end{tabular}}
 \end{center}
\end{table*}
 
\begin{table*}
 \begin{center}
 \caption{Mode parameters of KIC10454113, KIC6106415, KIC10963065, and KIC6116048. \label{tab_freq1}}
 \subfloat[KIC10454113]{
 \begin{tabular}{l c}
 \hline \hline
 \T \B $l$ & $\nu_{n,l}$ ($\mu$Hz) \\
 \hline
\T    0 & $1496.32 \pm  3.93$ \\
   0 & $1602.80 \pm  0.27$ \\
   0 & $1707.36 \pm  0.33$ \\
   0 & $1812.52 \pm  0.27$ \\
   0 & $1916.26 \pm  0.20$ \\
   0 & $2019.10 \pm  0.19$ \\
   0 & $2122.63 \pm  0.17$ \\
   0 & $2227.43 \pm  0.15$ \\
   0 & $2333.12 \pm  0.13$ \\
   0 & $2438.59 \pm  0.17$ \\
   0 & $2544.57 \pm  0.24$ \\
   0 & $2649.02 \pm  0.34$ \\
   0 & $2752.39 \pm  0.68$ \\
   0 & $2858.84 \pm  0.67$ \\
   1 & $1444.77 \pm  0.06$ \\
   1 & $1548.50 \pm  1.87$ \\
   1 & $1651.34 \pm  0.26$ \\
   1 & $1756.33 \pm  0.28$ \\
   1 & $1861.29 \pm  0.25$ \\
   1 & $1964.46 \pm  0.16$ \\
   1 & $2067.72 \pm  0.15$ \\
   1 & $2171.77 \pm  0.13$ \\
   1 & $2276.98 \pm  0.12$ \\
   1 & $2382.78 \pm  0.13$ \\
   1 & $2488.56 \pm  0.16$ \\
   1 & $2594.04 \pm  0.19$ \\
   1 & $2699.78 \pm  0.27$ \\
   1 & $2805.14 \pm  0.42$ \\
   2 & $1905.90 \pm  0.36$ \\
   2 & $2009.16 \pm  0.41$ \\
   2 & $2112.75 \pm  0.29$ \\
   2 & $2217.43 \pm  0.29$ \\
   2 & $2323.66 \pm  0.26$ \\
   2 & $2429.23 \pm  0.24$ \\
   2 & $2535.26 \pm  0.38$ \\
\B    2 & $2640.07 \pm  0.62$ \\
 \hline
 \end{tabular}}
 \quad
 \subfloat[KIC6106415]{
 \begin{tabular}{l c}
 \hline \hline
 \T \B $l$ & $\nu_{n,l}$ ($\mu$Hz) \\
 \hline
\T    0 & $1394.65 \pm  0.26$ \\
   0 & $1497.51 \pm  0.10$ \\
   0 & $1601.68 \pm  0.10$ \\
   0 & $1705.21 \pm  0.09$ \\
   0 & $1807.41 \pm  0.08$ \\
   0 & $1909.94 \pm  0.07$ \\
   0 & $2013.02 \pm  0.07$ \\
   0 & $2117.27 \pm  0.06$ \\
   0 & $2221.51 \pm  0.05$ \\
   0 & $2325.66 \pm  0.07$ \\
   0 & $2429.77 \pm  0.09$ \\
   0 & $2533.85 \pm  0.13$ \\
   0 & $2639.05 \pm  0.20$ \\
   1 & $1440.56 \pm  0.14$ \\
   1 & $1545.33 \pm  0.11$ \\
   1 & $1649.38 \pm  0.10$ \\
   1 & $1752.24 \pm  0.11$ \\
   1 & $1854.62 \pm  0.06$ \\
   1 & $1957.25 \pm  0.07$ \\
   1 & $2061.46 \pm  0.06$ \\
   1 & $2165.88 \pm  0.05$ \\
   1 & $2270.34 \pm  0.06$ \\
   1 & $2374.61 \pm  0.07$ \\
   1 & $2479.01 \pm  0.09$ \\
   1 & $2584.07 \pm  0.12$ \\
   1 & $2689.13 \pm  0.17$ \\
   2 & $1595.06 \pm  0.16$ \\
   2 & $1697.65 \pm  0.18$ \\
   2 & $1800.17 \pm  0.13$ \\
   2 & $1903.35 \pm  0.16$ \\
   2 & $2005.64 \pm  0.10$ \\
   2 & $2110.21 \pm  0.07$ \\
   2 & $2214.51 \pm  0.09$ \\
   2 & $2318.84 \pm  0.10$ \\
   2 & $2422.97 \pm  0.19$ \\
   2 & $2527.80 \pm  0.21$ \\
\B    2 & $2633.61 \pm  0.34$ \\
 \hline
 \end{tabular}}
 \quad
 \subfloat[KIC10963065]{
 \begin{tabular}{l c}
 \hline \hline
 \T \B $l$ & $\nu_{n,l}$ ($\mu$Hz) \\
 \hline
\T    0 & $1478.99 \pm  0.29$ \\
   0 & $1582.20 \pm  0.19$ \\
   0 & $1684.20 \pm  0.14$ \\
   0 & $1785.41 \pm  0.10$ \\
   0 & $1886.52 \pm  0.08$ \\
   0 & $1989.06 \pm  0.08$ \\
   0 & $2092.25 \pm  0.07$ \\
   0 & $2195.65 \pm  0.08$ \\
   0 & $2298.49 \pm  0.10$ \\
   0 & $2401.47 \pm  0.15$ \\
   0 & $2504.96 \pm  0.22$ \\
   0 & $2607.33 \pm  0.66$ \\
   0 & $2712.16 \pm  0.69$ \\
   0 & $2818.49 \pm  0.10$ \\
   1 & $1423.31 \pm  0.17$ \\
   1 & $1526.09 \pm  0.23$ \\
   1 & $1628.86 \pm  0.20$ \\
   1 & $1730.42 \pm  0.14$ \\
   1 & $1831.67 \pm  0.10$ \\
   1 & $1933.42 \pm  0.08$ \\
   1 & $2036.58 \pm  0.08$ \\
   1 & $2140.32 \pm  0.07$ \\
   1 & $2243.37 \pm  0.08$ \\
   1 & $2347.04 \pm  0.09$ \\
   1 & $2450.68 \pm  0.14$ \\
   1 & $2554.54 \pm  0.20$ \\
   1 & $2657.84 \pm  0.50$ \\
   1 & $2764.31 \pm  0.59$ \\
   1 & $2867.79 \pm  0.08$ \\
   2 & $1574.01 \pm  0.29$ \\
   2 & $1675.74 \pm  0.25$ \\
   2 & $1777.46 \pm  0.21$ \\
   2 & $1879.59 \pm  0.17$ \\
   2 & $1981.05 \pm  0.16$ \\
   2 & $2084.70 \pm  0.15$ \\
   2 & $2188.59 \pm  0.17$ \\
   2 & $2291.99 \pm  0.18$ \\
   2 & $2395.29 \pm  0.26$ \\
\B    2 & $2498.09 \pm  0.32$ \\
 \hline
 \end{tabular}}
 \quad
 \subfloat[KIC6116048]{
 \begin{tabular}{l c}
 \hline \hline
 \T \B $l$ & $\nu_{n,l}$ ($\mu$Hz) \\
 \hline
\T    0 & $1550.29 \pm  0.14$ \\
   0 & $1649.73 \pm  0.13$ \\
   0 & $1748.25 \pm  0.09$ \\
   0 & $1847.89 \pm  0.07$ \\
   0 & $1948.34 \pm  0.06$ \\
   0 & $2049.41 \pm  0.07$ \\
   0 & $2149.98 \pm  0.07$ \\
   0 & $2250.59 \pm  0.09$ \\
   0 & $2351.51 \pm  0.16$ \\
   0 & $2452.94 \pm  0.24$ \\
   0 & $2554.57 \pm  0.33$ \\
   1 & $1495.40 \pm  0.21$ \\
   1 & $1595.03 \pm  0.15$ \\
   1 & $1694.20 \pm  0.11$ \\
   1 & $1793.37 \pm  0.08$ \\
   1 & $1893.71 \pm  0.07$ \\
   1 & $1994.88 \pm  0.06$ \\
   1 & $2095.77 \pm  0.06$ \\
   1 & $2196.76 \pm  0.07$ \\
   1 & $2297.72 \pm  0.09$ \\
   1 & $2399.22 \pm  0.13$ \\
   1 & $2501.20 \pm  0.20$ \\
   1 & $2604.16 \pm  0.29$ \\
   1 & $2705.90 \pm  0.56$ \\
   2 & $1542.08 \pm  0.32$ \\
   2 & $1642.49 \pm  0.33$ \\
   2 & $1741.30 \pm  0.19$ \\
   2 & $1841.33 \pm  0.14$ \\
   2 & $1941.79 \pm  0.09$ \\
   2 & $2043.01 \pm  0.12$ \\
   2 & $2143.99 \pm  0.12$ \\
   2 & $2245.02 \pm  0.14$ \\
   2 & $2345.88 \pm  0.26$ \\
   2 & $2447.24 \pm  0.51$ \\
\B    2 & $2549.07 \pm  0.60$ \\
 \hline
 \end{tabular}}
 \end{center}
\end{table*}
 
\begin{table*}
 \begin{center}
 \caption{Mode parameters of KIC5184732, KIC3656476, KIC7296438, and KIC4914923. \label{tab_freq2}}
 \subfloat[KIC5184732]{
 \begin{tabular}{l c}
 \hline \hline
 \T \B $l$ & $\nu_{n,l}$ ($\mu$Hz) \\
 \hline
\T    0 & $1472.86 \pm  0.16$ \\
   0 & $1568.80 \pm  0.16$ \\
   0 & $1663.01 \pm  0.09$ \\
   0 & $1756.67 \pm  0.07$ \\
   0 & $1851.12 \pm  0.07$ \\
   0 & $1946.62 \pm  0.06$ \\
   0 & $2042.28 \pm  0.06$ \\
   0 & $2138.19 \pm  0.06$ \\
   0 & $2233.48 \pm  0.08$ \\
   0 & $2329.00 \pm  0.14$ \\
   0 & $2424.91 \pm  0.28$ \\
   1 & $1325.47 \pm  0.89$ \\
   1 & $1420.86 \pm  0.26$ \\
   1 & $1517.10 \pm  0.31$ \\
   1 & $1612.28 \pm  0.14$ \\
   1 & $1706.21 \pm  0.08$ \\
   1 & $1800.53 \pm  0.08$ \\
   1 & $1895.58 \pm  0.06$ \\
   1 & $1991.53 \pm  0.06$ \\
   1 & $2087.41 \pm  0.06$ \\
   1 & $2183.35 \pm  0.07$ \\
   1 & $2279.11 \pm  0.08$ \\
   1 & $2375.01 \pm  0.12$ \\
   1 & $2471.78 \pm  0.24$ \\
   2 & $1750.52 \pm  0.18$ \\
   2 & $1844.63 \pm  0.11$ \\
   2 & $1940.27 \pm  0.10$ \\
   2 & $2036.30 \pm  0.10$ \\
   2 & $2132.44 \pm  0.09$ \\
   2 & $2227.93 \pm  0.11$ \\
   2 & $2323.99 \pm  0.23$ \\
\B    2 & $2419.83 \pm  0.37$ \\
 \hline
 \end{tabular}}
 \quad
 \subfloat[KIC3656476]{
 \begin{tabular}{l c}
 \hline \hline
 \T \B $l$ & $\nu_{n,l}$ ($\mu$Hz) \\
 \hline
\T    0 & $1443.43 \pm  0.14$ \\
   0 & $1535.05 \pm  0.08$ \\
   0 & $1626.75 \pm  0.06$ \\
   0 & $1719.22 \pm  0.04$ \\
   0 & $1812.37 \pm  0.04$ \\
   0 & $1905.65 \pm  0.05$ \\
   0 & $1998.75 \pm  0.04$ \\
   0 & $2091.73 \pm  0.07$ \\
   0 & $2185.34 \pm  0.18$ \\
   0 & $2278.90 \pm  0.31$ \\
   1 & $1391.24 \pm  0.16$ \\
   1 & $1483.85 \pm  0.08$ \\
   1 & $1575.41 \pm  0.06$ \\
   1 & $1667.59 \pm  0.05$ \\
   1 & $1760.77 \pm  0.04$ \\
   1 & $1854.20 \pm  0.04$ \\
   1 & $1947.47 \pm  0.04$ \\
   1 & $2040.90 \pm  0.04$ \\
   1 & $2134.38 \pm  0.07$ \\
   1 & $2228.39 \pm  0.14$ \\
   1 & $2321.97 \pm  0.28$ \\
   2 & $1529.53 \pm  0.18$ \\
   2 & $1621.06 \pm  0.11$ \\
   2 & $1713.78 \pm  0.09$ \\
   2 & $1807.29 \pm  0.06$ \\
   2 & $1900.95 \pm  0.05$ \\
   2 & $1994.25 \pm  0.05$ \\
   2 & $2087.83 \pm  0.10$ \\
\B    2 & $2182.10 \pm  0.21$ \\
 \hline
 \end{tabular}}
 \quad
 \subfloat[KIC7296438]{
 \begin{tabular}{l c}
 \hline \hline
 \T \B $l$ & $\nu_{n,l}$ ($\mu$Hz) \\
 \hline
\T    0 & $1366.74 \pm  0.10$ \\
   0 & $1454.09 \pm  0.17$ \\
   0 & $1540.75 \pm  0.14$ \\
   0 & $1628.60 \pm  0.12$ \\
   0 & $1717.22 \pm  0.08$ \\
   0 & $1805.96 \pm  0.09$ \\
   0 & $1894.69 \pm  0.11$ \\
   0 & $1983.17 \pm  0.15$ \\
   0 & $2071.89 \pm  0.35$ \\
   0 & $2159.28 \pm  0.69$ \\
   1 & $1317.44 \pm  0.28$ \\
   1 & $1405.31 \pm  0.07$ \\
   1 & $1492.54 \pm  0.21$ \\
   1 & $1579.64 \pm  0.14$ \\
   1 & $1667.91 \pm  0.11$ \\
   1 & $1757.06 \pm  0.07$ \\
   1 & $1845.67 \pm  0.09$ \\
   1 & $1934.64 \pm  0.11$ \\
   1 & $2023.46 \pm  0.14$ \\
   1 & $2112.82 \pm  0.27$ \\
   2 & $1447.83 \pm  0.30$ \\
   2 & $1534.95 \pm  0.20$ \\
   2 & $1623.06 \pm  0.19$ \\
   2 & $1711.67 \pm  0.09$ \\
   2 & $1800.88 \pm  0.13$ \\
   2 & $1889.66 \pm  0.13$ \\
   2 & $1978.11 \pm  0.20$ \\
\B    2 & $2067.55 \pm  0.51$ \\
 \hline
 \end{tabular}}
 \quad
 \subfloat[KIC4914923]{
 \begin{tabular}{l c}
 \hline \hline
 \T \B $l$ & $\nu_{n,l}$ ($\mu$Hz) \\
 \hline
\T    0 & $1188.00 \pm  0.23$ \\
   0 & $1277.01 \pm  0.12$ \\
   0 & $1365.19 \pm  0.11$ \\
   0 & $1452.31 \pm  0.11$ \\
   0 & $1539.06 \pm  0.08$ \\
   0 & $1626.98 \pm  0.08$ \\
   0 & $1715.53 \pm  0.06$ \\
   0 & $1804.24 \pm  0.05$ \\
   0 & $1892.77 \pm  0.06$ \\
   0 & $1981.16 \pm  0.09$ \\
   0 & $2069.89 \pm  0.18$ \\
   0 & $2157.84 \pm  0.82$ \\
   1 & $1138.89 \pm  0.31$ \\
   1 & $1227.08 \pm  0.19$ \\
   1 & $1315.83 \pm  0.14$ \\
   1 & $1403.42 \pm  0.15$ \\
   1 & $1490.58 \pm  0.10$ \\
   1 & $1577.81 \pm  0.08$ \\
   1 & $1666.36 \pm  0.06$ \\
   1 & $1755.32 \pm  0.05$ \\
   1 & $1844.03 \pm  0.05$ \\
   1 & $1932.77 \pm  0.06$ \\
   1 & $2021.72 \pm  0.09$ \\
   1 & $2110.85 \pm  0.17$ \\
   1 & $2200.92 \pm  0.37$ \\
   2 & $1446.28 \pm  0.33$ \\
   2 & $1533.30 \pm  0.18$ \\
   2 & $1621.39 \pm  0.13$ \\
   2 & $1710.08 \pm  0.10$ \\
   2 & $1798.86 \pm  0.10$ \\
   2 & $1887.60 \pm  0.10$ \\
   2 & $1976.26 \pm  0.12$ \\
   2 & $2065.38 \pm  0.27$ \\
\B    2 & $2155.31 \pm  0.81$ \\
 \hline
 \end{tabular}}
 \end{center}
\end{table*}
 
\begin{table*}
 \begin{center}
 \caption{Mode parameters of KIC12009504, KIC8938364, KIC7680114, and KIC10516096. \label{tab_freq3}}
 \subfloat[KIC12009504]{
 \begin{tabular}{l c}
 \hline \hline
 \T \B $l$ & $\nu_{n,l}$ ($\mu$Hz) \\
 \hline
\T    0 & $1171.23 \pm  0.42$ \\
   0 & $1258.48 \pm  0.29$ \\
   0 & $1346.06 \pm  0.17$ \\
   0 & $1433.89 \pm  0.15$ \\
   0 & $1520.05 \pm  0.27$ \\
   0 & $1606.34 \pm  0.13$ \\
   0 & $1693.75 \pm  0.09$ \\
   0 & $1782.04 \pm  0.11$ \\
   0 & $1870.59 \pm  0.12$ \\
   0 & $1958.89 \pm  0.14$ \\
   0 & $2047.11 \pm  0.24$ \\
   0 & $2135.46 \pm  0.20$ \\
   0 & $2224.77 \pm  0.66$ \\
   0 & $2311.82 \pm  0.69$ \\
   1 & $1212.32 \pm  0.34$ \\
   1 & $1297.80 \pm  0.31$ \\
   1 & $1386.16 \pm  0.17$ \\
   1 & $1472.96 \pm  0.17$ \\
   1 & $1559.37 \pm  0.16$ \\
   1 & $1646.04 \pm  0.11$ \\
   1 & $1733.96 \pm  0.10$ \\
   1 & $1822.51 \pm  0.10$ \\
   1 & $1911.55 \pm  0.12$ \\
   1 & $1999.81 \pm  0.13$ \\
   1 & $2088.56 \pm  0.22$ \\
   1 & $2177.01 \pm  0.27$ \\
   1 & $2266.12 \pm  0.50$ \\
   1 & $2356.78 \pm  0.94$ \\
   2 & $1338.08 \pm  0.32$ \\
   2 & $1428.82 \pm  0.47$ \\
   2 & $1514.36 \pm  0.88$ \\
   2 & $1599.63 \pm  0.38$ \\
   2 & $1687.35 \pm  0.38$ \\
   2 & $1775.95 \pm  0.17$ \\
   2 & $1864.21 \pm  0.26$ \\
   2 & $1953.06 \pm  0.25$ \\
   2 & $2041.30 \pm  0.65$ \\
   2 & $2129.10 \pm  0.83$ \\
\B    2 & $2217.15 \pm  1.05$ \\
 \hline
 \end{tabular}}
 \quad
 \subfloat[KIC8938364]{
 \begin{tabular}{l c}
 \hline \hline
 \T \B $l$ & $\nu_{n,l}$ ($\mu$Hz) \\
 \hline
\T    0 & $1070.79 \pm  0.16$ \\
   0 & $1155.98 \pm  0.13$ \\
   0 & $1241.47 \pm  0.12$ \\
   0 & $1325.85 \pm  0.14$ \\
   0 & $1409.73 \pm  0.06$ \\
   0 & $1494.84 \pm  0.06$ \\
   0 & $1580.77 \pm  0.04$ \\
   0 & $1666.10 \pm  0.05$ \\
   0 & $1751.59 \pm  0.07$ \\
   0 & $1837.64 \pm  0.11$ \\
   0 & $1923.46 \pm  0.33$ \\
   1 & $1106.63 \pm  0.11$ \\
   1 & $1192.49 \pm  0.14$ \\
   1 & $1276.72 \pm  0.09$ \\
   1 & $1360.84 \pm  0.09$ \\
   1 & $1445.63 \pm  0.05$ \\
   1 & $1531.00 \pm  0.04$ \\
   1 & $1616.80 \pm  0.04$ \\
   1 & $1702.58 \pm  0.05$ \\
   1 & $1788.25 \pm  0.06$ \\
   1 & $1874.50 \pm  0.10$ \\
   1 & $1961.06 \pm  0.34$ \\
   2 & $1235.19 \pm  0.17$ \\
   2 & $1319.82 \pm  0.19$ \\
   2 & $1403.84 \pm  0.09$ \\
   2 & $1489.09 \pm  0.08$ \\
   2 & $1575.18 \pm  0.05$ \\
   2 & $1660.87 \pm  0.07$ \\
   2 & $1746.60 \pm  0.08$ \\
\B    2 & $1833.03 \pm  0.13$ \\
 \hline
 \end{tabular}}
 \quad
 \subfloat[KIC7680114]{
 \begin{tabular}{l c}
 \hline \hline
 \T \B $l$ & $\nu_{n,l}$ ($\mu$Hz) \\
 \hline
\T    0 & $1142.76 \pm  0.34$ \\
   0 & $1227.87 \pm  0.06$ \\
   0 & $1312.07 \pm  0.12$ \\
   0 & $1395.49 \pm  0.07$ \\
   0 & $1479.12 \pm  0.06$ \\
   0 & $1564.01 \pm  0.06$ \\
   0 & $1649.29 \pm  0.06$ \\
   0 & $1734.38 \pm  0.07$ \\
   0 & $1819.51 \pm  0.07$ \\
   0 & $1904.84 \pm  0.14$ \\
   1 & $1094.54 \pm  0.08$ \\
   1 & $1179.52 \pm  0.15$ \\
   1 & $1264.40 \pm  0.11$ \\
   1 & $1348.35 \pm  0.11$ \\
   1 & $1431.92 \pm  0.07$ \\
   1 & $1516.30 \pm  0.08$ \\
   1 & $1601.41 \pm  0.05$ \\
   1 & $1686.75 \pm  0.05$ \\
   1 & $1772.04 \pm  0.07$ \\
   1 & $1857.24 \pm  0.08$ \\
   1 & $1943.20 \pm  0.13$ \\
   1 & $2029.52 \pm  0.37$ \\
   2 & $1306.30 \pm  0.31$ \\
   2 & $1389.44 \pm  0.14$ \\
   2 & $1473.55 \pm  0.19$ \\
   2 & $1558.59 \pm  0.08$ \\
   2 & $1644.13 \pm  0.09$ \\
   2 & $1729.47 \pm  0.07$ \\
   2 & $1814.80 \pm  0.14$ \\
\B    2 & $1900.00 \pm  0.16$ \\
 \hline
 \end{tabular}}
 \quad
 \subfloat[KIC10516096]{
 \begin{tabular}{l c}
 \hline \hline
 \T \B $l$ & $\nu_{n,l}$ ($\mu$Hz) \\
 \hline
\T    0 & $1213.89 \pm  0.33$ \\
   0 & $1296.79 \pm  0.14$ \\
   0 & $1379.58 \pm  0.11$ \\
   0 & $1462.53 \pm  0.09$ \\
   0 & $1546.40 \pm  0.07$ \\
   0 & $1631.19 \pm  0.07$ \\
   0 & $1715.71 \pm  0.08$ \\
   0 & $1799.91 \pm  0.08$ \\
   0 & $1884.57 \pm  0.17$ \\
   1 & $1165.66 \pm  0.33$ \\
   1 & $1249.78 \pm  0.20$ \\
   1 & $1333.01 \pm  0.12$ \\
   1 & $1415.60 \pm  0.09$ \\
   1 & $1499.19 \pm  0.08$ \\
   1 & $1583.71 \pm  0.07$ \\
   1 & $1668.42 \pm  0.07$ \\
   1 & $1752.97 \pm  0.07$ \\
   1 & $1837.27 \pm  0.10$ \\
   1 & $1922.57 \pm  0.16$ \\
   1 & $2007.99 \pm  0.26$ \\
   2 & $1373.87 \pm  0.24$ \\
   2 & $1456.44 \pm  0.19$ \\
   2 & $1540.82 \pm  0.14$ \\
   2 & $1625.92 \pm  0.12$ \\
   2 & $1710.26 \pm  0.10$ \\
   2 & $1794.54 \pm  0.17$ \\
\B    2 & $1879.67 \pm  0.30$ \\
 \hline
 \end{tabular}}
 \end{center}
\end{table*}
 
\begin{table*}
 \begin{center}
 \caption{Mode parameters of KIC7206837, KIC8176564, KIC8694723, and KIC12258514. \label{tab_freq4}}
 \subfloat[KIC7206837]{
 \begin{tabular}{l c}
 \hline \hline
 \T \B $l$ & $\nu_{n,l}$ ($\mu$Hz) \\
 \hline
\T    0 & $1117.32 \pm  0.31$ \\
   0 & $1194.49 \pm  0.39$ \\
   0 & $1272.54 \pm  0.41$ \\
   0 & $1353.05 \pm  0.37$ \\
   0 & $1431.78 \pm  0.21$ \\
   0 & $1508.67 \pm  0.20$ \\
   0 & $1586.66 \pm  0.20$ \\
   0 & $1665.42 \pm  0.23$ \\
   0 & $1745.23 \pm  0.21$ \\
   0 & $1825.38 \pm  0.29$ \\
   0 & $1905.09 \pm  0.44$ \\
   0 & $1983.44 \pm  0.41$ \\
   0 & $2065.35 \pm  0.48$ \\
   0 & $2143.97 \pm  1.10$ \\
   1 & $1077.41 \pm  0.26$ \\
   1 & $1153.16 \pm  0.28$ \\
   1 & $1231.35 \pm  0.32$ \\
   1 & $1310.72 \pm  0.29$ \\
   1 & $1389.45 \pm  0.36$ \\
   1 & $1468.29 \pm  0.22$ \\
   1 & $1545.33 \pm  0.19$ \\
   1 & $1623.77 \pm  0.19$ \\
   1 & $1702.96 \pm  0.20$ \\
   1 & $1783.29 \pm  0.19$ \\
   1 & $1863.29 \pm  0.28$ \\
   1 & $1943.89 \pm  0.42$ \\
   1 & $2023.59 \pm  0.40$ \\
   1 & $2103.88 \pm  0.52$ \\
   2 & $1111.66 \pm  0.32$ \\
   2 & $1187.97 \pm  1.16$ \\
   2 & $1266.02 \pm  0.79$ \\
   2 & $1348.24 \pm  1.30$ \\
   2 & $1426.94 \pm  0.54$ \\
   2 & $1501.36 \pm  0.51$ \\
   2 & $1580.42 \pm  0.58$ \\
   2 & $1658.80 \pm  0.60$ \\
   2 & $1740.30 \pm  0.63$ \\
   2 & $1820.92 \pm  0.53$ \\
   2 & $1901.66 \pm  1.44$ \\
   2 & $1987.48 \pm  0.94$ \\
\B   2 & $2059.86 \pm  1.63$ \\
 \hline
 \end{tabular}}
 \quad
 \subfloat[KIC8176564]{
 \begin{tabular}{l c}
 \hline \hline
 \T \B $l$ & $\nu_{n,l}$ ($\mu$Hz) \\
 \hline
\T    0 & $1114.81 \pm  0.08$ \\
   0 & $1191.21 \pm  0.16$ \\
   0 & $1267.24 \pm  0.34$ \\
   0 & $1343.57 \pm  0.27$ \\
   0 & $1421.49 \pm  0.15$ \\
   0 & $1499.36 \pm  0.20$ \\
   0 & $1577.14 \pm  0.19$ \\
   0 & $1655.19 \pm  0.42$ \\
   0 & $1732.48 \pm  0.64$ \\
   0 & $1812.60 \pm  0.77$ \\
   1 & $1147.19 \pm  0.04$ \\
   1 & $1224.02 \pm  0.14$ \\
   1 & $1299.30 \pm  0.26$ \\
   1 & $1376.65 \pm  0.20$ \\
   1 & $1454.79 \pm  0.14$ \\
   1 & $1532.75 \pm  0.16$ \\
   1 & $1610.21 \pm  0.22$ \\
   1 & $1688.59 \pm  0.34$ \\
   1 & $1767.37 \pm  0.68$ \\
   1 & $1845.21 \pm  2.06$ \\
   2 & $1184.98 \pm  0.23$ \\
   2 & $1261.04 \pm  0.48$ \\
   2 & $1338.15 \pm  0.49$ \\
   2 & $1415.72 \pm  0.23$ \\
   2 & $1494.47 \pm  0.30$ \\
   2 & $1571.68 \pm  0.25$ \\
   2 & $1650.41 \pm  0.84$ \\
   2 & $1726.69 \pm  1.47$ \\
\B    2 & $1804.62 \pm  1.07$ \\
 \hline
 \end{tabular}}
 \quad
 \subfloat[KIC8694723]{
 \begin{tabular}{l c}
 \hline \hline
 \T \B $l$ & $\nu_{n,l}$ ($\mu$Hz) \\
 \hline
\T    0 & $ 917.70 \pm  0.41$ \\
   0 & $ 990.17 \pm  0.23$ \\
   0 & $1064.26 \pm  0.19$ \\
   0 & $1139.36 \pm  0.16$ \\
   0 & $1212.49 \pm  0.13$ \\
   0 & $1285.85 \pm  0.11$ \\
   0 & $1359.99 \pm  0.11$ \\
   0 & $1435.52 \pm  0.10$ \\
   0 & $1510.88 \pm  0.11$ \\
   0 & $1586.78 \pm  0.15$ \\
   0 & $1661.90 \pm  0.17$ \\
   0 & $1737.52 \pm  0.26$ \\
   1 & $ 876.47 \pm  0.27$ \\
   1 & $ 949.06 \pm  0.24$ \\
   1 & $1022.30 \pm  0.18$ \\
   1 & $1096.64 \pm  0.16$ \\
   1 & $1170.92 \pm  0.13$ \\
   1 & $1243.96 \pm  0.11$ \\
   1 & $1317.56 \pm  0.09$ \\
   1 & $1392.34 \pm  0.09$ \\
   1 & $1467.99 \pm  0.09$ \\
   1 & $1543.63 \pm  0.10$ \\
   1 & $1619.20 \pm  0.12$ \\
   1 & $1694.49 \pm  0.15$ \\
   1 & $1770.34 \pm  0.21$ \\
   1 & $1846.78 \pm  0.33$ \\
   2 & $1058.81 \pm  0.43$ \\
   2 & $1132.90 \pm  0.35$ \\
   2 & $1206.25 \pm  0.26$ \\
   2 & $1279.75 \pm  0.20$ \\
   2 & $1353.96 \pm  0.20$ \\
   2 & $1429.90 \pm  0.18$ \\
   2 & $1504.84 \pm  0.21$ \\
   2 & $1580.49 \pm  0.28$ \\
   2 & $1655.81 \pm  0.31$ \\
\B    2 & $1731.50 \pm  0.46$ \\
 \hline
 \end{tabular}}
 \quad
 \subfloat[KIC12258514]{
 \begin{tabular}{l c}
 \hline \hline
 \T \B $l$ & $\nu_{n,l}$ ($\mu$Hz) \\
 \hline
\T    0 & $ 997.07 \pm  0.08$ \\
   0 & $1071.81 \pm  0.12$ \\
   0 & $1146.47 \pm  0.08$ \\
   0 & $1219.91 \pm  0.09$ \\
   0 & $1293.05 \pm  0.07$ \\
   0 & $1367.18 \pm  0.06$ \\
   0 & $1442.17 \pm  0.05$ \\
   0 & $1517.31 \pm  0.05$ \\
   0 & $1592.19 \pm  0.06$ \\
   0 & $1666.77 \pm  0.10$ \\
   0 & $1741.88 \pm  0.16$ \\
   0 & $1816.24 \pm  0.23$ \\
   1 & $ 956.10 \pm  0.07$ \\
   1 & $1029.78 \pm  0.08$ \\
   1 & $1104.80 \pm  0.24$ \\
   1 & $1178.72 \pm  0.07$ \\
   1 & $1251.88 \pm  0.07$ \\
   1 & $1325.55 \pm  0.07$ \\
   1 & $1400.21 \pm  0.06$ \\
   1 & $1475.47 \pm  0.06$ \\
   1 & $1550.59 \pm  0.05$ \\
   1 & $1625.53 \pm  0.08$ \\
   1 & $1700.64 \pm  0.09$ \\
   1 & $1776.33 \pm  0.14$ \\
   1 & $1852.28 \pm  0.24$ \\
   2 & $1066.67 \pm  0.24$ \\
   2 & $1141.19 \pm  0.12$ \\
   2 & $1214.84 \pm  0.14$ \\
   2 & $1288.17 \pm  0.12$ \\
   2 & $1362.39 \pm  0.10$ \\
   2 & $1437.18 \pm  0.08$ \\
   2 & $1512.63 \pm  0.07$ \\
   2 & $1587.59 \pm  0.11$ \\
   2 & $1662.62 \pm  0.13$ \\
\B    2 & $1737.79 \pm  0.28$ \\
 \hline
 \end{tabular}}
 \end{center}
\end{table*}
 
\begin{table*}
 \begin{center}
 \caption{Mode parameters of KIC6933899, KIC11244118, KIC7510397, and KIC8228742. \label{tab_freq5}}
 \subfloat[KIC6933899]{
 \begin{tabular}{l c}
 \hline \hline
 \T \B $l$ & $\nu_{n,l}$ ($\mu$Hz) \\
 \hline
\T    0 & $ 965.04 \pm  0.17$ \\
   0 & $1036.81 \pm  0.07$ \\
   0 & $1107.52 \pm  0.09$ \\
   0 & $1177.78 \pm  0.08$ \\
   0 & $1249.22 \pm  0.06$ \\
   0 & $1321.71 \pm  0.05$ \\
   0 & $1393.90 \pm  0.05$ \\
   0 & $1465.94 \pm  0.07$ \\
   0 & $1538.36 \pm  0.09$ \\
   0 & $1610.54 \pm  0.15$ \\
   0 & $1683.36 \pm  0.27$ \\
   1 & $ 853.72 \pm  0.05$ \\
   1 & $ 924.42 \pm  0.06$ \\
   1 & $ 996.35 \pm  0.16$ \\
   1 & $1067.31 \pm  0.09$ \\
   1 & $1137.61 \pm  0.08$ \\
   1 & $1208.21 \pm  0.07$ \\
   1 & $1280.12 \pm  0.05$ \\
   1 & $1352.48 \pm  0.05$ \\
   1 & $1424.59 \pm  0.06$ \\
   1 & $1496.68 \pm  0.07$ \\
   1 & $1569.11 \pm  0.11$ \\
   1 & $1641.89 \pm  0.14$ \\
   1 & $1714.96 \pm  0.22$ \\
   2 & $1031.76 \pm  0.16$ \\
   2 & $1101.74 \pm  0.22$ \\
   2 & $1172.90 \pm  0.15$ \\
   2 & $1243.97 \pm  0.10$ \\
   2 & $1316.70 \pm  0.11$ \\
   2 & $1389.00 \pm  0.10$ \\
   2 & $1461.04 \pm  0.11$ \\
   2 & $1533.51 \pm  0.15$ \\
   2 & $1606.40 \pm  0.22$ \\
\B    2 & $1679.33 \pm  0.35$ \\
 \hline
 \end{tabular}}
 \quad
 \subfloat[KIC11244118]{
 \begin{tabular}{l c}
 \hline \hline
 \T \B $l$ & $\nu_{n,l}$ ($\mu$Hz) \\
 \hline
\T    0 & $ 958.68 \pm  0.08$ \\
   0 & $1030.05 \pm  0.11$ \\
   0 & $1100.39 \pm  0.06$ \\
   0 & $1169.91 \pm  0.06$ \\
   0 & $1240.85 \pm  0.06$ \\
   0 & $1312.24 \pm  0.04$ \\
   0 & $1383.76 \pm  0.05$ \\
   0 & $1455.28 \pm  0.06$ \\
   0 & $1526.83 \pm  0.08$ \\
   0 & $1598.59 \pm  0.16$ \\
   0 & $1670.48 \pm  0.43$ \\
   1 & $ 918.96 \pm  0.08$ \\
   1 & $ 990.27 \pm  0.08$ \\
   1 & $1060.38 \pm  0.08$ \\
   1 & $1130.43 \pm  0.05$ \\
   1 & $1200.23 \pm  0.05$ \\
   1 & $1271.01 \pm  0.05$ \\
   1 & $1342.64 \pm  0.04$ \\
   1 & $1414.12 \pm  0.05$ \\
   1 & $1485.23 \pm  0.06$ \\
   1 & $1556.73 \pm  0.09$ \\
   1 & $1629.13 \pm  0.13$ \\
   1 & $1701.26 \pm  0.33$ \\
   2 & $ 953.57 \pm  0.11$ \\
   2 & $1024.56 \pm  0.13$ \\
   2 & $1094.35 \pm  0.10$ \\
   2 & $1164.27 \pm  0.11$ \\
   2 & $1231.35 \pm  0.11$ \\
   2 & $1307.42 \pm  0.07$ \\
   2 & $1379.07 \pm  0.06$ \\
   2 & $1450.55 \pm  0.07$ \\
   2 & $1521.97 \pm  0.12$ \\
\B    2 & $1593.91 \pm  0.17$ \\
 \hline
 \end{tabular}}
 \quad
 \subfloat[KIC7510397]{
 \begin{tabular}{l c}
 \hline \hline
 \T \B $l$ & $\nu_{n,l}$ ($\mu$Hz) \\
 \hline
\T    0 & $ 700.30 \pm  0.09$ \\
   0 & $ 759.74 \pm  0.24$ \\
   0 & $ 819.83 \pm  0.14$ \\
   0 & $ 881.24 \pm  0.16$ \\
   0 & $ 943.76 \pm  0.17$ \\
   0 & $1004.77 \pm  0.16$ \\
   0 & $1065.52 \pm  0.13$ \\
   0 & $1127.00 \pm  0.10$ \\
   0 & $1189.59 \pm  0.12$ \\
   0 & $1252.50 \pm  0.13$ \\
   0 & $1314.89 \pm  0.17$ \\
   0 & $1377.12 \pm  0.20$ \\
   0 & $1440.24 \pm  0.31$ \\
   0 & $1503.38 \pm  0.35$ \\
   0 & $1564.50 \pm  0.49$ \\
   1 & $ 666.49 \pm  0.08$ \\
   1 & $ 726.61 \pm  0.08$ \\
   1 & $ 785.83 \pm  0.12$ \\
   1 & $ 846.54 \pm  0.12$ \\
   1 & $ 908.60 \pm  0.12$ \\
   1 & $ 970.42 \pm  0.12$ \\
   1 & $1031.35 \pm  0.11$ \\
   1 & $1091.92 \pm  0.10$ \\
   1 & $1154.18 \pm  0.08$ \\
   1 & $1216.93 \pm  0.10$ \\
   1 & $1279.70 \pm  0.11$ \\
   1 & $1342.42 \pm  0.13$ \\
   1 & $1405.39 \pm  0.17$ \\
   1 & $1468.07 \pm  0.24$ \\
   1 & $1532.05 \pm  0.28$ \\
   1 & $1594.75 \pm  0.53$ \\
   2 & $ 754.80 \pm  0.21$ \\
   2 & $ 813.10 \pm  0.25$ \\
   2 & $ 877.13 \pm  0.25$ \\
   2 & $ 939.21 \pm  0.24$ \\
   2 & $ 999.46 \pm  0.26$ \\
   2 & $1061.33 \pm  0.17$ \\
   2 & $1122.75 \pm  0.11$ \\
   2 & $1184.93 \pm  0.15$ \\
   2 & $1248.01 \pm  0.21$ \\
   2 & $1310.65 \pm  0.22$ \\
   2 & $1372.63 \pm  0.39$ \\
   2 & $1436.67 \pm  0.51$ \\
\B    2 & $1497.95 \pm  0.46$ \\
 \hline
 \end{tabular}}
 \quad
 \subfloat[KIC8228742]{
 \begin{tabular}{l c}
 \hline \hline
 \T \B $l$ & $\nu_{n,l}$ ($\mu$Hz) \\
 \hline
\T    0 & $ 760.18 \pm  0.20$ \\
   0 & $ 820.64 \pm  0.17$ \\
   0 & $ 882.28 \pm  0.12$ \\
   0 & $ 943.80 \pm  0.13$ \\
   0 & $1004.07 \pm  0.13$ \\
   0 & $1064.82 \pm  0.10$ \\
   0 & $1126.70 \pm  0.08$ \\
   0 & $1189.35 \pm  0.10$ \\
   0 & $1251.53 \pm  0.11$ \\
   0 & $1313.75 \pm  0.17$ \\
   0 & $1375.67 \pm  0.24$ \\
   0 & $1438.63 \pm  0.38$ \\
   0 & $1500.86 \pm  0.91$ \\
   1 & $ 726.20 \pm  0.35$ \\
   1 & $ 786.06 \pm  0.29$ \\
   1 & $ 847.24 \pm  0.13$ \\
   1 & $ 908.73 \pm  0.12$ \\
   1 & $ 969.92 \pm  0.13$ \\
   1 & $1030.22 \pm  0.09$ \\
   1 & $1091.61 \pm  0.08$ \\
   1 & $1153.47 \pm  0.08$ \\
   1 & $1216.09 \pm  0.09$ \\
   1 & $1278.58 \pm  0.09$ \\
   1 & $1340.58 \pm  0.13$ \\
   1 & $1403.71 \pm  0.18$ \\
   1 & $1466.36 \pm  0.23$ \\
   1 & $1529.47 \pm  0.39$ \\
   2 & $ 815.31 \pm  0.52$ \\
   2 & $ 877.51 \pm  0.28$ \\
   2 & $ 939.18 \pm  0.35$ \\
   2 & $ 999.39 \pm  0.20$ \\
   2 & $1060.38 \pm  0.26$ \\
   2 & $1122.08 \pm  0.15$ \\
   2 & $1184.76 \pm  0.16$ \\
   2 & $1247.10 \pm  0.18$ \\
   2 & $1309.43 \pm  0.23$ \\
   2 & $1371.50 \pm  0.37$ \\
   2 & $1434.91 \pm  0.50$ \\
\B    2 & $1498.87 \pm  1.77$ \\
 \hline
 \end{tabular}}
 \end{center}
\end{table*} 

\section{Polynomial fit to the observed $r_{010}$ ratios \label{app_polyfit}}

We denote as $\vect{y}=(y_1,\hdots,y_n)$ the set of $n$ observed $r_{010}$ ratios, and as $\vect{x}=(x_1,\hdots,x_n)$ the corresponding frequencies. We would like to fit a $2^{\rm nd}$-order polynomial to the functional $\vect{y}(\vect{x})$. Since we would like to interpret individual coefficients of the polynomial regression, we need them to be independent from one another. We are thus required to use orthogonal polynomials. For this purpose, we fit polynomials of the type
\begin{equation}
P(\nu) = a_0 + a_1 (\nu-\beta) + a_2(\nu-\gamma_1)(\nu-\gamma_2)
\end{equation}
to the observed ratios $\vect{y}$, where $\beta$, $\gamma_1$, and $\gamma_2$ need to be determined so as to ensure the independence of the $a_k$ coefficients. The values taken by the polynomial $P(\nu)$ at each point of vector $\vect{x}$ can be written in a vectorial way as 
\begin{equation}
P(x_i) = (\vect{J}\vect{a})_i
\end{equation}
where $\vect{a}=(a_0,a_1,a_2)$, and $\vect{J}$ is an $n\times3$ matrix such that
\begin{align}
J_{i,0} & = 1 \\
J_{i,1} & = x_i - \beta \\
J_{i,2} & = (x_i - \gamma_1) (x_i - \gamma_2)
\end{align}

In our case, the observables $\vect{y}$ are combinations of mode frequencies and are thus highly correlated. We denote as $\vect{C}$ the covariance matrix of the observables. At first, let us assume that $\vect{C}$ is invertible and denote its inverse as $\vect{W}$. In this case, the $\chi^2$ function to be minimized can be written as
\begin{equation}
\chi^2 = (\vect{J}\vect{a}-\vect{y})^{T}  \vect{W}  (\vect{J}\vect{a}-\vect{y})
\end{equation}
where the exponent $T$ indicates matrix transposition. The gradient of this function is given by
\begin{equation}
\nabla\chi^2 = 2\vect{J}^T \vect{W} \vect{J} \vect{a} - 2\vect{J}^T \vect{W} \vect{y}
\end{equation}
and the optimal set of coefficients is then
\begin{equation}
\vect{a}_{\rm min} = \left(\vect{J}^T \vect{W} \vect{J}\right)^{-1} \left(\vect{J}^T \vect{W} \vect{y}\right)
\label{eq_amin}
\end{equation}
where $\left(\vect{J}^T \vect{W} \vect{J}\right)^{-1}$ corresponds to the error matrix for the coefficients $a_k$. To ensure that these coefficients are uncorrelated, we must thus require that the non-diagonal coefficients of the error matrix vanish. This yields the set of equations
\begin{align}
0 & = \sum_{i,j=1}^n (x_i-\beta)W_{i,j} \label{eq_beta} \\
0 & = \sum_{i,j=1}^n (x_j-\gamma_1)(x_j-\gamma_2)W_{i,j}  \label{eq_gamma1} \\
0 & = \sum_{i,j=1}^n x_i (x_j-\gamma_1)(x_j-\gamma_2)W_{i,j}  \label{eq_gamma2}
\end{align}
which can be solved to yield $\beta$, $\gamma_1$, and $\gamma_2$. 

When fitting polynomials to the $r_{010}$ ratios of models, the errors on the mode frequencies are assumed to be zero, so the covariance matrix and its inverse $\vect{W}$ are the identity, which simplifies the set of equations \ref{eq_beta} to \ref{eq_gamma2}. In particular, $\beta$ reduces the mean of the vector of frequencies $\vect{x}$.

For polynomial fits to the observations, one complication arises. As mentioned in Sect. \ref{sect_fit_r010}, the determinant of the covariance matrix $\vect{C}$ of the observed $r_{010}$ ratios is vanishingly small. As a result, $\vect{C}$ is almost non-invertible and numerical problems arise when trying to invert it. To remedy this, we resorted to a truncated SVD (singular value decomposition) approach. We computed the eigenvalues of matrix $\vect{C}$, further denoted $\lambda_1\geqslant\hdots\geqslant\lambda_n$, and its eigenvectors denoted as matrix $\vect{P}$, where $\vect{P}_k$ is the eigenvector corresponding to the $k^{\rm th}$ eigenvalue. The covariance matrix $\vect{C}$ can thus be rewritten as $\vect{P}^T\vect{D}\vect{P}$, where $\vect{D}$ is the diagonal matrix whose diagonal coefficients are the $\lambda_i$.

The conditioning of matrix $\vect{C}$, given by the ratio $\lambda_1/\lambda_n$, is very large (of the order of $10^6$). We thus 
%project the observables $\vect{y}$ onto a subset of the $m$ eigenvectors $\vect{P}_k$, $k=1,m$ that correspond to the $m$ largest eigenvalues. 
truncate the $n-m$ smallest eigenvalues in order to improve the conditioning. For this purpose, the covariance matrix is approximated by the matrix $\widetilde{\vect{C}} \equiv \widetilde{\vect{P}}^T\widetilde{\vect{D}}\widetilde{\vect{P}}$, where $\widetilde{\vect{D}}$ is a diagonal matrix whose diagonal coefficients are the $m$ largest $\lambda_i$ and $\widetilde{\vect{P}}$ is an $(m,n)$ matrix composed of the $m$ eigenvectors corresponding to the retained eigenvalues. For each star, we suppressed eigenvalues until the impact of suppressing an eigenvalue on the results of the polynomial fit is negligible. We found that suppressing $n-m=5$ eigenvalues is generally enough to ensure the latter condition.

The $\chi^2$ function to be minimized becomes
\begin{equation}
\chi^2 = (\vect{J}\vect{a}-\vect{y})^{T} \widetilde{\vect{W}}  (\vect{J}\vect{a}-\vect{y}) 
\end{equation}
where $\widetilde{\vect{W}} \equiv \widetilde{\vect{P}}^T\widetilde{\vect{D}}^{-1}\widetilde{\vect{P}}$. In this case, the values of $\beta$, $\gamma_1$, $\gamma_2$ required to ensure the independence of the $a_k$ are given by Eq. \ref{eq_beta} to \ref{eq_gamma2} where the coefficients of matrix $\vect{W}$ are to be replaced by those of matrix $ \widetilde{\vect{W}}$. The best-fit coefficients $a_k$ are obtained by doing the same thing with Eq. \ref{eq_amin}.

\end{appendix}

\end{document}